\documentclass[11pt]{article}

\usepackage{geometry}
\usepackage{fullpage}
\usepackage{graphicx}
\usepackage{pdflscape}
\usepackage{multirow}
\usepackage{ragged2e}

\usepackage{dcolumn}
\usepackage{booktabs,calc}
\usepackage{longtable}
\usepackage{array}
\usepackage{paralist}
\usepackage{verbatim}
\usepackage{subfig}
\usepackage{setspace}
\usepackage{amsmath}
\usepackage{amssymb}
\usepackage{amsthm}
\usepackage{natbib}
\usepackage[linktocpage=true, colorlinks=true, linkcolor=blue, citecolor=black]{hyperref}
\usepackage{enumerate}
\usepackage{authblk}
\usepackage{placeins}
\usepackage{multicol}
\usepackage{array}
\usepackage[usenames, dvipsnames]{xcolor}
\usepackage{url}
\usepackage{standalone}
\usepackage{geometry}
\usepackage{rotating}
\usepackage{pdflscape}
\usepackage{flexisym}
\usepackage{breqn}
\usepackage{epsfig}
\usepackage[mathscr]{eucal}
\usepackage{makeidx}
\numberwithin{equation}{section}
\usepackage{psfrag}
\usepackage{afterpage}
\usepackage{pdfpages}
\usepackage{ifpdf}
\usepackage{tikz}

\newcolumntype{J}[1]{>{\justifying\arraybackslash}p{#1}}
\newcolumntype{P}[1]{>{\centering\arraybackslash}m{#1}}
\newcommand{\sym}[1]{\rlap{#1}}

\def\sym#1{\ifmmode^{#1}\else\(^{#1}\)\fi}

\begin{document}

\title{Impact Evaluations in Data Poor Settings: The Case of Stress-Tolerant Rice Varieties in Bangladesh\thanks{Correspondence to \href{mailto:jdmichler@arizona.edu}{jdmichler@arizona.edu}. A pre-analysis plan for this research has been filed with Open Science Framework (OSF): \href{https://doi.org/10.17605/OSF.IO/YE7PV}{https://osf.io/ye7pv}. We gratefully acknowledge funding from the Standing Panel on Impact Assessment (SPIA), the Gates Foundation, and the CGIAR Research Program on Rice. We are especially grateful to Aileen Maunahan, Jorrel Aunario, Pavan Yeggina, and Renaud Mathieu for their work on the early stages of EO data generation, model building, and creating training datasets. We also greatly appreciate the work on Donald Villanueva and Humnath Bhandari in the 2022 data collection effort as well as Donald, Rose San Valentin, and Rowell Dikitanan for initial data cleaning and database construction. This paper has been shaped by conversations with participants at the Center for Environmental Economics and Sustainability Policy (CEESP) seminar at Arizona State University, the $7^{th}$ African Conference of Agricultural Economists in Durban, the $6^{th}$ International Rice Congress in Manila, and the the $32^{nd}$ International Conference of Agricultural Economists in New Delhi. An earlier version of this paper was presented at the AAEA annual meeting in Anaheim. Kyle Emerick, Tom Evans, Ricardo Labarta, Travis Lybbert, Valerie Mueller, Jonathan Sullivan, and
} }

\author[1]{Jeffrey D. Michler}
\author[2]{Dewan Abdullah Al Rafi}
\author[3]{Jonathan Giezendanner}
\author[1]{Anna Josephson}
\author[4]{Valerien O. Pede}
\author[5]{Elizabeth Tellman}
	\affil[1]{\small \emph{Department of Agricultural and Resource Economics, University of Arizona}}
	\affil[2]{\small \emph{Department of Agricultural and Applied Economics, University of Georgia}}
	\affil[3]{\small \emph{International Rice Research Institute (IRRI)}}
	\affil[4]{\small \emph{Earth Intelligence Lab, Massachusetts Institute of Technology (MIT)}}
	\affil[5]{\small \emph{Nelson Institute for Environmental Studies, University of Wisconsin-Madison}}

\date{June 2025}
\maketitle

\thispagestyle{empty}

\begin{center}\begin{abstract}
     \noindent New technologies are sometimes introduced at times or in places that lack the necessary data to conduct a well-identified impact evaluation. We develop a methodology that combines Earth observation (EO) data and advances in machine learning with administrative and survey data so as to allow researchers to conduct impact evaluations when traditional economic data is missing. To demonstrate our method, we study stress tolerant rice varieties (STRVs) first introduced to Bangladesh 15 years ago. Using EO data on rice production and flooding for the entire country, spanning two decades, we find evidence of STRV effectiveness. We highlight how the nature of the technology, which is only effective under a specific set of circumstances, creates a Goldilocks Problem that EO data is particularly well suited to addressing. Our findings speak to the promises and challenges of using EO data to conduct impact evaluations in data poor settings.
	\end{abstract}\end{center}

	{\small \noindent\emph{JEL Classification}: C51, C81, D83, O13, Q12, Q54 }

\newpage
\onehalfspacing


\section{Introduction} \label{sec:intro}

Obtaining a sufficient amount of relevant data has always been a challenge in answering economic questions, whether theoretical or empirical. Karl Marx struggled to obtain firm-level production data for use in developing his theory of value \citep{Marx67}. John Maynard Keynes lamented the lack of sufficient data to definitively disprove Arthur Pigou's theory of unemployment \citep{Keynes16}. Simon Kuznets, in part, won his Nobel Prize for creating standardized and regularly collected data on national accounts \citep{Rockoff20}. One reason why the earliest advances in econometrics were made in the field of agricultural economics was because George F. Warren convinced the U.S. Census Bureau to conduct a separate national census of agriculture at more frequent intervals than the decennial census \citep{Stanton07}. This meant that agricultural economists had much more data for empirical research than those working on macro or other micro questions.\footnote{Nearly as important as having any data is having sufficient data. Well know empirical findings, such as the Philips Curve \citep{Phillips58} and the deterrence effect of capital punishment \citep{Ehrlich75}, have proven much less robust than initially thought once additional data were added to the analysis \citep{HazellEtAl22, Nagin13}.}

The last quarter century has seen an explosion in the availability of data, both economic and otherwise. Data, in combination with the personal computer, the internet, mobile information, and communication technology, has created a data rich world. This plethora of data has allowed economists to answer a host of new questions on anything from the historic effects of colonial institutions \citep{Dell10} to the current effects of misinformation \citep{AllcottGentzkow17} to the future effects of climate change \citep{DescheneGreenstone07}. Economists have also become adept at integrating traditional socioeconomic data with new types of data from a variety of sources. In particular, remotely sensed Earth observation (EO) data has become a favorite for use in economic research \citep{DellEtAl14, DonaldsonStoreygard16}. Economists have used EO data to examine questions about economic growth, agricultural productivity, land use, population growth, poverty, child mortality, and governance \citep{BenYishayEtAl17}.

Yet the recency of so much of this data means that there is at least one place that remains relatively data poor: the past. As \cite{BurkeEtAl21} show, the increase in the spatial and temporal resolution of EO data has primarily occurred since 2010, with the most substantial improvements ($<10$m, daily) occurring in the last decade. While EO data has been used to answer many research questions about recent or contemporaneous events, its usefulness in answering economic questions about the (relatively recent) past has, thus far, been limited. The requirements for ground truth or training data to improve the accuracy of EO derived outcomes also places perceived limits on where and when EO data can be used to answer a research question.

In this paper, we develop a methodology to overcome the recency bias in high quality EO data so that it can be used to answer economic research questions in settings that remain - and are likely to remain - data poor. We do this in the context of a long-term, large-scale impact evaluation on the effectiveness of stress tolerant rice varieties (STRVs) in mitigating yield loss to flooding in Bangladesh. The flood-tolerant rice variety, Swarna-Sub1, was first released in Bangladesh in 2010 and other varieties quickly followed \citep{MackillEtAl12}. While a randomized control trial (RCT) was conducted in 128 villages in Odisha, India in 2011 \citep{emerick_technological_2016} and household panel data on varietal-specific rice production exists in South Asia starting in 2014 \citep{rms-2017}, no data exist that would allow for a before/after comparison of the impacts of adoption on a large-scale. The relatively long time since the introduction of varieties with the \emph{Sub1} gene means that the high spatial and temporal resolution EO data now being used in impact evaluations \citep{CarlsonEtAl17, JayachandranEtAl17, JainEtAl19} also does not exist for use as a baseline.

Further impoverishing the datasetting is the fact that STRVs, like many new agricultural technologies, have a stochastic treatment effect. For the technology to work and, by extension, be identifiable from space, it requires not too much flooding but also not too little flooding. This creates a Goldilocks Problem in which the treatment effect of the technology is only be observable under a specific set of circumstances. For varieties with the \emph{Sub1} gene, that set of circumstances is complete submergence for longer than five days but less than 20. Outside that 5-20 day window, STRVs have yields statistically indistinguishable from non-STRVs. Thus, in any existing data for use in an impact evaluation, one would need precise information about the location and duration of flooding. Given the now well-known problem of measurement error in self-reported agricultural data \citep{Wollburg21, AbayEtAl23}, households may be unable to retrospectively report on flooding with the precision necessary to capture the agronomically relevant level of inundation required for STRVs to be effective. As we show using Monte Carlo simulations and agronomic field trial data, identifying these higher order treatment effects is highly sensitive to classical measurement error.

In order to conduct our impact evaluation in this data poor environment we combine data from a variety of sources (EO, administrative, household survey) and leverage recent advances in machine learning (ML). Our first task is to build maps for where rice is grown in Bangladesh. The standard approach is to collect GPS data on rice and non-rice plots and use that as training data for an ML algorithm to predict rice area at a larger scale \citep{burke_satellite-based_2017, AzzariEtAl21}. While we have this type of ground truth data for 2021 and 2022, it does not exist for previous years. To overcome this lack of ground truth data, we hand-annotate thousands of Google Earth images to build a training dataset going as far back as 2002. Having identified rice area, we next build maps of where rice was flooded. The traditional method for identifying floods is to use an existing databases of large-scale floods based on optical sensors that might be blocked by clouds \citep{tellman-2021} or to handcraft flood measures using EO data \citep{GuiterasEtAl15, ChenEtAl17}. Both methods miss many small flood events or have cloud blockage and thus under-measure the extent of flooding \citep{SaunderEtAl23}. Substantial improvements in accuracy can be achieved using ML methods similar to those used in rice mapping but here again the problem is a lack of ground truth data. To overcome this, we use radar data from Sentinel-1, a satellite with high spatial and temporal resolution not blocked by clouds, launched in 2014, as training data for a Convolutional Neural Network - Long Short-Term Memory (CNN-LSTM) model that then predicts past flooding in lower spatial resolution historical data using the MODIS sensor and topographic variables \citep{GiezendannerEtAl23, ThomasEtAl23}.

Our primary approach to estimating the impacts of STRVs combines these rice and flood maps with the EO derived enhanced vegetative index (EVI), a common proxy for yields, and administrative data on STRV seed release. This produces a panel dataset that covers the entire country for the period 2002 through 2021. We use a variety of econometric methods (event study, difference-in-difference, two-way fixed effects), each with its own strengths and weaknesses regarding identifying assumptions. In order to highlight the severity of the Goldilocks Problem in our setting, we vary the level of inundation required to be considered a flood. And we vary the length and duration of the window in which a flood occurs. This inherently agnostic approach to what constitutes a flood produces a large number $(>600)$ of flood metrics that are potentially ``just right'' for identifying the treatment effects of STRVs. While a majority of these flood metrics are too little or too much flooding to identify significant impacts, we do find a set of flood metrics that produce positive and significant results. These Goldilocks floods produce consistent results across variation in how we measure EVI, our model specification, and if we use households panel data, though not to different spatial aggregations of the data.

We make three contributions to the literature. First, we contribute to the growing set of geospatial impact evaluations. A geospatial impact evaluation uses georeferenced EO data to measure either the intervention, the outcome, or both \citep{BenYishayEtAl17}. This geospatial data can be used as part of an RCT, as in \cite{JayachandranEtAl17}, \cite{JainEtAl19}, and \cite{HuangEtAl21}, but is more frequently used as part of a quasi-experimental approach to establishing the causal effects of an intervention. Geospatial impact evaluations have been used to study the impacts of governance \citep{BurgessEtAl12}, anti-poverty programs \citep{Alix-GarciaEtAl13}, conditional cash transfers \citep{FerraroEtAl20}, and the impact of insecticide treated bed nets \citep{DolanEtAl19}, to name a few. Many of these geospatial impact evaluations combine EO data with survey data on specific individual outcomes in specific areas. However, a growing proportion of studies, like \cite{BurgessEtAl12} and \cite{FerraroEtAl20}, rely solely on EO and administrative data. In this paper, our primary results relying solely on EO and administrative data. We then show that these results are robust to using a combination of EO and household survey data. We believe this multi-pronged approach provides increased reliability in our estimates by providing evidence from a variety of data sources at a variety of scales.

Second, we make a methodological contribution to the broader literature that looks to combine EO data with socioeconomic data. Both the availability and quality of EO data is changing rapidly. So too is the sophistication of model building that uses both optical and radar EO data to predict specific outcomes of interest. Early economic studies that integrated EO data typically relied on what \cite{BurkeEtAl21} term ``handcrafted features'' for use in shallow predictive models. One commonly used example of a handcrafted metric is the Modified Normalized Difference Water Index (MNDWI) \citep{GuiterasEtAl15, ChenEtAl17, ChenMueller18}. The MNDWI is simply a ratio of the differences in surface reflectance such that positive values are inferred to be water and negative values are inferred to be land \citep{Xu06}. An example of a shallow predictive model is using a simple linear regression to predict values of pixels that might be missing or obscured by clouds \citep{WeissEtAl20}. We leverage recently developed techniques from ML to fuse optical and radar data, including models that use the spatial and temporal structure of the data, to improve predictive ability and reduce noise in our estimates of where rice is grown and where flooding occurs. 

Finally, we add to the literature on the impact of agricultural technology by studying a stochastic technology that generates higher order treatment effects. The aim of the first-generation of Green Revolution technologies was to improve on the low yields of landrace varieties \citep{EvensonGollin03}. Newer generations of high yielding varieties now also embody genetics to allow the plant to tolerate biotic and abiotic stresses \citep{hossain_adoption_2006}. This creates a challenge for the researcher trying to measure impact, as the effectiveness of these technologies only manifests within a very specific set of circumstances that is often a function of a stochastic event. Identifying the treatment effects of these stochastic technologies therefore is highly sensitivity to mismeasurement, misallocation, or mis-classification. The existence of measurement error in both self-reported data \citep{Wollburg21, AbayEtAl23} and in EO data \citep{Alix-GarciaEtAl23, JosephsonEtAl25}, means that there are many settings in which data might exist but not at a sufficient level of precision to be useful for capturing higher order treatment effects. The Goldilocks Problem created by stochastic technologies expands the scope and scale of data poor settings. Our paper provides new tools to improve the precision of existing data so as to allow researchers to answer these type of questions.


\section{Study Context} \label{sec:lit} 

\subsection{From the Green Revolution to the Goldilocks Problem} \label{sec:goldi}

Green Revolution technologies played a critical role in improving food security and reducing poverty in many developing countries \citep{EvensonGollin03}. Early in the Green Revolution, many technologies were developed for and disseminated in agronomically favorable environments, bypassing environments where biotic and abiotic stresses resulted in low or uncertain yields \citep{KhushVirk00, GollinEtAl21}. These earlier technologies often embodied unconditional yield effects. Simply planting them resulted in higher yields under existing growing conditions and farming practices, though even greater yields could be achieved when high yielding seeds were bundled with other modern inputs.

Stress tolerant seed varieties, including stress tolerant rice varieties (STRVs), were developed to address the rising challenges of climate change-induced stresses, such as salinity, extreme temperature, drought, or submergence \citep{lichtenthaler-1998, sevanthi_2019}. In South and Southeast Asia, where seasonal flooding is critical to rice cultivation, farmers face a substantial risk of crop loss due to early and sustained flooding \citep{mishra_abiotic_2015}. To address this, researchers isolated submergence tolerant traits (the \emph{Sub1} gene) in wild rice \citep{XuMackill96} and cross-bred it with a high yielding variety (Swarna) to create Swarna-Sub1 \citep{yamauchi_2013}. 

\begin{figure}[!htbp]
	\begin{minipage}{\linewidth}		
		\caption{Yield of Swarna and Swarna-Sub1 in Field Trials}
		\label{fig:field_trial}
		\begin{center}
			\includegraphics[width=\linewidth,keepaspectratio]{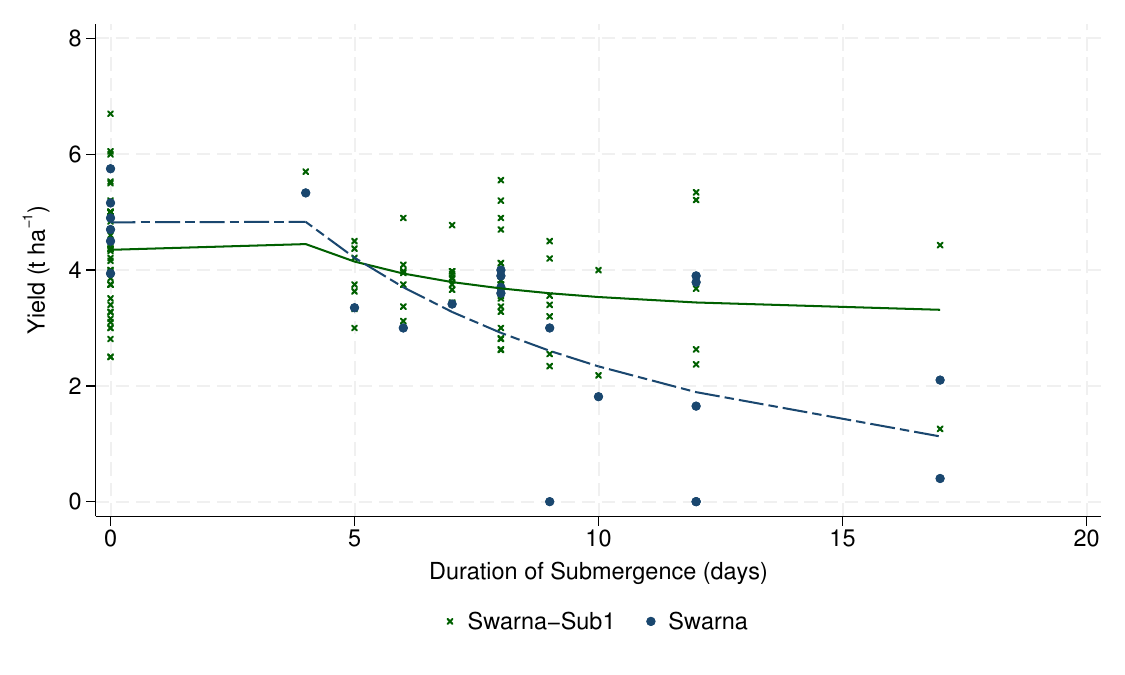}
		\end{center}
		\footnotesize  \textit{Note}: Data are from field trials testing Swarna and Swarna-Sub1 under variable duration of submergence conducted in the Philippines and India between 2005 and 2011. We add non-linear lines of best fit to the data points by variety type. In total, the figure presents 133 observations as reported in three different publications \citep{SarkarEtAl09, SinghEtAl09, SinghEtAl11} and unpublished data from the Stress Tolerant Rice for Africa and South Asia (STRASA) project.
	\end{minipage}	
\end{figure}

Swarna-Sub1 has no yield penalty in the absence of flooding. If and when flooding occurs, Swarna-Sub1 can survive complete submergence anywhere from between five and 20 days, in up to 25 cm of water, with little yield penalty. But, while Swarna-Sub1 and other \emph{Sub1} varieties outperform non-STRVs, the yield effect is not unconditional. Figure~\ref{fig:field_trial} traces the yield effect in relation to the duration of flooding. There is a clear region that can be considered the ``just right'' amount of flooding for \emph{Sub1} and other STRVs to produce large yield gains. But what exactly is ``just right'' flooding varies in both the start and end day of the flood window (length) as well as the days within that window (duration). Some studies report the window as seven to 15, others as five to 17, and still others five to 20. This means there is disagreement about not only when the window starts (five, seven, etc.) and ends (15, 20, etc.) but the number of days that make up that window. Thus, the effectiveness of STRVs is a function of an essentially random event (the length and duration of flooding) making STRVs a stochastic technology and meaning attempts to quantify their impact requires measuring a higher order treatment effect. If flooding is outside some window of day that varies between five and 20, then we would expect STRVs to have a treatment effect equal to zero.\footnote{Additional details on \emph{Sub1} varieties is available in Online Appendix~\ref{sec:Sub1_agron_app}.}

\subsection{Assessing Sensitivity} \label{sec:MC}

The manifestation of the Goldilocks Problem for STRVs means that there is variation around what flooding may be ``just right.'' Yields will vary based on the duration of submergence but also the start and end of the flood window. Beyond the length of time in the flood window and the size of the flood window, there are issues regarding the timing of floods in the growing season and the intensity or depth of flooding. This variation around what exactly characterizes a Goldilocks flood means that the researcher has a large set of potential ways to define a flood.\footnote{To avoid this problem, we pre-specified our analysis and archived a pre-analysis plan on Open Science Foundation (OSF) \citep{PAP}. Results from the original, pre-specified approach are publicly available in our populated pre-analysis plan \citep{popPAP}. None of the results in this paper were pre-specified, meaning all results can be considered exploratory. Information on our rationale for the pre-analysis plan as well as its implementation can be found in Online Appendix~\ref{sec:framework}.}

To demonstrate the degree to which measuring STRV impact is sensitive to capturing just the right amount of flooding (and to mismeasurement in general), we conduct a series of Monte Carlo simulations using experimental data that forms the basis of \cite{dar_flood-tolerant_2013} and \cite{emerick_technological_2016}.\footnote{Though the two papers rely on the same source of data, the regressions run in \cite{dar_flood-tolerant_2013}, which closely match tests for efficacy in the agronomic field trial literature, are not the specifications used in \cite{emerick_technological_2016}. We are not replicating any specific result in \cite{emerick_technological_2016}. Rather we are using the full two years of data to reproduce results that match \cite{dar_flood-tolerant_2013} in magnitude, sign, and significance.} We start by calculating the mean and standard deviation of flood duration in the RCT data ($\mu = 6.03$; $\sigma = 4.99$) . Next, we draw, with replacement, from a normal distribution with mean zero and standard deviation equal to one percent of the standard deviation in the data $(0.01\sigma = 0.0499)$. For each observation of flood duration, we add one of these randomly drawn values and then estimate the impact of STRVs during flooding by regressing the yield data on indicators for variety type, the new duration of flooding, and the interaction of variety with the new duration of flooding. We repeat this process 10,000 times. We then increase the noise by one percent, so that the new distribution that we draw from has mean zero and a standard deviation equal to two percent of the original standard deviation $(0.02\sigma)$. We add this new noise level to the original data and rerun the regression for 10,000 draws from this new distribution. We conduct this process at one percent intervals from $0.00\sigma$ up to $0.20\sigma$, meaning we run 2,000,000 simulations.

\begin{landscape}
\begin{figure}[!htbp]
	\begin{minipage}{\linewidth}		
		\caption{Monte Carlo Simulations of Sensitivity in \cite{dar_flood-tolerant_2013} and \cite{emerick_technological_2016}}
		\label{fig:mc}
		\begin{center}
			\includegraphics[width=.4\linewidth,keepaspectratio]{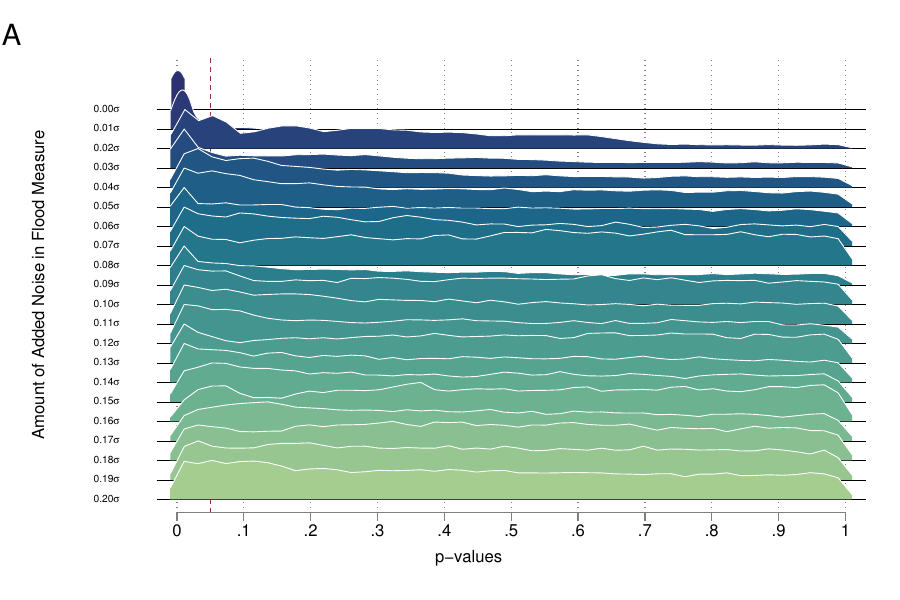}
			\includegraphics[width=.4\linewidth,keepaspectratio]{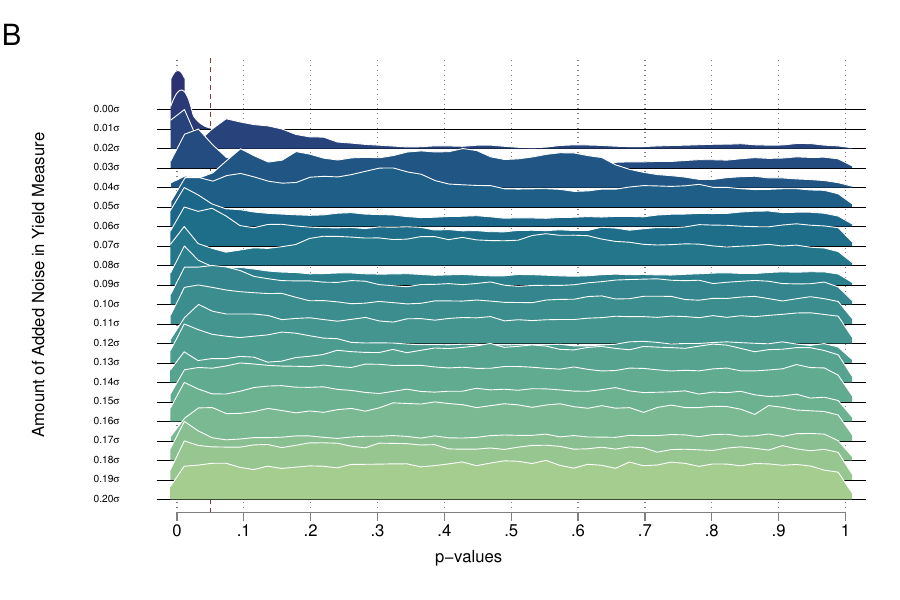}
			\includegraphics[width=.4\linewidth,keepaspectratio]{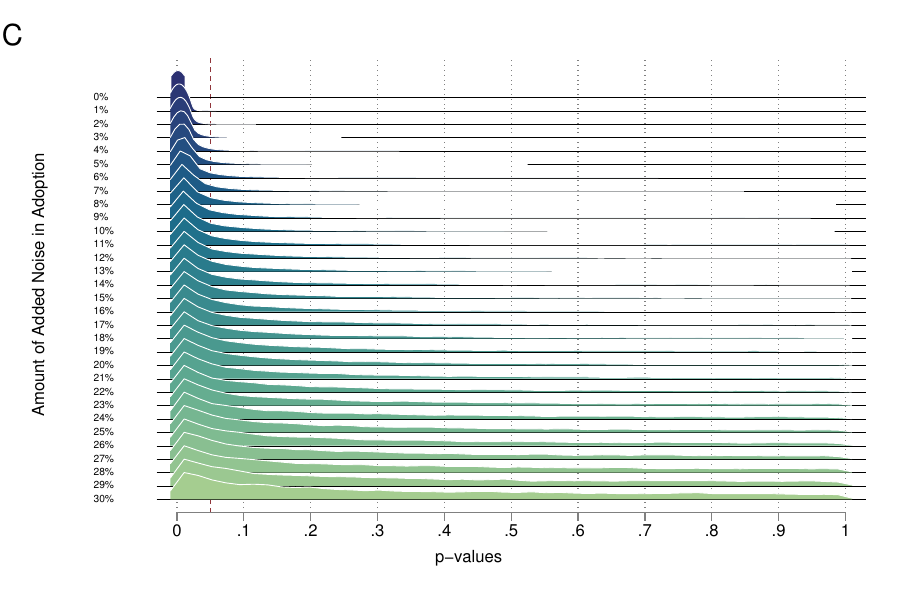}
		\end{center}
		\footnotesize  \textit{Note}: The figure presents the distribution of $p$-values from Monte Carlos simulations of regressions run to test the impact of STRVs and flooding on yields. The top two panels present distributions from adding fractions of noise to either flood (Panle A) or yield (Panel B). The fraction of noise is drawn from a normal distribution with mean zero and the standard deviation calculated as $x\% \sigma$, where $x\%$ ranges from 0 to 20 in one percent increments and $\sigma$ is the true standard deviation. At each increment, we conduct 10,000 simulations. Panel (C) presents the distribution of $p$-values from mis-classifying a random subset of adopters and non-adopters. We calculate the percentage to be re-classified based on the true share of adopters in the data and then repeat this at one percent intervals up to $50\%$. At each increment we conduct 10,000 simulations. Data for simulations come from \cite{dar_flood-tolerant_2013} and \cite{emerick_technological_2016}.
	\end{minipage}	
\end{figure}
\end{landscape}

We conduct the Monte Carlo simulations first for flood duration and then, independently, for yield. Results for both are presented in the top two panels of Figure~\ref{fig:mc}, which graph the distribution of $p$-values on the interaction of STRV adoption and flooding for each level of added noise. The vertical dashed red line marks $p < 0.05$. In the top row for both panels, all $p$-values from the 10,000 simulations are statistically significant. This is because we are just replicating \cite{dar_flood-tolerant_2013} 10,000 times. When we add $0.01\sigma$ to the true flood data, $99\%$ of $p$-values are still significant. But adding just two percent of the true standard deviation as noise to the flood data results in only $14\%$ of regressions producing significant results. Adding any more than six percent of the true standard deviation to the true flood data results in less than $10\%$ of $p$-values being significant - the same amount of $p$-values one would expect to be significant due to random chance with a $90\%$ critical value. Results are similar for yield, but the loss of significance happens slightly slower, with $26\%$ of $p$-values significant after adding $0.02\sigma$ to the flood data. 

We follow a similar process for simulating mis-classification in the adoption data. However, as adoption is a binary indicator, it is not useful to use the value of the standard deviation to add noise to the data. Rather, we calculate the percentage of adopters $(42\%)$ in the data and then re-assign adopters and non-adopters at random. So, for one percent of the data we draw at random four households, determine if they are an adopter or non-adopter, and then switch their adoption status. We repeat this reassignment $10,000$ times. As before, we start the process at zero, so that we replicate the original results, and then move upwards at one percent intervals until we are re-assigning half of all observations.

Results from the mis-classification simulations are graphed in the bottom panel of Figure~\ref{fig:mc}. It is clear that the results are much more robust to mis-classification of adoption status. Mis-classifying up to five percent of observations has no meaningful impact on the outcomes. Mis-classifying $14\%$ of outcomes results in about half of all regressions producing significant results. Only if one mis-classified $37\%$ of observations does the share of significant $p$-values fall below the ten percent threshold, where their significance could be wholly ascribed to random chance.

The take-away from this exercise is that even results on the efficacy of STRVs using experimental data can be very sensitive to mis-measurement (though less so to mis-classification). Perturbing the true flood data by adding to each value a random number drawn from a distribution $\mathcal{N}(0,0.10)$, or two percent of the standard deviation, produces non-significant results in $86\%$ of regressions. Significance goes away adding slightly more noise to the yield data. In terms of what this means for our study context, the Monte Carlo simulations demonstrate the need for not only sufficient data, but sufficiently precise data, if one is to identify the stochastic treatment effects of STRVs. Not only do we need data to establish a pre-2010 baseline for our study, but we need that data to contain precise information on flooding, something \cite{dar_flood-tolerant_2013} and \cite{emerick_technological_2016} were able to collect in a small-scale experimental setting, but which is rare in nationally representative observational data or census data. Only through highly refined EO data could we ever hope to measure the relevant flooding that occurred more than 20 years ago.

\subsection{A Data Poor Setting}

Data exists on both the adoption and impact of STRVs, however no data exists that would allow us to study the long-term, large-scale impact of adopting STRVs on rice production. Previous data collection efforts, including an RCT in India \citep{dar_flood-tolerant_2013} and a nationally representative survey in Bangladesh \citep{KretzschmarEtAl18}, provided useful information on STRVs. But, they each provide a circumscribed picture that is limited by the purpose and methodology of each program. While the RCT has strong internal validity for measuring the impact of STRVs, it has weak external validity as it focused on a small set of villages in a single Indian state and as treated farmers were given the seeds. And, while the survey provides strong external validity for the breadth of adoption across Bangladesh, it has weak internal validity as it started collecting observational data four years after the initial introduction of STRVs.

With all of this taken together, it is clear that we are operating in a data poor environment. No historical data exist that provide the necessary information: 1) a panel, 2) observations before 2010 and well after 2010 (to allow time for dissemination and adoption), 3) representativeness at the national level or at least representativeness of rice cultivation, and 4) precise data on length of flooding, yields, and rice varieties grown. And current or future data collection efforts will not be useful as STRVs were introduced over a decade ago, meaning no relevant baseline can be constructed.\footnote{There are a number of nation-wide panel surveys of households in Bangladesh that might appear to be good candidates for filling this data gap. In Online Appendix~\ref{sec:data_app} we discuss a variety of available datasets and why they are insufficient to answer our research question.}

As an alternative, we use EO data to reconstruct the past. EO data provides a repeated time series for a given location (panel data), provides observations long before and after 2010, provides coverage of an entire country, and provides spectral information that can be used to identify floods and yields. What EO data lacks is information of varieties grown, which we can proxy for with administrative data on dissemination efforts. We can also enrich and validate the EO-based analysis using existing, if incomplete, survey data. Although, there is no way to generate the data one would want if one were planning a prospective impact evaluation of a new agricultural technology, our method provides a way for researchers to answer important questions about the impact of technology adoption in data poor settings, particularly the past.

As a test case for our method, we focus on the adoption of Swarna-Sub1 and subsequent submergence tolerant varieties (which we collectively term STRVs) in flood-prone rice-growing regions of Bangladesh during the \emph{Aman} (wet) season. In the \emph{Aman} season, farmers face more crop loss due to natural disasters like flooding \citep{BairagiEtAl21, khan_2020}. The intuition behind our identification strategy is simple: for a given rice-growing location, if no flooding occurs, EVI provides a consistent signal of crop growth up until harvest. For that location, prior to 2010, a Goldilocks flood of amount $x$ would cause a consistent signal response in which EVI values abruptly drop. After the introduction of STRVs, if there is no flooding, there would be no change in the EVI signal in that location relative to before 2010. And, if there is flooding of $x$ amount and no adoption of STRVs, there would again be no change in the signal response in that location relative to before 2010. Further, if there is flooding and adoption of STRVs, but that flooding is not $x$, meaning it is outside the relevant flood window, there would also be no change in the signal response. Only if there is the ``just right'' amount of flooding $x$ and adoption of STRVs would we expect a change in the signal response. If we see such a change in the signal response in a given location to a Goldilocks flood post-2010 relative to a Goldilocks flood pre-2010, the only explanation would be the adoption of STRVs, as no other technology to reduce the negative impact of flooding on rice was concurrently introduced.


\section{Data}\label{sec:data}

Data availability is the key challenge to establishing the long-term, large-scale impacts of agricultural technology in international development. As \cite{BurkeEtAl21} point out, the relative infrequency of in-person surveys in many developing countries can potentially be overcome using remotely sensed EO data. In the case of STRVs in South Asia, where no comprehensive baseline data was collected, EO data allows us to look back into the past and reconstruct outcomes before and after the technology was disseminated.

Construction of our EO dataset starts by identifying where rice was cultivated in Bangladesh. From the time series of rice maps, we extract EVI values. We then construct flood maps for areas where rice was grown. All of this pixel-level data is aggregated to the district-level (there are 64 districts in Bangladesh) and then combined with district-level administrative data on seed dissemination. In this section, we provide a non-technical, intuitive summary of the methods used to generate the EO data.\footnote{Online Appendix~\ref{sec:rs_app} provides the technical details sufficient to implement our procedures in alternative contexts.}


\subsection{Rice Area Mapping}

As early as \cite{XiaoEtAl06}, rice pixels have been identified by detecting surface water followed by a sudden growth of vegetation. The detection of surface water is critical for identification as it represents agronomic flooding, which is unique in rice cultivation and the most common practice for cultivating rice in Bangladesh.\footnote{In recent years there has been a push by IRRI and other development organizations to promote alternate wetting and drying (AWD) and/or direct seeded rice (DSR), both of which would substantially reduce or completely eliminate the need for agronomic flooding. However, promotion of both methods has mainly focused on use during the \emph{Boro} (dry) season when groundwater irrigation is necessary for cultivation. To date, there is a dearth of reliable data to measure adoption of either technology in Bangladesh. See \cite{AlauddinEtAl20} for a recent overview of AWD in Bangladesh.} In early applications of remote sensing, which relied on handcrafted, indice-based features, agronomic flooding is signaled when the value of vegetation indices (e.g., the Enhanced Vegetation Index [EVI]), are lower than the value of water indices (e.g., Land Surface Water Index [LSWI]). The sudden growth of vegetation is determined if the vegetation index reaches half of the maximum value within 40 days immediately after agronomic flooding.

If our goal was to simply construct contemporaneous rice area maps, then we could follow more recent methods similar to \cite{lobell_eyes_2020}. This would involve taking GPS measures of the area of a number of rice and non-rice plots and using them as ground truth data for training a machine learning classification algorithm. The algorithm could then use the ground truth training data to predict rice and non-rice areas for the rest of the country for that year. However, our goal is to build rice area maps for the 20 year period of 2002-2021, providing us with a large $t$ in both the pre- and post-STRV release period. No traditional ground truth data exists for predicting where rice is grown 20 year ago.

To address this data deficit, we take two approaches designed to complement one other. Our first approach involves collecting traditional ground truth data. In 2021, we conducted a small survey to collect ground truth data from households, including plot GPS coordinates, crop varieties, and planting dates $(n = 787)$. This effort was augmented in 2022 by a third round of the Rice Monitoring Survey (RMS), which included household socioeconomic information plus plot production information and GPS locations $(n = 3,219)$. All plot location data was collected following guidance in \cite{AzzariEtAl21} and allows for creating rice area maps in 2021 and 2022 following standard procedures as in \cite{lobell_eyes_2020}.

Our second approach re-constructs ``ground truth'' through visual inspection of high resolution Google Earth images. This visual reconstruction using optical imagery is particularly useful for capturing difficult to measure activities, like illegal fishing \citep{ParkEtAl20}. In our case, it is difficult to capture rice cultivation spanning back two decades. To build our historical reference dataset, we selected three districts in Bangladesh (Barisal, Kurigam, and Rajshahi) that experience different hydrological characteristics and thus different distributions of riceland. We derived grids from the MODIS pixels and selected a stratified random sample of 150 points (75 points each for rice and non-rice areas) in each district for a total of 450 points (see Panel A of Figure~\ref{fig:rice_flood}). This was completed for the nine years for which Google Earth imagery exists: 2002, '04, '06, '09, '15, '16, '18, '19 and '21. The MODIS grids were overlaid on the Google Earth imagery as the basis for visually determining the land cover of each grid for the particular season being validated. MODIS pixels were categorized as rice if, in the Google Earth imagery, more than $70\%$ of the grid was visually determined to be rice. Pixels were categorized as non-rice otherwise.

We then classify MODIS pixels for all years for the entire country as rice/not-rice by training a random forest (RF) algorithm on the ground truth data. The input data to the model is a combination of surface reflectance and EVI extracted from the MODIS eight- and sixteen-day composites at 500m resolution, as well as topographical indicators including elevation and slope, derived from the Digital Elevation Model (DEM) product FABDEM~\citep{hawker2022}. The RF model consists of 1,000 trees each with five leaves. We use a standard leave-one-out cross validation scheme in which we rotate the districts used as training, leaving one district out each time. This ensures generalizability of the model over space. To ensure generalizability over time, we reserve 2020 as an out-of-sample test set. Our model is accurate at predicting rice and non-rice pixels in 2020 $82\%$ of the time.\footnote{Online Appendix~\ref{sec:rs_app:rice_mapping} contains more details on the methods, including results of accuracy tests for the model.} The end result is a time series of rice and not-rice pixels for the entire country for all years spanning 2002 to 2022.


\subsection{Flood Area Mapping}

As with the rice maps, building reliable flood maps that look into the past raises data availability challenges. Until recently, two common approaches to using EO data for generating flood maps before 2015 (when public radar imagery and higher spatial resolution data from Sentinel-1/2 became available) was either to rely on the historic flood data base curated by the Dartmouth Flood Observatory (DFO) \citep{Brakenridgend} or to use MODIS to construct the Modified Normalized Difference Water Index (MNDWI) \citep{GuiterasEtAl15, ChenEtAl17, ChenMueller18}. 

The archive from the DFO, which extends back to 1985, has been highly cited but suffers from several well-known data deficiencies. Most importantly, it only includes large flood events that members of the observatory were able to identify in news stories and governmental sources. Many small floods, that might destroy crops for some farmers but result in no deaths and generally do not make news, are not covered in the archive. Recently, \cite{tellman-2021} develop the Global Flood Database (GFD), which uses 12,719 scenes from Terra and Aqua MODIS sensors to produce daily images of flooding at 250m resolution from 2000 to 2018. The GFD builds on the flood events reported in the DFO data and is validated on 30m resolution Landsat data. While the GFD is based on MODIS 250m, it provides only a binary flood and not-flood indicator at that resolution and only for events cataloged by the DFO.

In terms of using MNDWI and MODIS, the MNDWI is a threshold method for identifying flooding developed by \cite{Xu06}. Water and non-water are identified based on surface reflectance and a pixel is defined as water if MNDWI is above zero. As with the handcrafted feature for identifying rice, this handcrafted flood feature has been popular as it requires no ground truth data and is simple to construct. However, as \cite{BurkeEtAl21} point out, methods that combine EO data with ground truth data and methods from deep learning can make substantial improvements to accuracy relative to previous approaches, like EO-only determined thresholds. Furthermore, using only MODIS, or any optical sensor, to map floods can cause large underestimations in inundation due to cloud cover. Thus, current remote sensing methods to measure flooding are likely to be highly imprecise in predicting the vary variable critical to overcoming the Goldilocks Problem in STRVs.

Since 2017, Sentinel-1 has provided a numbers of improvements over MODIS and the DFO data and it addresses the issues of low resolution and underestimation. Specifically, Sentinel-1 comes at a spatial resolution of 10m and uses a synthetic aperture radar (SAR) sensor, which allows it to penetrate cloud cover. The limitation to Sentinel-1, for our purposes, is its relative recency. To overcome the lack of historic Sentinel-1 data, we train a Convolutional Neural Network – Long Short-Term Memory (CNN–LSTM) on Sentinel-1 fractional flood data to predict historic MODIS satellite data and then project back into the past, prior to the launch of Sentinel-1. The method is described in detail in \cite{GiezendannerEtAl23}.\footnote{Online Appendix~\ref{sec:rs_app:flood} provides additional details and context about the model and tests of the model against traditional methods.} The main reason this model is able to overcome underestimation of inundation by using only MODIS imagery is because of the temporal nature of the CNN-LSTM. The model is trained not just on concurrent MODIS and Sentinel-1 imagery, but on the 10 weeks preceding the observation as well as topographic variables. This allows the LSTM portion of the model to learn the flood dynamics in Bangladesh and accurately interpolate when a cloud covers a pixel. It does this because the model has learned, for example, what a rising flood wave looks like in time. It has also learned in space that if pixel $i$ is flooded and it is at a similar or higher elevation to neighboring pixel $j$, then pixel $j$ is also likely flooded. Using the 2022 flood in Sylhet, Bangladesh as a test case, \cite{SaunderEtAl23} show that handcrafted MODIS measures, like MNDWI, indeed fail to predict major floods but that our ML approach with MODIS data replicates the inundated area and temporal flood peak captured by Sentinel-1 with $72\%$ accuracy.

The workflow for our flood model is summarized in Panel B of Figure~\ref{fig:rice_flood}. To detect water in Sentinel-1 data we use a dynamic thresholding algorithm tailored to Bangladesh by \cite{ThomasEtAl23}. This produces a binary flood indicator for the 10m pixel. We then upscale the 10m pixel to create a nearly continuous fractional index at 500m resolution, which matches the resolution in MODIS's eight-day composite images. Next we feed this Sentinel-1 derived fractional index, along with hydrologically relevant topographical measures (elevation, slope, and height above nearest drainage), and MODIS 500m eight-day composite data into a CNN-LSTM to predict fractional flood area all the way back to 2002. The data (features) from a single time step are fed into a first CNN which produces a single pixel of spatially contextualized data. This is repeated for each time step $(t)$ to produce a time series of length $T$ for that pixel. We then feed a time series of length $T-1$ into the LSTM to produce a single temporal value. The output of the LSTM is then combined with the output of the CNN for time $T$ and feed into a second CNN to produce a fractional value for the extent of flooding in a given pixel for a given time step. As with the RF model for rice area, the CNN-LSTM is trained using leave-one-out cross validation, where a time step is withheld for validation one at a time. While \cite{SaunderEtAl23} achieves $72\%$ accuracy in their context, our model is accurate $90\%$ of the time. The result is a time series of the fraction of a 500m pixel that was flooded, covering all of Bangladesh from 2002 to 2022.


\subsection{Creation of District-Level EO Dataset} \label{sec:EOdata}

\begin{figure}[!htbp]
	\begin{minipage}{\linewidth}		
		\caption{Workflow for Building EO dataset}
		\label{fig:eo_method}
		\begin{center}
			\includegraphics[width=\linewidth,keepaspectratio]{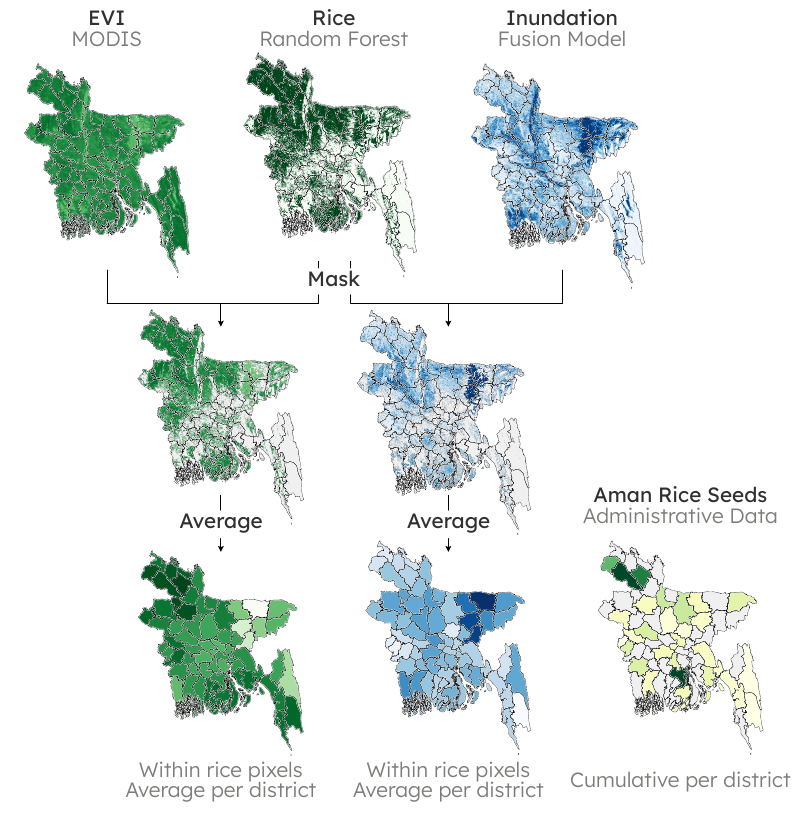}
		\end{center}
		\footnotesize  \textit{Note}: The figure shows the workflow used to construct the various maps (rice area, flooding) based on EO data as well as how these maps are aggregated and combined to create the district-level panel dataset.
	\end{minipage}	
\end{figure}

The creation of the district-level panel data follows the workflow summarized in Figure~\ref{fig:eo_method}. We start by using the rice area maps for \emph{Aman} season for each year to mask the flood maps, so that we only consider inundation levels for rice area. We also use the rice maps to mask EVI data from MODIS so that we only consider EVI values for rice area.

\begin{figure}[!htbp]
	\begin{minipage}{\linewidth}		
		\caption{Before/After Introduction of STRVs with Different EVI Measures}
		\label{fig:evi_ba}
		\begin{center}
			\includegraphics[width=.49\linewidth,keepaspectratio]{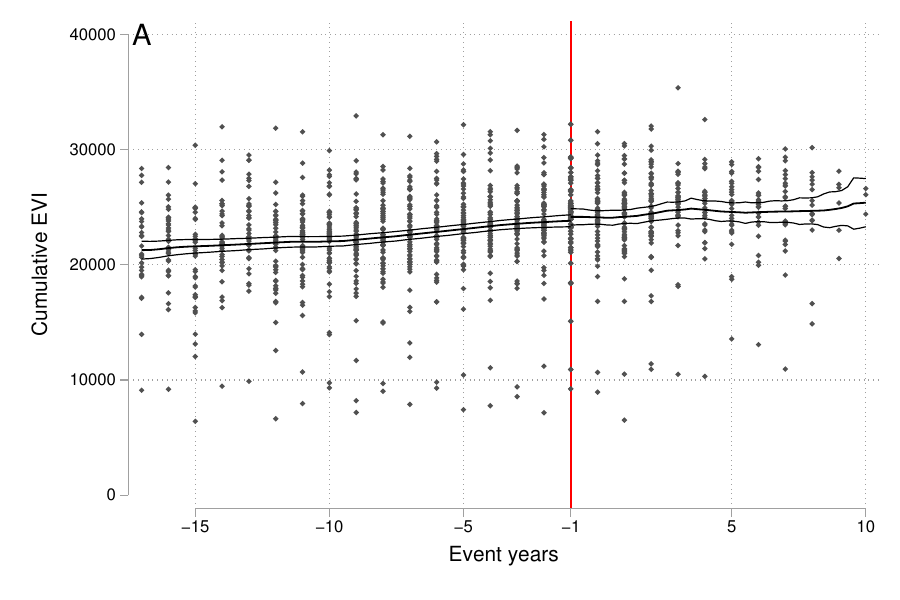}
			\includegraphics[width=.49\linewidth,keepaspectratio]{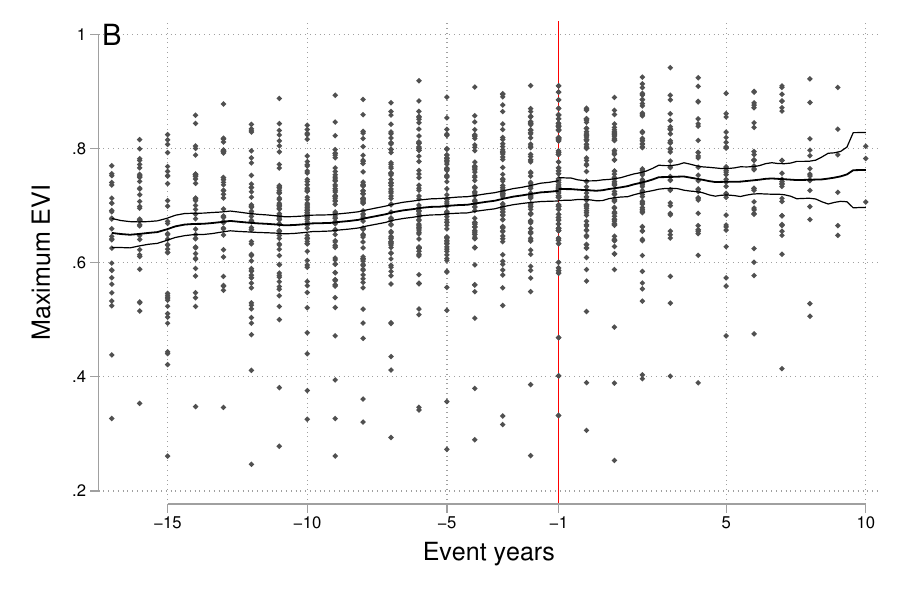}
			\includegraphics[width=.49\linewidth,keepaspectratio]{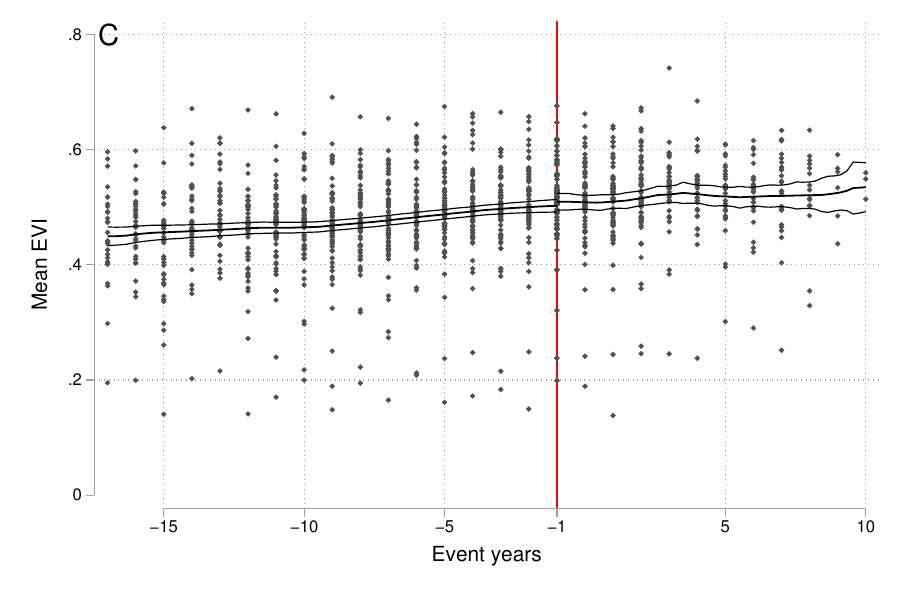}
			\includegraphics[width=.49\linewidth,keepaspectratio]{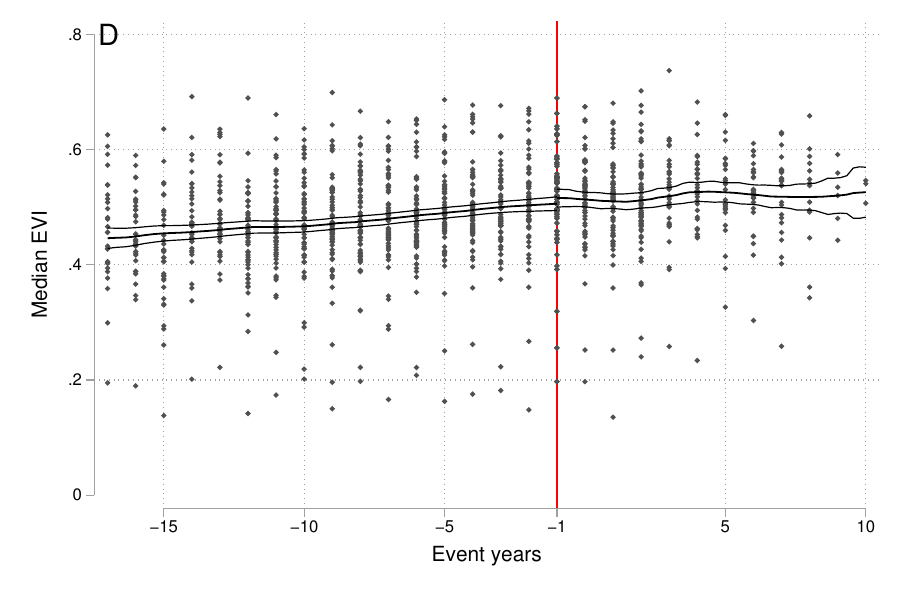}
		\end{center}
		\footnotesize  \textit{Note}: All panels show district-level values for the four EVI measures (cumulative, max, mean, median) by the number of years before of after the introduction of STRV seed in that district. Year 0 is the year the seeds first became available in the district. We then plot non-parametric regressions to both the before and after data with $95\%$ confidence intervals.
	\end{minipage}	
\end{figure}

Given the resolution of the data, it is not possible to measure EVI on a rice field for use in predicting yields as in \cite{lobell_eyes_2020}. The rice maps exist at a spatial resolution of 250m while the mean size of a rice plot in Bangladesh is around two thirds of a hectare. Nor is it possible to measure the exact, agronomically relevant inundation on a given rice field since the flood maps exist at 500m resolution. To overcome this data limitation, we calculate a variety of EVI and flood metrics to ensure our proxies for relevant yield and flooding are robust. For EVI, we calculate the maximum value of EVI for the season, the mean and median values, and the cumulative EVI (sum of EVI throughout the season). Figure~\ref{fig:evi_ba} summarizes how each of these values change in a district before and after the introduction of STRVs into the district.

As flooding is the more sensitive metric (relative to adoption) based on the field trial data and our Monte Carlo simulations, we generate a large number of flood metrics. We create variation in the flood maps by 1) varying the amount of water above baseflow that we consider to be a flood and 2) varying the start and end day of the flood window (length) as well as the days within that window (duration) necessary to be considered a Goldilocks flood. To vary the amount of flooding, we draw the distribution of flooding across the entire time series. We then define a pixel as being flooded if its fractional value is in the top five percent of values ($95\%$ of the time series is below that value) and zero otherwise. We repeat this process for every five percentile increment up to the $95^{th}$ percentile ($5\%$ of the time series is below that value) to create 19 different flood quantiles.

\begin{figure}[!htbp]
	\begin{minipage}{\linewidth}		
		\caption{Distribution of Flooding by Quantile}
		\label{fig:new_flood}
		\begin{center}
			\includegraphics[width=.89\linewidth,keepaspectratio]{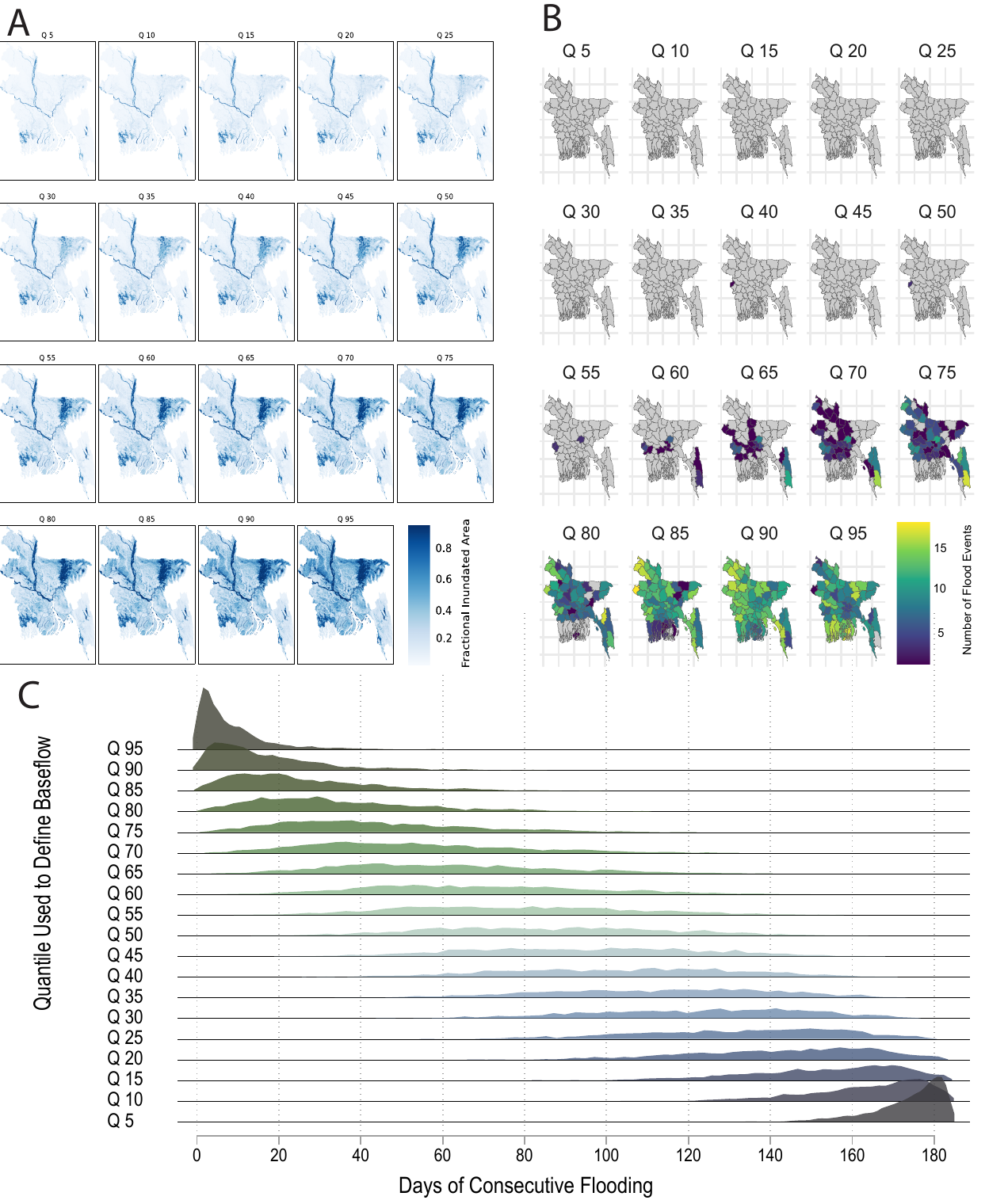}
		\end{center}
		\footnotesize  \textit{Note}: The figure presents visual representations of flooding as defined by quantiles above baseflow. Panel A maps the value of fractional inundated area at the pixel-level at which a given quantile is below that value. The lowest value is Q 5 ($95\%$ of the data is above that value) while the highest value is Q 95 ($5\%$ of the data is above that value). Panel B aggregates that data to the district-level and calculates how many years that district experienced a flood in the 5-20 day window at that quantile. Panel C draws the distribution of the district-level days of consecutive flooding at each quantile.
	\end{minipage}	
\end{figure}

Panel A of Figure~\ref{fig:new_flood} shows the results for all of Bangladesh. Flooding at Q 5 ($5^{th}$ percentile) represents a very low threshold for flooding while flooding at Q 95 ($95^{th}$ percentile) represents extreme flooding, which is why the map at Q 95 is very blue - almost all the pixel has to be full (i.e., the fraction index is close to one) before it registers as a flood. Panel B aggregates the pixel-level fractional flood measure to the district-level and, for each quantile, calculates how many years that district experienced a flood in the largest possible flood window (5-20 days). For flooding at quantiles between Q 5 and Q 50, no district experiences a flood event in the flood window. This is because the threshold is so low districts look flooded for greater than 20 days. At high thresholds, like Q 95, where only extreme floods get counted, many districts experience numerous floods within the flood window. In fact, at Q 95 the coastal area and Northeast region appear to experience flood events in nearly every one of the 20 years of study. Panel C draws the distribution of the district-level days of consecutive flooding at each quantile. At low quantiles (Q 5), almost any amount of water above baseflow is considered a flood, so the length of consecutive days of flooding stretches for almost the entire rice season. At high quantiles (Q 95), inundation has to be extremely severe to register as a flood, meaning few of these occur and when they do they last for a very small consecutive number of days.

To vary the length of flooding, we create a set of potentially relevant flood windows that differ in start and end data and thus differ in length. The flood window can start at any day between five and 15 days after onset and can end at any day between 10 and 20 days after onset. We then bound the number of days that can be in the flood window between five and 15, so the largest flood window is the 15 day window starting on day five and ending on day 20. There are 11 equally small flood windows, each of five day length, ranging in start date from day five to day 15.  So, for example, one flood window starts at day six and end at day 11 for a length of five days. Another starts at day five but ends at day 18 for a length of 13 days. And yet another starts at day 12 and ends at day 20 for a length of eight days. Systematically varying the window like this produces 66 potentially relevant flood windows.

We then intersect the number of days of consecutive flooding at each quantile with the flood window. This intersection produces 1,254 possible measures of the ``just right'' amount of flooding. However, many of the intersections are empty, in that no district experiences flooding that falls within a given window. As the lower panel of Figure~\ref{fig:new_flood} shows, at lower quantiles, where inundation needs to be minimal in order to be considered a flood, the duration of a flood lasts almost the entire season. Eliminating these empty intersections gives us 656 flood windows in which there is variation across space and time in the districts.

The final piece of data that we integrate into our district-level panel is administrative data on seed dissemination. From IRRI and the Bangladesh Rice Research Institute (BRRI), we developed an exhaustive list of all seed production organizations, both private and public. While both IRRI and BRRI have been instrumental in developing submergence tolerant seeds varieties, neither is engaged in seed multiplication, which is conducted by a number of public and private entities. Investment in the seed system helped promote and disseminate STRVs from 2010 through 2019. We visited each seed producer/distributor and obtained disaggregate (district) level data on 1) the STRV variety names being multiplied or sold/distributed in the district, 2) the amount (tons) of each variety being multiplied in the district, and 3) the amount (tons) of each variety being sold/distributed in the district. Figure~\ref{fig:seed_trends} aggregates data from the 64 districts up to the division level and graphs the cumulative amount of seed available over time.  We use the cumulative amount of seed available in each district in each year to account for the growing availability of STRVs and the fact that farmers recycle seed or obtain seed from outside their district.


\subsection{Data Limitations} \label{sec:data_lim}

Before moving to our empirical method, it is important to note and discuss the data limitations in the study. In an ideal situation, researchers would have collected baseline data from rice growing households prior to the release of STRV seeds and would have followed-up with these households periodically over the years. Unfortunately, no such panel with pre-release baseline data exists. This motivates our use of EO data to reconstruct how the EVI signal responds to flooding pre-STRV release and then to examine how that signal changes post release.

In terms of ideal EO data, one would like data from a sensor that is at both a high spatial and temporal resolution and that can penetrate clouds. Additionally, one would like to have ground truth data in terms of both flooding and rice area/yields across both space and time. As highlighted in the previous sections, such data does not exist, mainly because high resolution, cloud penetrating EO sensors, like Sentinel-1, did not exist in the public domain prior to the introduction of STRVs. Thus, there is no way to construct a baseline at both a high spatial and temporal resolution. Additionally, baseline ground truth data do not exist because no baseline survey was conducted. This poverty of ideal data motivates our approach to constructing rice area maps using Google Earth images and flood maps using our CNN-LSTM model.

In developing our approach, we prioritized products with a high temporal resolution (MODIS) at the expense of products with a high spatial resolution (Landsat) because of the importance of capturing transitory flood events and EVI values immediately prior to harvest. We are aware of this trade-off and the limitation it creates: our EO dataset is constructed at 500m, which is an area equal to 25ha. In a cropping system like rice in Bangladesh, where the average rice area for a household is less than one hectare, this spatial resolution is not ideal. However, considering the types of measurement error introduced by the trade-off leads us to believe that prioritizing temporal over spatial resolution is the preferred approach. We have explored conducting the analysis using Landsat 30m data. However, Landsat's temporal resolution is 16 days, meaning it frequently misses known flood events and EVI readings at harvest are frequently missed. Given this temporal resolution, it is possible to under count by missing a flood or harvest but it is not possible to over count. This means that the measurement error introduced is non-classical and while the direction of the bias is known, the size of the bias is unknown. By contrast, the relatively course spatial resolution of MODIS, combined with the ground truth data, and our machine learning classification approach means that we may misallocate a pixel to rice/not-rice and to flood/not-flood, but that misallocation is likely random. This means the measurement error introduced should be classical - increasing the variance or noise in the data but not introducing any systematic bias. Essentially, the spatial fuzziness of MODIS allows us to make guided inference on whether or not a pixel is rice/not-rice or flooded/not-flooded while the missing temporal images of Landsat are not amenable to making inference about what happened in those missing pixels. This spatial fuzziness can be measured in terms of accuracy metrics for our classification algorithms (RF and CNN-LSTM), which are in the range of $80$-$90\%$.

Beyond the issue of coarse spatial resolution, we also lack objectively measured harvest information for training an ML algorithm on to predict yields. This forces us to rely on off-the-shelf EVI measures as proxies for yields, though this approach is not without its critics. A large body of research demonstrates strong correlation between EVI, and particularly maximum EVI, and yields \citep{MkhabelaEtAl11, LiuEtAl20, lobell_eyes_2020}. However, the accuracy of using EVI as a proxy for yields varies substantially by location \citep{AzzariEtAl17} and crop, with EVI performing particularly poorly for rice in the U.S. \citep{Johnson16}. For rice in Bangladesh, the correlation between EO derived metrics and yield can vary from $55$-$91\%$ \citep{IslamEtAl21}. However, \cite{IslamEtAl21} relied on government reported area yield statistics, a notoriously inaccurate source for measures of yield in low-income countries \citep{KilicEtAl21}, where statistical departments are often under-resourced and politicized. We are unaware of tests of accuracy of EVI with rice yields in Bangladesh that rely on objective measures of yield.

While these data limitations are not unique to our study, and would be present in any attempt to measure impact of an agricultural technology introduced over a decade ago, they should be kept in mind when interpreting our results. 


\section{Empirical Method} \label{sec:method}

Our empirical method follows that laid out in \cite{GollinEtAl21} to estimate the impact of Green Revolution technologies on yields. Our empirical strategy relies on the release dates of various varieties and the staggered accumulation of seed in each district, which we argue is plausibly exogenous to rice yields. We estimate the effects of STRVs on EVI as a proxy for yields using \emph{Aman} season data at the district-level. We use a variety of econometric methods (event study, difference-in-difference, two-way fixed effects), each with its own strengths and weaknesses regarding identifying assumptions and the potential for attenuation bias due to mismeasurement.

We start by estimating a simple event study model:

\begin{equation}
    \textrm{EVI}_{dt} = \sum_{j \in T} \alpha_{j} \cdot \textbf{1}_{t}^{t = \tau + j} + \phi \textrm{flood} + \mu_{d} + \mu_{t} + \epsilon_{dt} \label{eq:event}
\end{equation}

\noindent where $d$ indexes districts and $t$ indexes time. The two terms $\mu_{d}$ and $\mu_{t}$ denote district and time fixed effects, meaning that only within-district time variation in relative EVI remains. The district fixed effects control for all district-specific time-invariant variation and the time fixed effects control for all time-specific district-invariant variation. We also include a control for the percentage of a district that was flooded, which varies by district and time.

We expect the introduction of STRVs in a district to increase EVI relative to that district's EVI before the introduction. To capture this effect in the regression, we include an indicator function $\textbf{1}_{t}^{t = \tau + j}$ that takes a value of one $j$ years after the introduction of the first STRV, which we denote as $\tau$. As $\epsilon_{dt}$ captures district-specific trends in relative yields, $\alpha_{j}$ measures by how much the relative yield in the average district has changed $j$ years after the introduction of STRVs relative to the benchmark year. We set the benchmark year in each district as the year immediately prior to the introduction of STRVs in that district. Our hypothesis is that $\alpha_{j} > 0$ after the introduction of STRVs and that $\alpha_{j} = 0$ before the introduction of STRVs.

There are two shortcomings to the event study approach. One is that STRVs reach different districts in different years. It is plausible that STRV seeds reached more flood-prone districts, or districts with more advanced farming techniques, earlier than other districts. If these characteristics are time-invariant, our district fixed effects will pick them up. However, one can imagine situations where time-variant, district-specific events accelerated the first introduction of STRV seed. While we explicitly control for contemporaneous flooding, flood events in the recent past might make the year a district got STRVs no longer exogenous. A second shortcoming is that STRVs are likely to increase EVI only very slightly over time, as a change in EVI would only occur when there was a Goldilocks flood and sufficient seed was available. If a district did not experience such a flood in a year after STRVs were released, or had very low quantities of seed, we would expect no change in EVI. Thus, the event study will underestimate the effects of STRV release date on EVI.

To address these two issue, we next estimate a simple difference-in-difference (DID) model:

\begin{equation}
    \textrm{EVI}_{dt} = \beta \left( 1_{dt}^{\textrm{2010}} \cdot 1_{dt}^{\textrm{flood}} \right)+ \rho 1_{dt}^{\textrm{2010}} + \gamma 1_{dt}^{\textrm{flood}} + \mu_{d} + \mu_{t} + \epsilon_{dt} \label{eq:did}
\end{equation}

\noindent where $1_{t}^{\textrm{2010}}$ is an indicator equal to one in the years after 2010, they year the Bangladesh government approved STRVs, and zero otherwise. Similarly, $1_{t}^{\textrm{flood}}$ equals one in districts that are prone to flooding and zero in districts that are not prone to flooding. To define a district as flood-prone or not, we first calculate the median amount of flooding across all districts and all years. We then calculate the median flood value in each district across the time series. We categorize a district as flood-prone if its median value is above the median value in all the data. The remaining terms are as defined in Equation~\eqref{eq:event}.\footnote{Note that because we have converted both adoption and flood to binary indicators that do not vary over time, our DID estimator is not subject to the recently exposed potential biases of DID in complicated settings. For a summary of what settings complicate DID estimation and the numerous ways to correct for bias, see \cite{RothEtAl23}.}

We expect EVI in flood-prone districts to increase after 2010, which was when STRVs first became available in Bangladesh. In this set-up, there are no pure control districts. Rather, we compare the difference in the differences that exist between flood-prone/not and pre/post 2010 to establish a counterfactual. Our hypothesis is that $\beta > 0$ while $\rho = 0$ and $\gamma < 0$. The main identifying assumption is that if STRVs had not been released, yields in flood-prone districts would have followed the same trend as yields in not-flood-prone districts. Obviously, flood-prone districts are likely to have lower yields than not-flood-prone districts. Thus we assume a mean shift in yields, as without STRVs mean yields are likely to be lower in flood-prone districts than in districts that are not prone to flooding, though the existence of parallel trends remains unobservable.

As with the event study set-up, the DID model is not without its shortcomings. In particular, in the DID model we assume that farmers in all districts have access to STRVs starting in 2010. This assumption avoids endogeneity concerns about targeting of certain districts for seed but, as shown in Figure~\ref{fig:seed_trends}, we know that STRVs were not universally available in 2010. In fact, in our administrative data, only one district had seed locally produced and distributed that early. This suggests that, like any agricultural technology, for STRVs the process of roll out, dissemination, and adoption was gradual and took time. The end result is that our use of a dummy for seed release in 2010, while exogenous, likely overestimates the number of districts ``treated'' with STRV seed, attenuating the true treatment effect. A second shortcoming is that while a relative measure like flood-prone might drive adoption of STRVs and impact EVI, it is a coarse measure that does not vary over time.

Our third approach to modeling the district-level effects of STRV availability is a two way fixed effects (TWFE) approach that seeks to address the shortcoming of the event study and DID approaches. It explicitly accounts for seed availability and actual flooding in the district. The TWFE equation is essentially the same as the DID: 

    \begin{equation} \label{eq:twfe}
        \textrm{EVI}_{dt} = \beta(\textrm{seed}_{dt} \cdot \textrm{flood}_{dt}) + \rho \textrm{seed}_{dt} + \gamma \textrm{flood}_{dt} + \delta \textrm{rice} + \mu_{d} + \mu_{t} + \epsilon_{dt}.
    \end{equation}

\noindent One main difference is that the TWFE approach uses administrative data on cumulative seed available in each district in each year instead of a binary indicator for if the observation comes after 2010. The other main difference is that we use measures of actual flooding in a given year instead of a categorical variable for if the district is flood-prone or not. All other variables are the same as in the DID model.\footnote{The use of data on the staggered rollout of STRV seed means our TWFE model may be subject to the recently exposed potential biases of TWFE in complicated settings. However, as \cite{Wooldridge21} argues, the main issue with the older TWFE literature was that it did not account for heterogeneity that impacted both treatment effects and common trends. In our model we have explicitly accounted for heterogeneity by interacting treatment (STRV) with flooding in the belief that both treatment effect and common trends will vary by flooding. Thus, our models already accounts for the existence of heterogeneity in treatment and trends that many of the new methods are designed to address.}

The benefit of the TWFE approach is that it uses available data on seed distribution and flooding. In the event study, we controlled for flooding but our variable of interest $(\alpha_j)$ measured changes in EVI in each year after seeds were introduce not changes to EVI when seeds were introduced and flooding occurred. This likely underestimates the effect of STRVs. In the DID model, we interact indicators for flooding and the introduction of STRVs but our variable of interest $(\beta)$ considers all districts as having equal access to seeds in every year since 2010. This also likely underestimates the effect of STRVs. Our TWFE model addresses these issues but it comes with the caveat that if seed availability is a function of time-varying, district-level events it may be endogenous. Recall that we control for time-invariant district characteristics, like soil quality and propensity to flood, as well as district-invariant time events, like changes to national agricultural policy. So, the administrative seed data will only be endogenous if dissemination changed based on some time-varying, district-level event, such as targeting a district for increased dissemination in the year after an unexpectedly severe flood.

Our overall approach to estimating the long-term, large-scale effects of STRVs in Bangladesh is to balance the richness of the EO and administrative data with concerns about endogeneity. The event study and DID approach minimize concerns about endogeneity by using coarser indicators for the year STRVs were approved in Bangladesh, a clearly exogenous event. The TWFE approach uses data on actual flooding and seed availability, the later of which may be endogenous. By presenting all three approaches we attempt to provide a robust picture of the causal effects of STRVs in Bangladesh.


\section{Results} \label{sec:rslt}

In the following sections we discuss the results of our econometric analysis of the impacts of STRV adoption on rice production during floods. We first discuss results from the event study, DID, and TWFE models. We then summarize a set of robustness checks.

\subsection{Econometric Evidence}\label{subsec:district_econ}

\begin{figure}[!htbp]
	\begin{minipage}{\linewidth}		
		\caption{Growth of STRV Available by Division}
		\label{fig:seed_trends}
		\begin{center}
			\includegraphics[width=\linewidth,keepaspectratio]{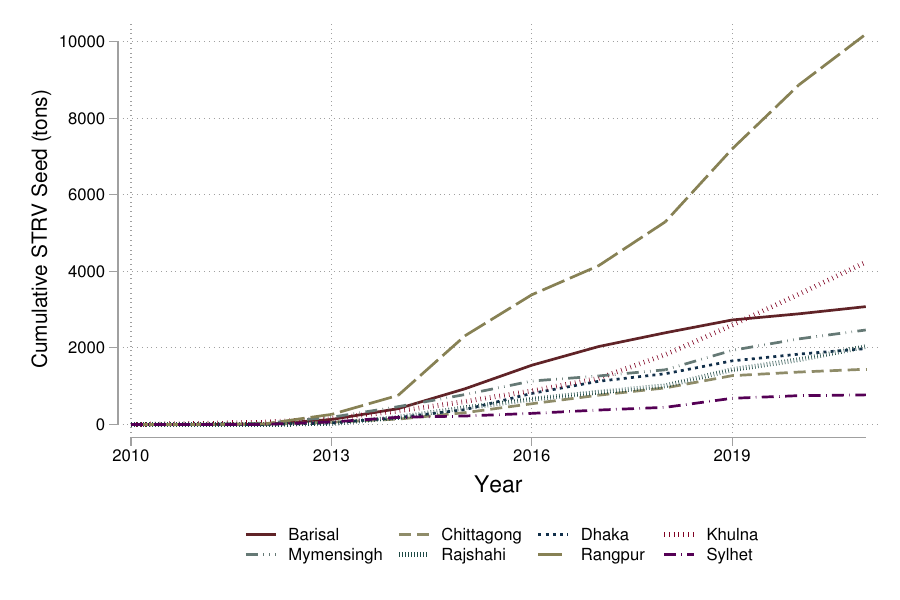}
		\end{center}
		\footnotesize  \textit{Note}: Figure presents the cumulative amount of STRV seed available in each division in each year. Seed availability comes from administrative data and is calculated as the total amount of flood tolerant seed (of any variety) produced and/or distributed by public and private enterprises within a district. As there are 64 districts, we sum up district-level values to obtain division-level values.
	\end{minipage}	
\end{figure}

\begin{figure}[!htbp]
	\begin{minipage}{\linewidth}		
		\caption{Event Study Results}
		\label{fig:event}
		\begin{center}
			\includegraphics[width=.49\linewidth,keepaspectratio]{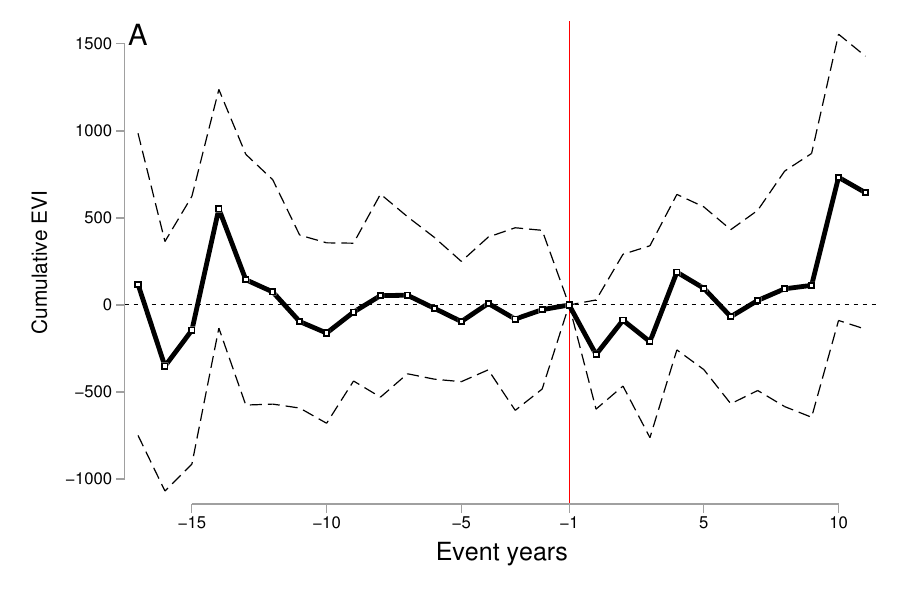}
			\includegraphics[width=.49\linewidth,keepaspectratio]{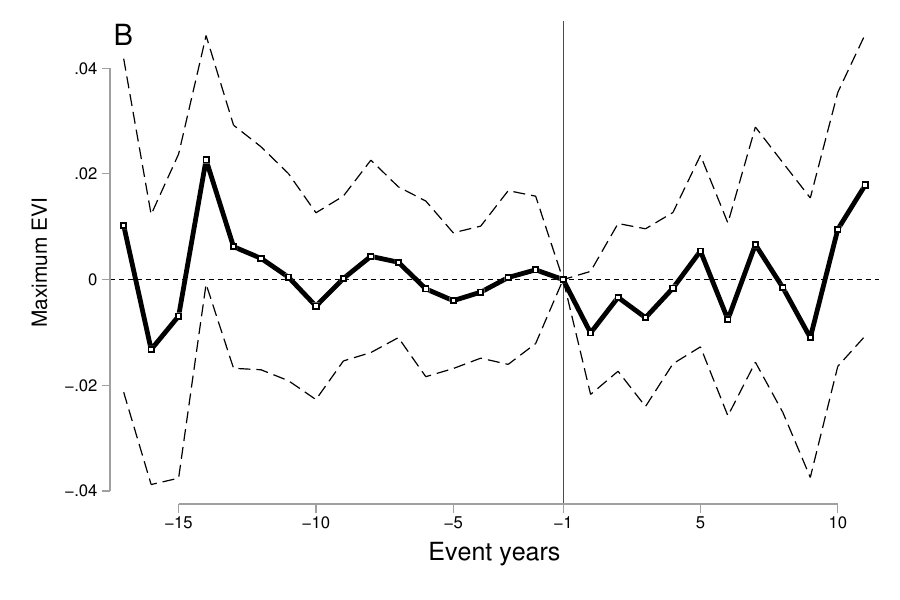}
			\includegraphics[width=.49\linewidth,keepaspectratio]{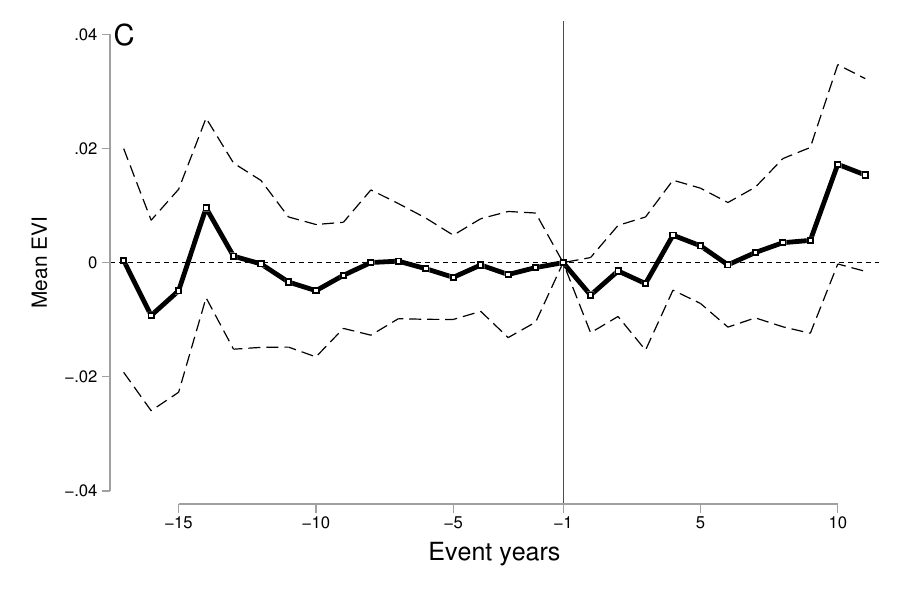}
			\includegraphics[width=.49\linewidth,keepaspectratio]{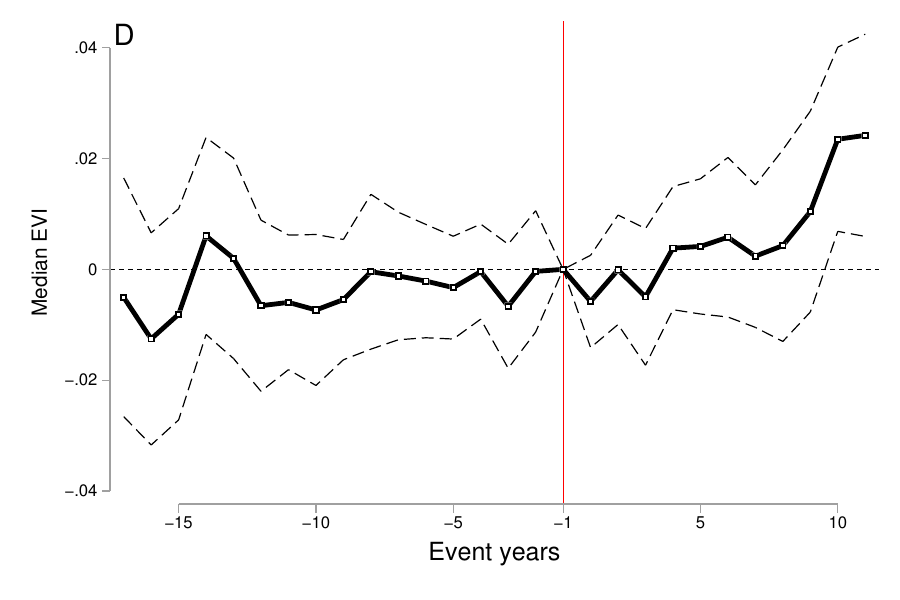}
		\end{center}
		\footnotesize  \textit{Note}:  Figure displays coefficients from district-level event study regressions. The dependent variable in each panel is a specific EVI measure (cumulative, max, mean, median). The solid line shows coefficient estimates from the model, with the event year (the year immediately prior to the district having access to STRVs, indicated as -1) as the excluded category. Dotted lines represent $95\%$ confidence intervals calculated using standard errors clustered at the district-level.
	\end{minipage}	
\end{figure}

Figure~\ref{fig:event} reports the results of the event study regression. Using administrative data on the date that a district first had access to STRV seed, we estimate the change in that district's EVI value before and after the introduction date. As we are looking at changes in district-wide EVI values, instead of changes in EVI values where flooding occurred, our event study is likely an underestimate of the effectiveness of STRVs in preventing flood damage. Regardless of which EVI measure we use, the event study results generally support our stated hypothesis. Prior to the introduction, EVI values were flat, fluctuating around zero change relative to the year before STRV availability (indicated as -1). In the years after STRVs were available in the district, EVI tends to increase relative to the value immediately before introduction. This pattern is strongest when we use median EVI and weakest when we use the maximum value of EVI. Regardless of the EVI metric, the longer the time period since introduction, the greater the increase in EVI. This is consistent with the slow but gradual increase in seed availability presented in Figure~\ref{fig:seed_trends}. As the availability and amount of STRVs in the seed system has grown, so too has EVI values. This suggests that STRVs are effective in reducing yield loss when flooding occurs, as before their introduction flooded rice would have died, resulting in a lower EVI value than after STRV introduction. 

We present results for our DID model (and our TWFE model) in specification charts. These charts allow us to compactly summarize the size and significance of our coefficient of interest as we test more than 600 potential flood windows to determine if STRVs are effective during a Goldilocks flood. Recall from Section~\ref{sec:goldi} that there is a range of potential flooding that could be considered ``just right'' for STRVs to be effective. And recall from Section~\ref{sec:EOdata} that we generate a large set of potentially relevant flood windows by varying the start and end day of the flood window as well as the total length of the flood window. We then intersect these windows with the number of days of consecutive flooding measured at various quantiles above baseflow. Along the horizontal axis, model specifications are organized from smallest coefficient to largest. One can then read up the vertical axis using the gray markers to determine exactly what combination of flood window and flood quantile was used for that specific regression. Across the top are coefficient sizes with $95\%$ confidence intervals. Red bars signify negative and significant coefficients; black bars signify insignificant coefficients; and blue bars signify positive and significant coefficients. Given that we did not pre-specify these flood metrics, and in the interests of transparency, we report results from all regressions for each flood metric and for each EVI metric, not just those that conform to our hypotheses.

\begin{landscape}
\begin{figure}[!htbp]
	\begin{minipage}{\linewidth}		
		\caption{Specification Charts of DID Results}
		\label{fig:flood_days_did}
		\begin{center}
			\includegraphics[width=.48\linewidth,keepaspectratio]{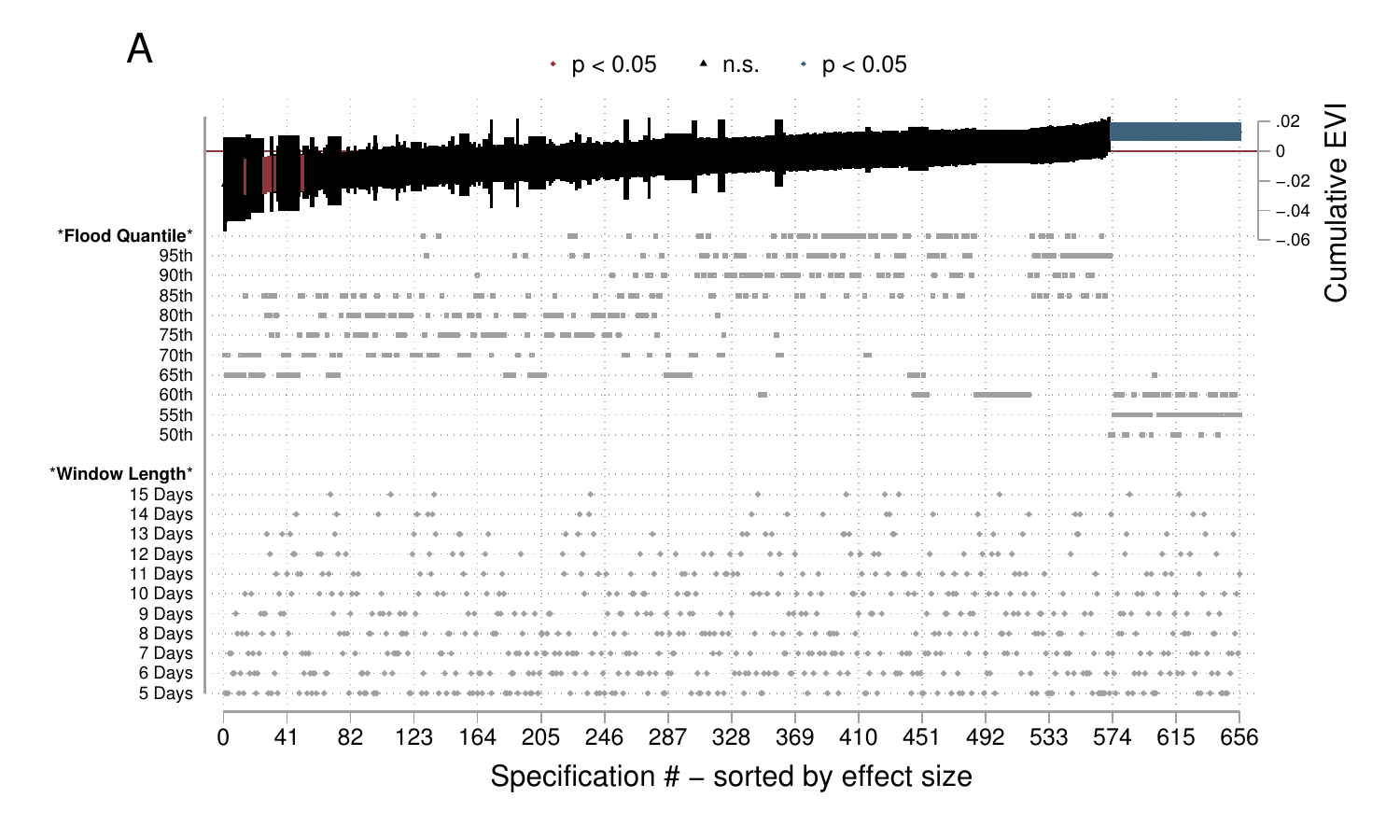}
			\includegraphics[width=.48\linewidth,keepaspectratio]{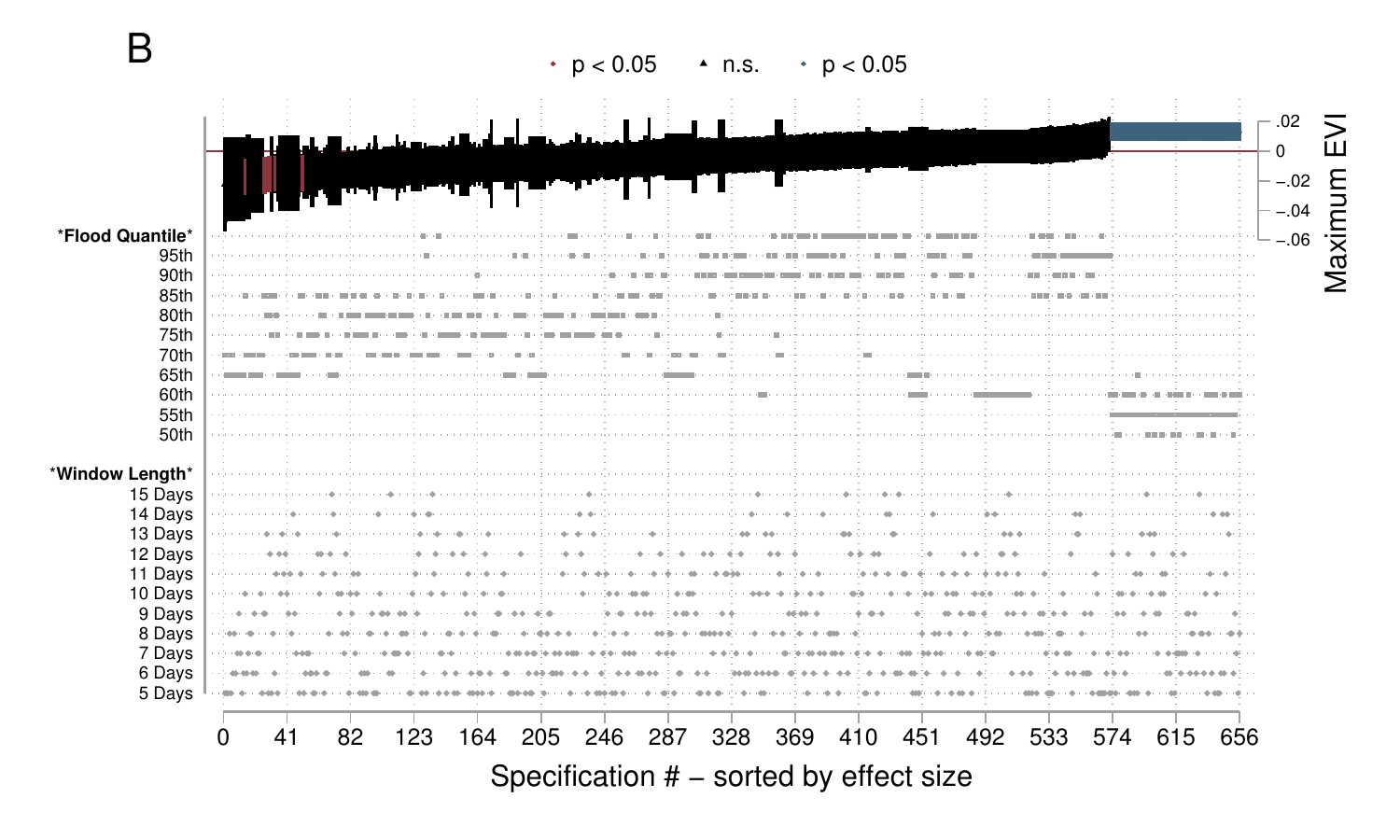}
			\includegraphics[width=.48\linewidth,keepaspectratio]{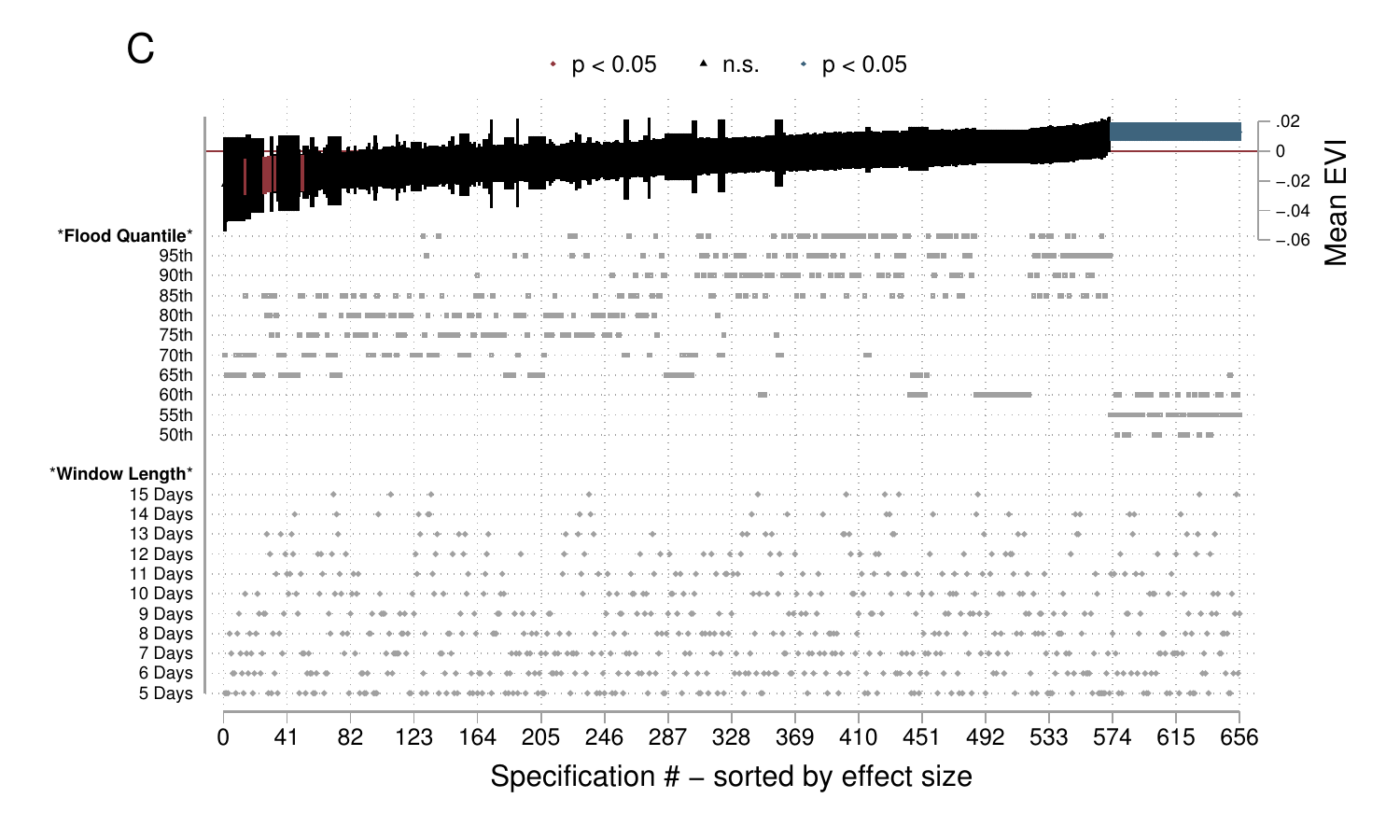}
			\includegraphics[width=.48\linewidth,keepaspectratio]{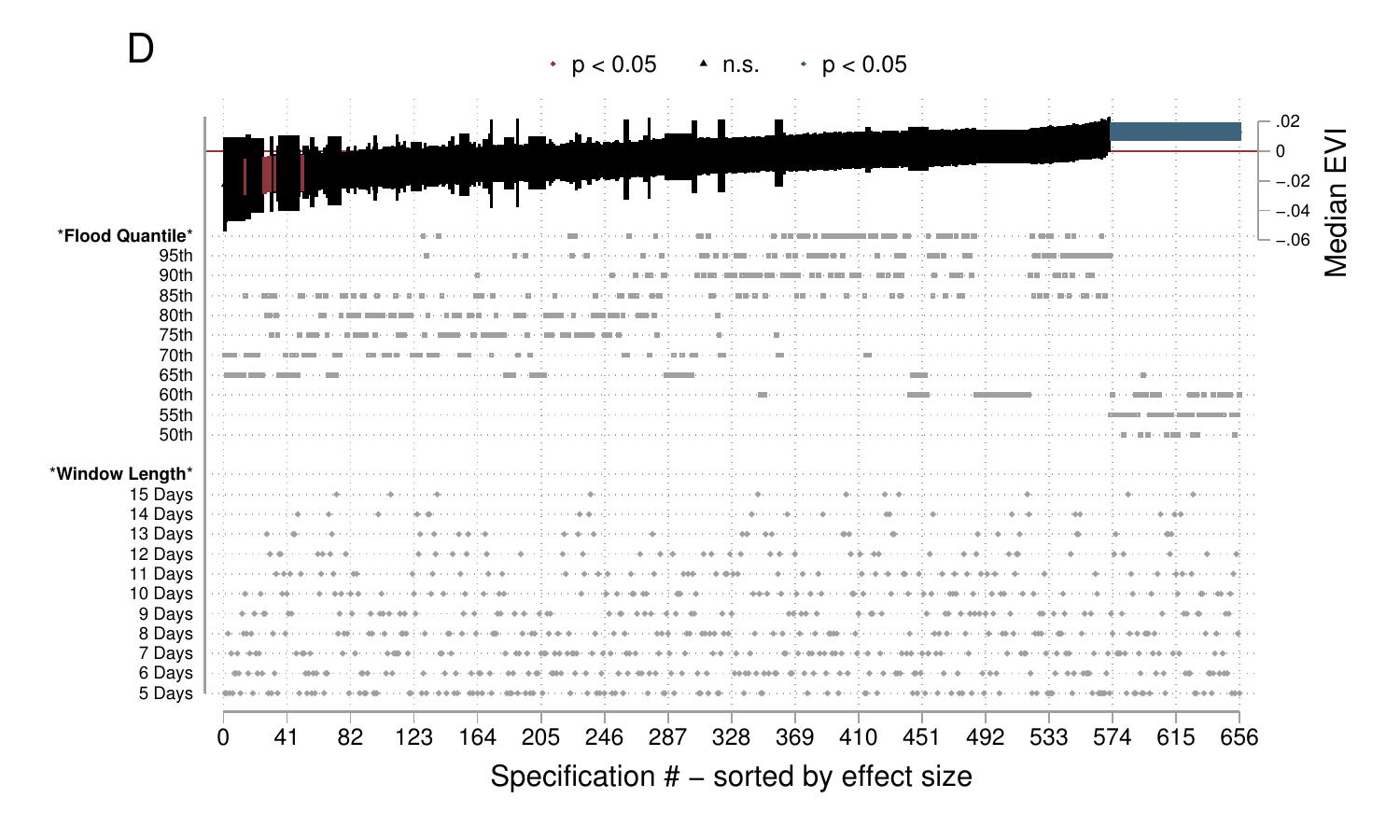}
		\end{center}
		\footnotesize  \textit{Note}: The figure presents results from district-level DID regressions. Each panel in the figure displays coefficient estimates and $95\%$ confidence intervals on the interaction of STRVs and a flood metric for a specific EVI measure (cumulative, max, mean, median). Specifications are sorted based on coefficient size from smallest (left) to largest (right). The gray diamonds below the coefficients indicate which combination of quantile and flood window was used in the regression.
	\end{minipage}	
\end{figure}
\end{landscape}

For our DID results, we use the various flood window-quantile combinations to determine median flooding in the data. When then classify a district as flood prone if it experiences flooding above the median level and its binary flood value is set to one. A district below the median value is classified as not flood prone and its binary flood value is set to zero. We include an indicator for if the EVI observation is from after 2009, meaning all districts are including in the ``treated'' group starting in 2010, even if STRV seeds were not available until several years later. Our coefficient of interest is on the interaction of these two binary indicators $\left( 1_{dt}^{\textrm{2010}} \cdot 1_{dt}^{\textrm{flood}} \right)$. Panel A of Figure~\ref{fig:flood_days_did} presents results from the DID specification with the cumulative EVI value as the dependent variable. We see that a large majority ($87\%$) of potential flood metrics produce negative or insignificant results. A small but not insubstantial set of flood metrics appear to be ``just right'' in that when we use one of those flood metrics to define flood prone and interact it with the post-2009 indicator, it significantly increases cumulative EVI.\footnote{Because the DID specification uses a binary indicator for flood prone, there ends up being almost no variation introduced by the different flood durations. If a district was classified as flood prone using the $60^{th}$ flood quantile at a window of five days, it was also classified as flood prone when the flood window was expanded to more days. This means that many of the 656 DID regressions we run are redundant. When we consider TWFE regressions, this is no longer the case.}

That $13\%$ of 656 regressions turn our to be positive and significant at the $95\%$-level clearly raises concerns that we are discovering significant results due to random chance. If STRVs are ineffective, meaning the true effect is zero, and we ran 100 regressions, we would expect a false positive in five of the regressions, given we have set our threshold for significance at $95\%$. Across all panels in Figure~\ref{fig:flood_days_did} we find a larger share of significant results than we would expect just by chance. There are also several regularities in our results that build confidence that they are not due to random chance. First and foremost is that fact that the significant results all come from flood measures at the $60^{th}$ to $50^{th}$ quantile. If results were truly random, we would expect some positive results at each quantile. Second, this consistent pattern is repeated across all four EVI measures. Again, if results were driven by random chance, then we would not expect consistency in which quantiles mattered as we changed how we measure the dependent variable.

\begin{landscape}
\begin{figure}[!htbp]
	\begin{minipage}{\linewidth}		
		\caption{Specification Charts of TWFE Results Using Binary if Flooding Was in Window}
		\label{fig:flood_days_bin}
		\begin{center}
			\includegraphics[width=.48\linewidth,keepaspectratio]{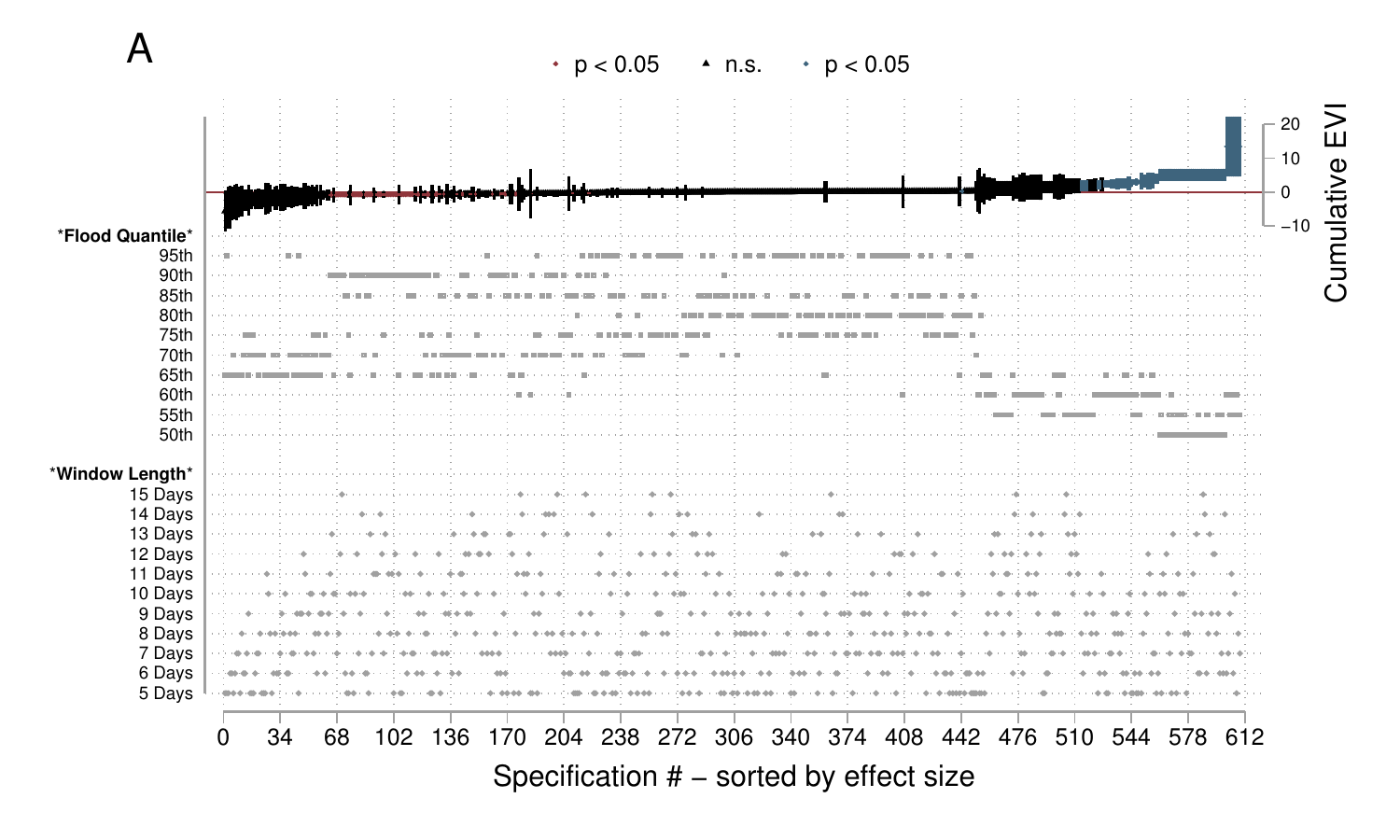}
			\includegraphics[width=.48\linewidth,keepaspectratio]{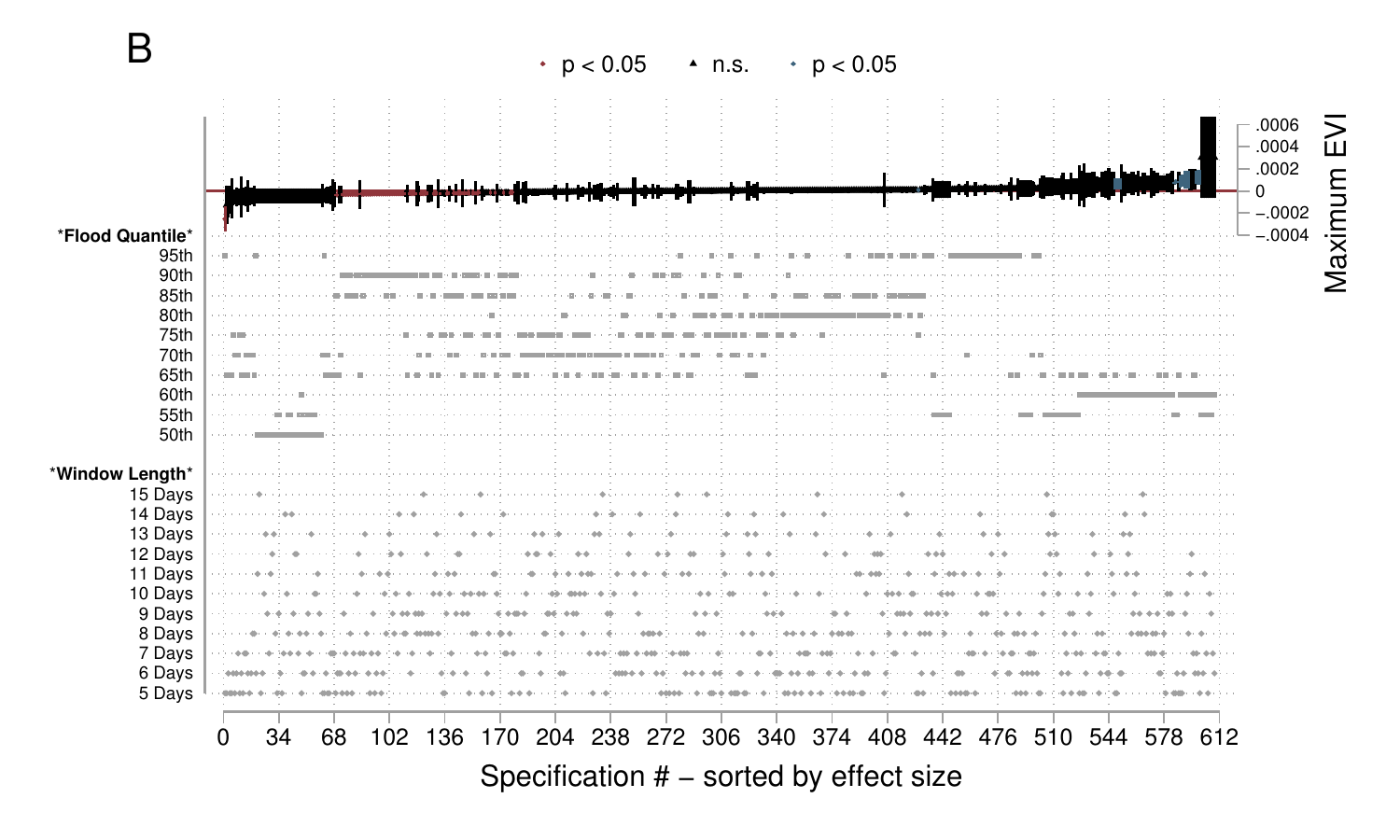}
			\includegraphics[width=.48\linewidth,keepaspectratio]{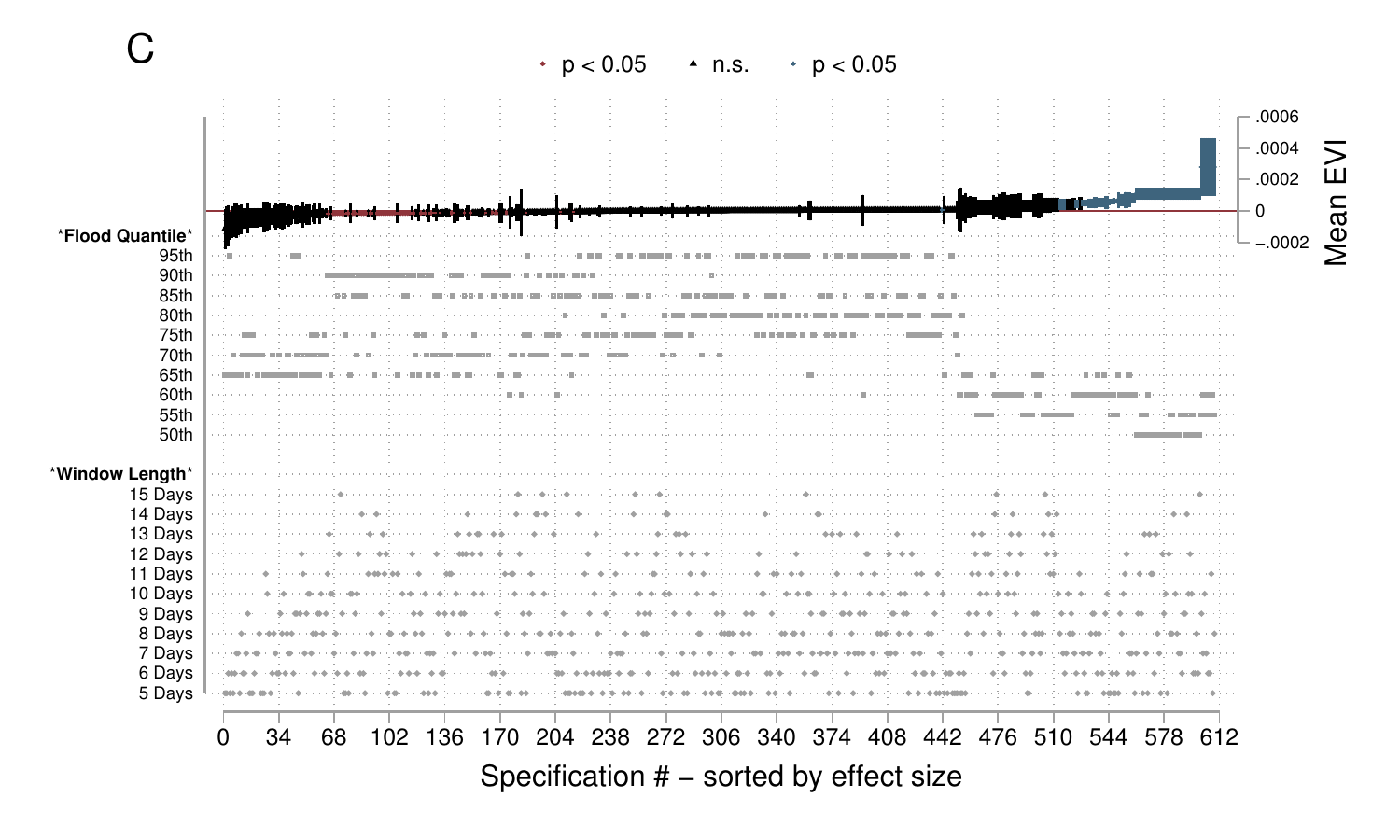}
			\includegraphics[width=.48\linewidth,keepaspectratio]{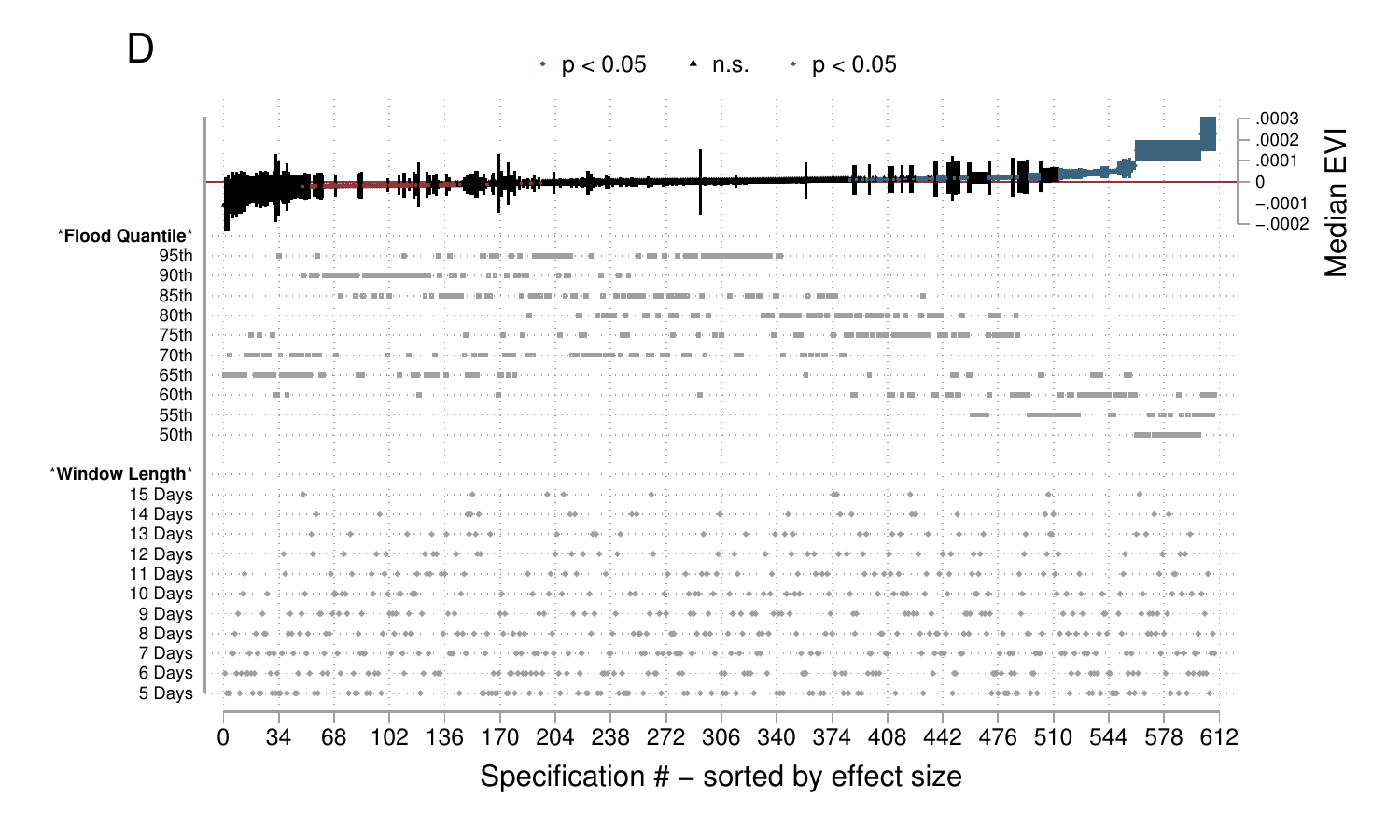}
		\end{center}
		\footnotesize  \textit{Note}: The figure presents results from district-level TWFE regressions. Each panel in the figure displays coefficient estimates and $95\%$ confidence intervals on the interaction of STRVs and a flood metric for a specific EVI measure (cumulative, max, mean, median). Specifications are sorted based on coefficient size from smallest (left) to largest (right). The gray diamonds below the coefficients indicate which combination of quantile and flood window was used in the regression.
	\end{minipage}	
\end{figure}
\end{landscape}

For our TWFE estimation, we generate flood metrics in two different ways. The first is to use a dummy variable to indicate if flooding at a given quantile fell within a specific flood window. The second is to use the actual number of days that flooding at a given quantile was within a specific flood window. Figure~\ref{fig:flood_days_bin} presents results from the TWFE with the binary indicators for flooding. A pattern very similar that in our DID results is clearly visible. Focusing on Panel A, the majority ($85\%$) of potential flood metrics produce null or negative results. Of the 612 regressions, 90 produce positive and significant results.\footnote{There are slightly fewer potential flood metrics than with the DID regressions because there are more empty sets at the intersection of quantile and flood window.} These ``just right'' floods occur when we consider flood measures produced at the $65^{th}$ to $50^{th}$ quantile, which substantially overlaps the quantiles that produced positive and significant results in the DID regressions. Again, these results are robust to using any of our four EVI measures. In terms of flood window, unlike in the DID regression, here a pattern begins to emerge. When the window is five to 10 days long, around $16\%$ of regressions produce positive and significant results. This falls to around $12\%$ of coefficients being significant when the flood window is longer than 10 days.

\begin{landscape}
\begin{figure}[!htbp]
	\begin{minipage}{\linewidth}		
		\caption{Specification Charts of TWFE Results Using Days in Flood Window}
		\label{fig:flood_days_win}
		\begin{center}
			\includegraphics[width=.48\linewidth,keepaspectratio]{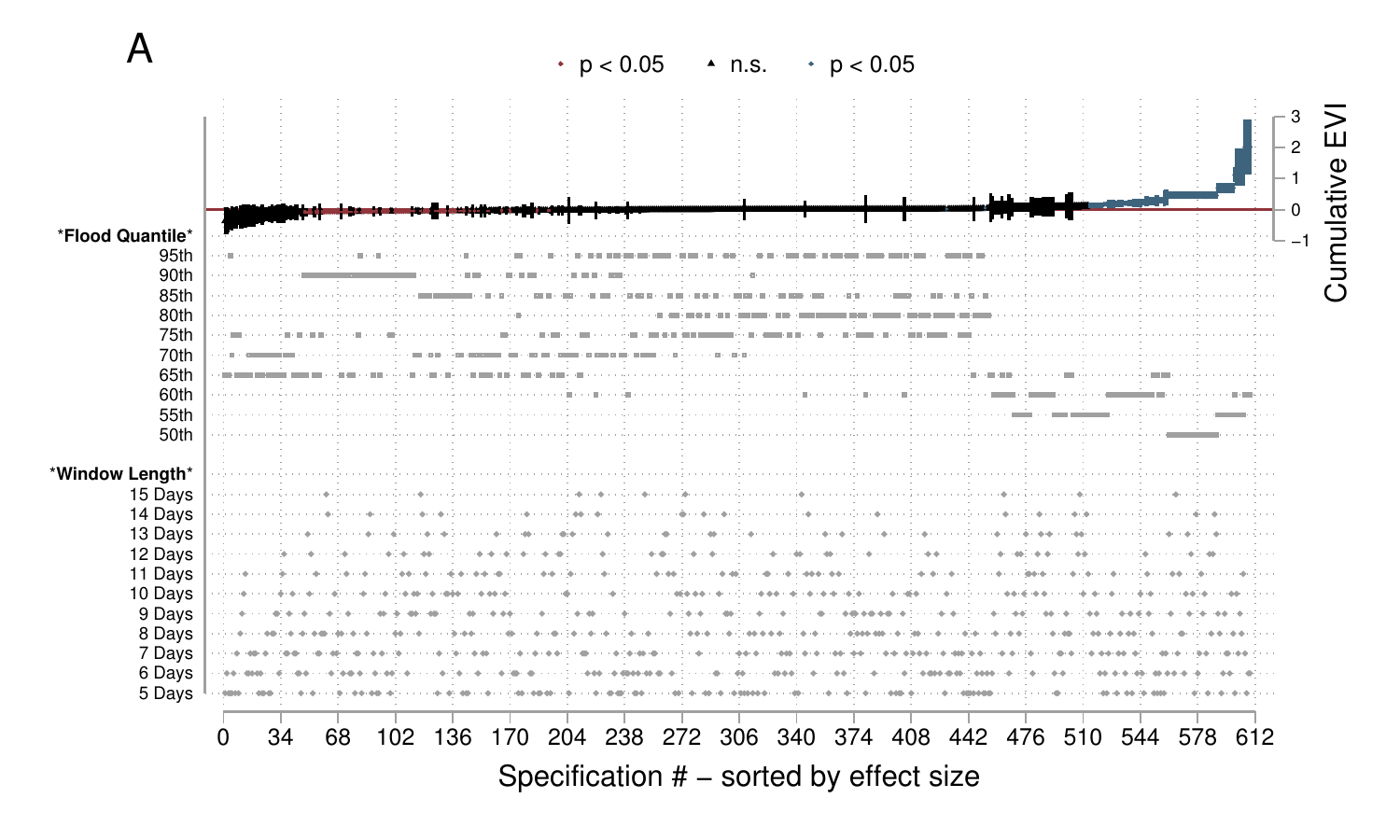}
			\includegraphics[width=.48\linewidth,keepaspectratio]{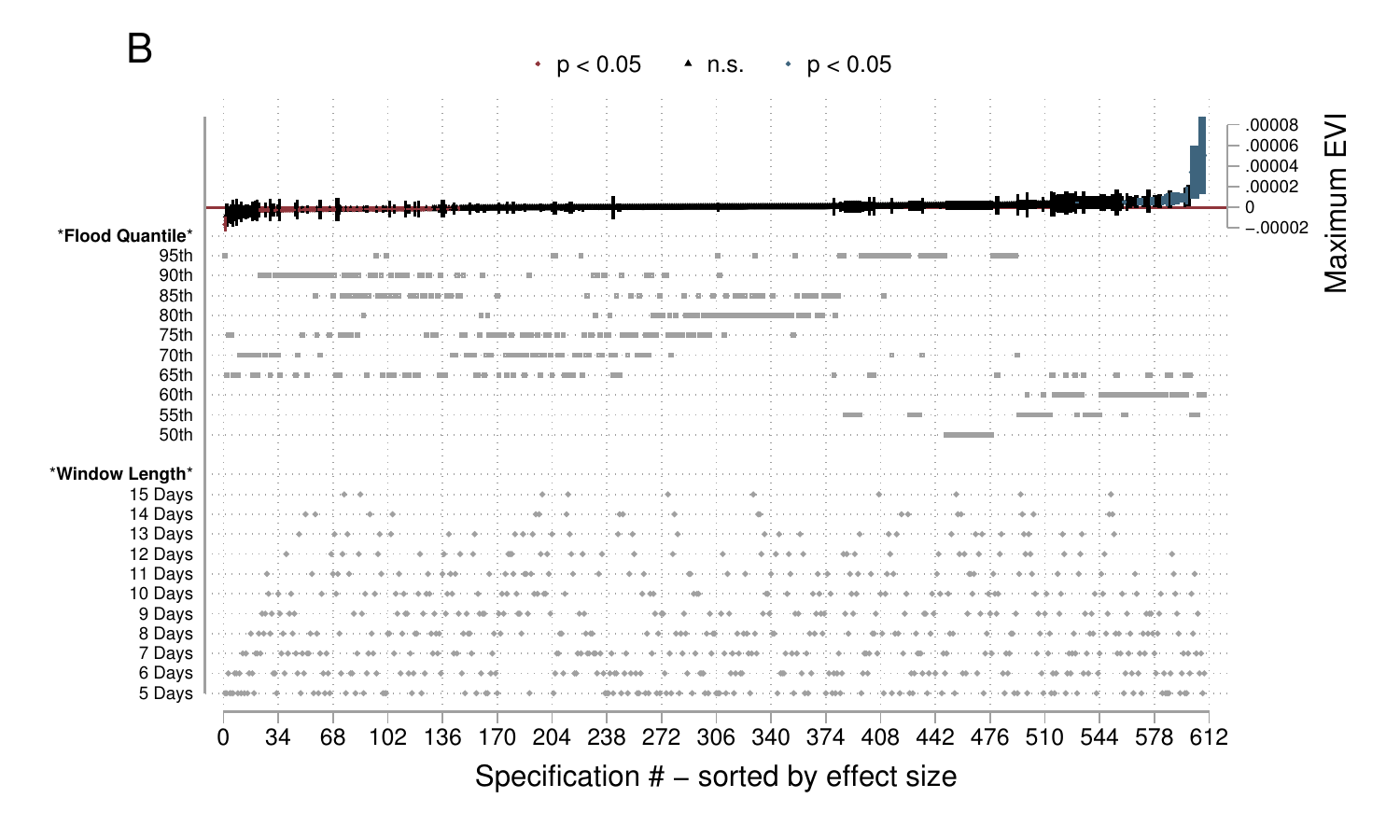}
			\includegraphics[width=.48\linewidth,keepaspectratio]{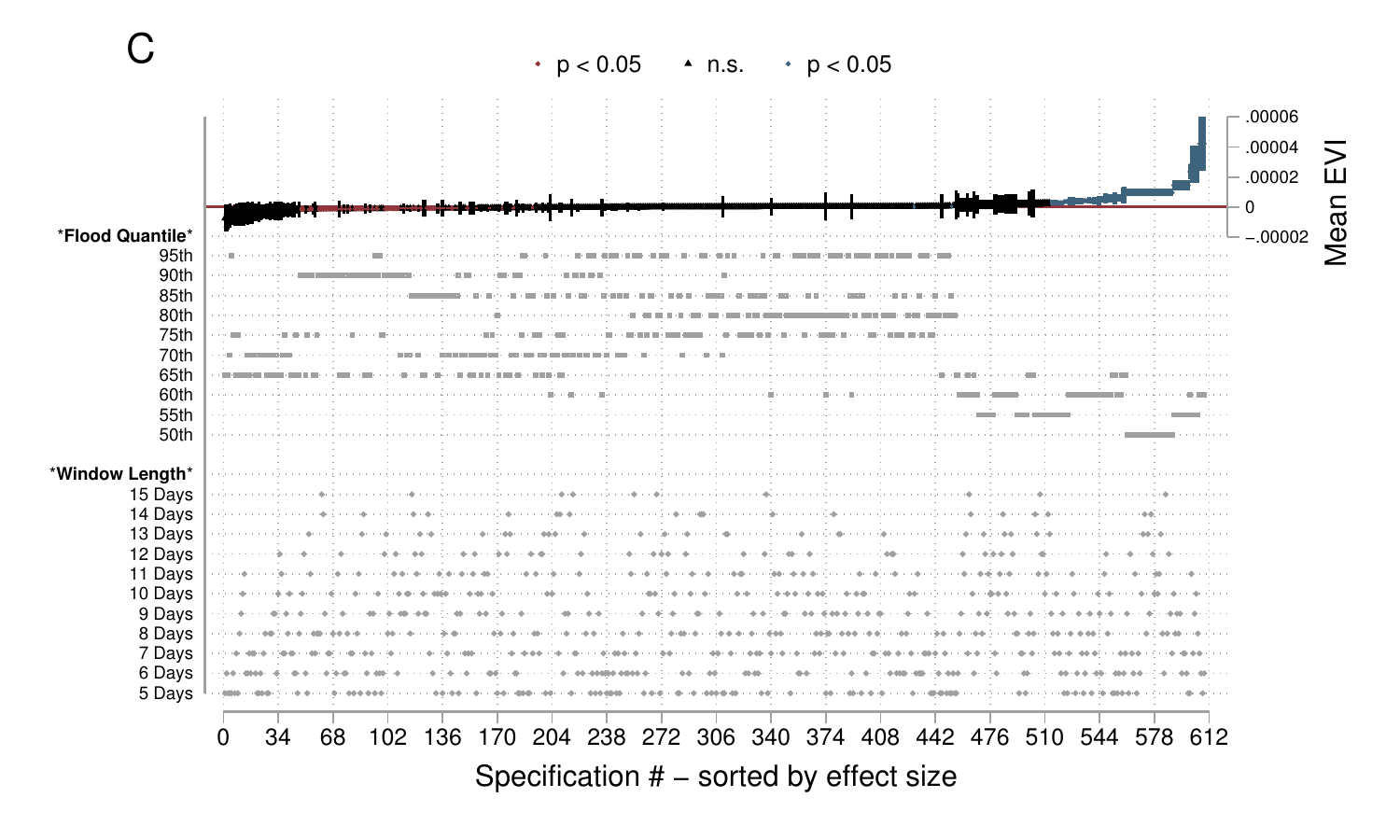}
			\includegraphics[width=.48\linewidth,keepaspectratio]{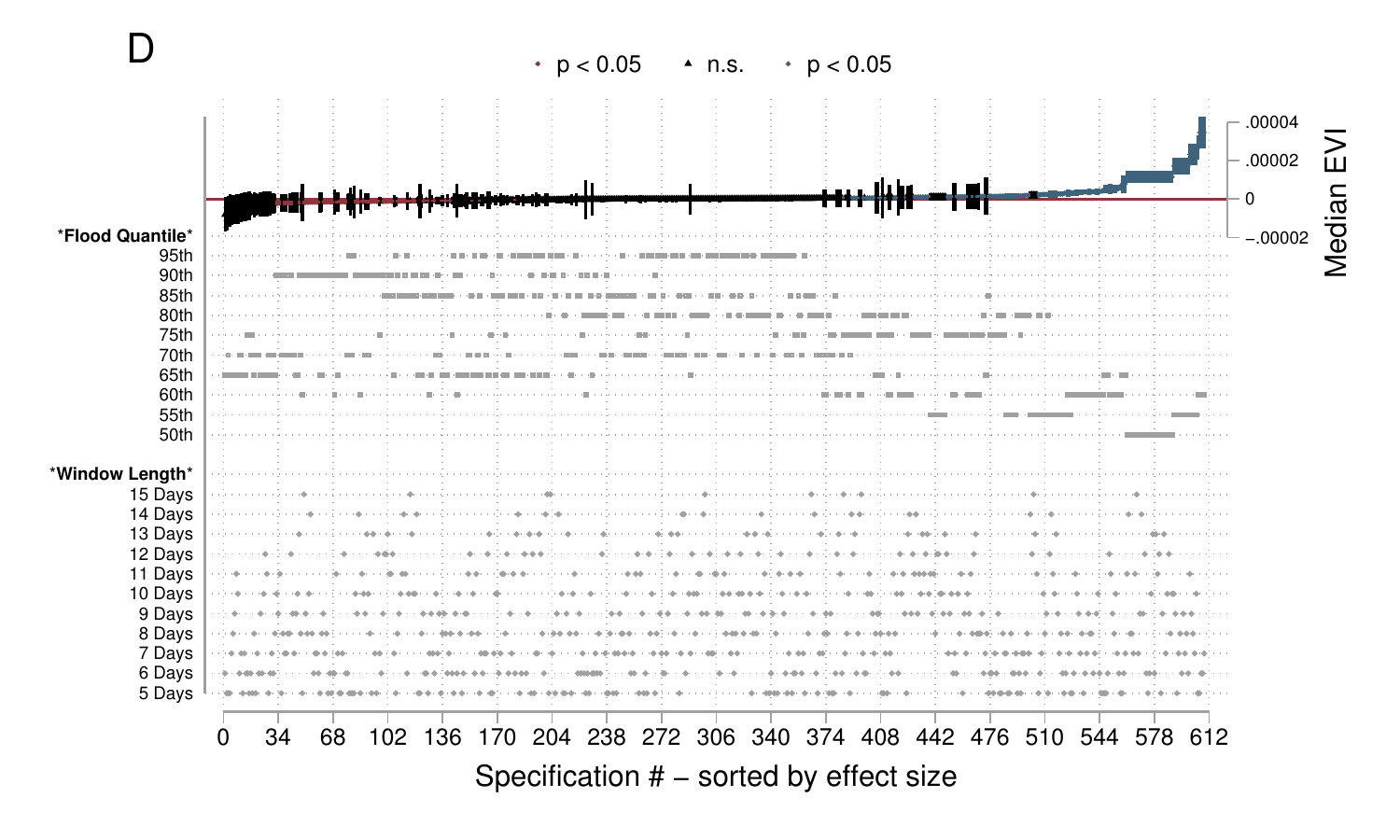}
		\end{center}
		\footnotesize  \textit{Note}: The figure presents results from district-level TWFE regressions. Each panel in the figure displays coefficient estimates and $95\%$ confidence intervals on the interaction of STRVs and a flood metric for a specific EVI measure (cumulative, max, mean, median). Specifications are sorted based on coefficient size from smallest (left) to largest (right). The gray diamonds below the coefficients indicate which combination of quantile and flood window was used in the regression.
	\end{minipage}	
\end{figure}
\end{landscape}

Turning to the TWFE results in Figure~\ref{fig:flood_days_win}, which uses the number of days in the window, we see the same pattern as previously in terms of quantiles that produce Goldilocks floods. And we see a stronger pattern in terms of which flood windows matter than previously evident. As before, results do not differ qualitatively when we use different EVI metrics, so we focus on results in Panel A. Of the 612 regressions, 98 ($16\%$) are significant, slightly larger than the share of positive and significant results in the DID and binary TWFE regressions. These positive and significant coefficients again come from the $65^{th}$ to $50^{th}$ quantile. In terms of which flood window matters, around $18\%$ of regressions are positive and significant when the window is between five and 10 days, with only around $10\%$ of regressions producing significant results when the window is greater than 10 days. When the window starts or ends, does not appear to matter much, with about $15\%$ of regressions producing positive and significant results regardless of whether the window starts at day five or day 15. This may suggest that, in terms of EO data at least, getting the correct intensity and duration of flooding is more important than getting correct the exact start and end day of the flood window (length).

Summarizing our results, we run 7,520 DID or TWFE regressions in which we interact different flood measures with proxies for adoption of STRVs. Across all specification and EVI measures, we find that $15\%$ of coefficients are positive and significant - many more than we would expect to find by chance if the true effect of STRVs were zero. Furthermore, the significant coefficients all come from floods categorized at the same quantiles above baseflow: between the $65^{th}$ and $50^{th}$ quantile. While not all flood windows at these flood quantiles produce positive and significant effects, no flood measured at a quantile above the $65^{th}$ or below the $50^{th}$ ever produces a positive and significant results. We take this as strong evidence that 1) STRVs have a positive and significant impact on rice production, as proxied by EVI, during flooding and 2) this effect only materializes during a Goldilocks flood. Both of these outcomes confirm the existing field trial and RCT evidence in that STRVs are effective in reducing yield loss to flooding and that the technology is stochastic, only becoming effective when flooding is ``just right.''

\subsection{Robustness Checks}\label{subsec:district_robust}

While our main results are consistent across specifications and changes to the dependent variable, there are a variety of concerns that other factors may be driving the results. Key among these are: 1) other abiotic stressors, 2) the level of aggregation, and 3) the use of EVI and administrative seed data as proxies for actual yield and adoption. In this section, we briefly summarize a series of checks on our results to determine if they are robust to these other factors.\footnote{Figures and tables related to these results along with additional discussion and data considerations are in Online Appendix~\ref{sec:gis_result_app} and~\ref{sec:plot_app}.}

First, our analysis might be picking up the impact of other abiotic stressors, as well as seed designed to mitigate those stresses, instead of the impact of STRVs. STRVs are just one example of a class of stress tolerant rice varieties, the next most important of which, in Bangladesh, is salinity tolerant rice. In coastal regions, flooding is often tidal, which carries with it (and leaves behind) salt. Potentially compounded this issue is work by \cite{ChenMueller18} that finds very little impact of flooding on rice yields, because farmers adjust to floods by delaying planting date, and instead finds significant impacts of soil salinity. To test if our results are picking up the impacts of salinity tolerant rice, and not submergence tolerant rice, we conduct our event study, DID, and TWFE analysis on our district-level data but exclude the 19 coastal districts where salinity can be a problem. Our event study results are more muted in their measured impact. But, our DID and TWFE tend to be stronger in that coefficients tend to be slightly larger and there are slightly more positive and significant outcomes. We take this as evidence that our results are driven by the effect of submergence tolerant rice on EVI during floods and not salinity tolerant rice.

Second, the choice regarding the unit of aggregation may be an important factor in determining our results. Our main results are at the district-level because that is the unit of observation for the administrative seed data. However, the district-scale measure of STRV seed availability is itself a noisy measure of actual adoption. Further, by aggregating flood and EVI, which is observed in pixels identified as rice, up to the district-level likely smooths over a lot of variation. To test if our results are driven by aggregation to the district-level, we conduct our event study, DID, and TWFE analysis at the upazila-level.\footnote{Bangladesh is divided into 553 upazilas. We combine many extremely small upazilas in the city and metro area of Dhaka, where rice is not grown, to get a final dataset of 503 upazilas.} Results from the event study are stronger in that EVI is larger and increases more rapidly after the introduction of STRVs than EVI does in the district-level panel. However, when we turn to our DID and TWFE results, we find very little evidence of significant impacts. In most of the specification charts, less than $5\%$ of regressions are significant, meaning we cannot discount that those significant results are purely due to chance; though the quantiles that do produce significant results are the same as those in the district-level analysis. One reason our non-event study results might not be robust to using upazila-level EO EVI and flood data is that the seed data is still at the district-level. This mismatch of units means that we are trying to explain intra-district variation in EVI using a single district-wide seed value. This type of mismatch can introduce noise into estimates and in fact we do find confidence intervals to be much larger in the upazila-level results. Coefficients of similar size in district-level and upazila-level specification charts have much larger confidence intervals in the upazila-level results than in the district-level results. However, this explanation is not totally satisfying since the upazila-level results are similar for both TWFE and DID regressions, the later of which do not use any seed data. Regardless of the explanation, what is clear is that our results tend not to be robust to disaggregating EVI and flood measures to the upazila.

Third, our results may be driven by our use of EVI as a proxy for yield and seed dissemination data as a proxy for adoption. Rice is not cultivated at the district-level, or event the upazila-level, but by individual households on small ($< 1$ ha) fields. District-level EO measures may be noisy or biased proxies of the on-the-ground household reality. While no household data exists that allows us to perfectly test the robustness of our results, we do have the three-year Rice Monitoring Survey (RMS) panel that is representative of rice cultivation in Bangladesh, though only includes data post-STRV introduction. We generate the same set of flood metrics as before but using household GPS coordinates. We then estimate the TWFE model using measured household rice yield instead of EVI and household adoption status instead of rice seed availability.\footnote{Online Appendix~\ref{sec:plot_app} discusses limitations in the data both in terms of models we can estimate (event study and DID) and endogeneity of the adoption decision.} Results are robust to the use of household-level data. Though only around $7\%$ of regressions produce positive and significant results, those results come from quantiles in the $65^{th}$ to $55^{th}$ quantile range. Unlike most of the district-level results, the flood window plays a much larger role in determining what exactly is a Goldilocks flood. Significant results only come from floods that fall in the five to ten day window. Floods longer than that never produce significant results. We take this as evidence that our results are not just the outcome of our use of aggregate proxies for yield, flooding, and adoption but that our EO results do in fact reflect the impacts of adopting STRVs by households in Bangladesh.



\section{Conclusion}

There are numerous research questions that economists would like to answer but are unable to because of a lack of data. Recent years have seen a proliferation in the availability of data, in particular remotely sensed earth observation (EO) data. This has allowed economists and other quantitative researchers to answer questions that were previously viewed as unanswerable. However, there is a recency bias in these new data, meaning the past is still a place that remains data poor.

Our study details a new method for overcoming this recency bias in EO data. With this method we conduct an impact evaluation in a data poor environment. To complicate matters, we demonstrate this methodology using a stochastic technology, STRVs, which only have a measurable impact when flooding is ``just right.'' We combine data from a variety of sources and use innovative approaches to generating ground truth data. Using recent high resolution EO data and deep learning algorithms, we are able to infer where flooding occurred in the past and generate EO datasets at a higher accuracy than previously existed. We combine these new flood and rice area maps, which cover the entire country of Bangladesh for 20 years, with administrative data on seed dissemination efforts in order to conduct a large-scale, long-term evaluation of the effectiveness of STRVs.

Taking an agnostic approach, we vary the length and duration of the window in which a flood occurs producing a large number $(>600)$ of floods that are potentially ``just right'' for identifying the treatment effects of STRVs. While a majority of these flood metrics are too little or too much flooding to identify significant impacts, we do find a set of flood metrics that produce positive and significant results. These Goldilocks floods produce consistent results across variation in how we measure EVI and our model specification. These results are also robust to excluding regions of the county that face other abiotic stresses and to using household panel data. However, they are not robust to spatially disaggregating the EO data to the smaller upazila-level. We hypothesize that this is because our primary treatment measure - administrative data on STRV seed availability - is measured only at the larger district-level.

This paper demonstrates the possibilities and challenges for conducting impact evaluations in data poor settings. It also reveals the added dimensionality of the challenge when trying to capture higher order treatment effects associated with stochastic technologies. As the recent economic and geospatial literature shows, and this paper reaffirms, the mismeasurement problem in data can be acute, often biasing or obscure a true signal. However, using the methods developed and deployed in this paper we are able to address the Goldilocks Problem and answer previously unanswerable economic questions, allowing us to reduce the number of places that remain data poor, including the past.


\newpage
\onehalfspace
\bibliographystyle{chicago} 
\bibliography{irri_ref}

\begin{thebibliography}{}

\bibitem[\protect\citeauthoryear{Abay, Wossen, Abate, Stevenson, Michelson, and
  Barrett}{Abay et~al.}{2023}]{AbayEtAl23}
Abay, K.~A., T.~Wossen, G.~T. Abate, J.~R. Stevenson, H.~Michelson, and C.~B.
  Barrett (2023).
\newblock Inferential and behavioral implications of measurement error in
  agricultural data.
\newblock {\em Annual Review of Resource Economics\/}~{\em 15}, 63--83.

\bibitem[\protect\citeauthoryear{Al~Rafi, Josephson, Michler, and Pede}{Al~Rafi
  et~al.}{2023}]{PAP}
Al~Rafi, D.~A., A.~Josephson, J.~D. Michler, and V.~Pede (2023).
\newblock Impact of stress tolerant rice varieties adoption in flood prone
  regions of {S}outh {A}sia.
\newblock OSF Registries. February 9. https://doi.org/10.17605/OSF.IO/YE7PV.

\bibitem[\protect\citeauthoryear{Alauddin, Rashid~Sarker, Islam, and
  Tisdell}{Alauddin et~al.}{2020}]{AlauddinEtAl20}
Alauddin, M., M.~A. Rashid~Sarker, Z.~Islam, and C.~Tisdell (2020).
\newblock Adoption of alternate wetting and drying ({AWD}) irrigation as a
  water-saving technology in {B}angladesh: Economic and environmental
  considerations.
\newblock {\em Land Use Policy\/}~{\em 91}, 104430.

\bibitem[\protect\citeauthoryear{Alix-Garcia, McIntosh, Sims, and
  Welch}{Alix-Garcia et~al.}{2013}]{Alix-GarciaEtAl13}
Alix-Garcia, J., C.~McIntosh, K.~R.~E. Sims, and J.~R. Welch (2013).
\newblock The ecological footprint of poverty alleviation: Evidence from
  {Mexico's} {Oportunidades} program.
\newblock {\em Review of Economics and Statistics\/}~{\em 95\/}(2), 417--435.

\bibitem[\protect\citeauthoryear{Alix-Garcia and Millimet}{Alix-Garcia and
  Millimet}{2023}]{Alix-GarciaEtAl23}
Alix-Garcia, J. and D.~L. Millimet (2023).
\newblock Remotely incorrect? accounting for nonclassical measurement error in
  satellite data on deforestation.
\newblock {\em Journal of the Association of Environmental and Resource
  Economists\/}~{\em 10\/}(5), 1335–1367.

\bibitem[\protect\citeauthoryear{Allcott and Gentzkow}{Allcott and
  Gentzkow}{2017}]{AllcottGentzkow17}
Allcott, H. and M.~Gentzkow (2017).
\newblock Social media and fake news in the 2016 election.
\newblock {\em Journal of Economic Perspectives\/}~{\em 31\/}(2), 211--236.

\bibitem[\protect\citeauthoryear{Asfawa, Battista, and Lipper}{Asfawa
  et~al.}{2016}]{Asfawa2016}
Asfawa, S., F.~D. Battista, and L.~Lipper (2016).
\newblock Agricultural technology adoption under climate change in the {Sahel}:
  Micro-evidence from {Niger}.
\newblock {\em Journal of African Economies\/}~{\em 25}, 637--669.

\bibitem[\protect\citeauthoryear{Azzari, Jain, and Lobell}{Azzari
  et~al.}{2017}]{AzzariEtAl17}
Azzari, G., M.~Jain, and D.~B. Lobell (2017).
\newblock Towards fine resolution global maps of crop yields: Testing multiple
  methods and satellites in three countries.
\newblock {\em Remote Sensing of Environment\/}~{\em 202}, 129--141.

\bibitem[\protect\citeauthoryear{Azzari, Jain, Jeffries, Kilic, and
  Murray}{Azzari et~al.}{2021}]{AzzariEtAl21}
Azzari, G., S.~Jain, G.~Jeffries, T.~Kilic, and S.~Murray (2021).
\newblock Understanding the requirements for surveys to support satellite-based
  crop type mapping: Evidence from {S}ub-{S}aharan {A}frica.
\newblock {\em Remote Sensing\/}~{\em 13\/}(23), 4749.

\bibitem[\protect\citeauthoryear{Bairagi, Bhandari, {Kumar Das}, and
  Mohanty}{Bairagi et~al.}{2021}]{BairagiEtAl21}
Bairagi, S., H.~Bhandari, S.~{Kumar Das}, and S.~Mohanty (2021).
\newblock Flood-tolerant rice improves climate resilience, profitability, and
  household consumption in bangladesh.
\newblock {\em Food Policy\/}~{\em 105}, 102183.

\bibitem[\protect\citeauthoryear{BenYishay, Runfola, Trichler, Dolan, Goodman,
  Parks, Tanner, Heuser, Batra, and Anand}{BenYishay
  et~al.}{2017}]{BenYishayEtAl17}
BenYishay, A., D.~Runfola, T.~Trichler, C.~Dolan, S.~Goodman, B.~Parks,
  J.~Tanner, S.~Heuser, G.~Batra, and A.~Anand (2017).
\newblock A primer on geospatial impact evaluation methods, tools, and
  applications.
\newblock {AIDDATA} Working Paper 44.

\bibitem[\protect\citeauthoryear{Brakenridge}{Brakenridge}{nd}]{Brakenridgend}
Brakenridge, G. (n.d.).
\newblock {G}lobal {A}ctive {A}rchive of {L}arge {F}lood {E}vents.
\newblock Dartmouth Flood Observatory, University of Colorado, USA.
  \url{http://floodobservatory.colorado.edu/ Archives/}.

\bibitem[\protect\citeauthoryear{Brodeur, Cook, and Heyes}{Brodeur
  et~al.}{2020}]{BrodeurEtAl18}
Brodeur, A., N.~Cook, and A.~Heyes (2020).
\newblock Methods matter: P-hacking and publication bias in causal analysis in
  economics.
\newblock {\em American Economic Review\/}~{\em 110\/}(11), 3634--3660.

\bibitem[\protect\citeauthoryear{Burgess, Hansen, Olken, Potapov, and
  Sieber}{Burgess et~al.}{2012}]{BurgessEtAl12}
Burgess, R., M.~Hansen, B.~A. Olken, P.~Potapov, and S.~Sieber (2012).
\newblock The political economy of deforestation in the tropics.
\newblock {\em Quarterly Journal of Economics\/}~{\em 127\/}(4), 1707--1754.

\bibitem[\protect\citeauthoryear{Burke, Driscoll, Lobell, and Ermon}{Burke
  et~al.}{2021}]{BurkeEtAl21}
Burke, M., A.~Driscoll, D.~B. Lobell, and S.~Ermon (2021).
\newblock Using satellite imagery to understand and promote sustainable
  development.
\newblock {\em Science\/}~{\em 371}, eabe8628.

\bibitem[\protect\citeauthoryear{Burke and Lobell}{Burke and
  Lobell}{2017}]{burke_satellite-based_2017}
Burke, M. and D.~B. Lobell (2017).
\newblock Satellite-based assessment of yield variation and its determinants in
  smallholder {African} systems.
\newblock {\em Proceedings of the National Academy of Sciences\/}~{\em
  114\/}(9), 2189--2194.

\bibitem[\protect\citeauthoryear{Carlson, Heilmayr, Gibbs, Noojipady, Burns,
  Morton, Walker, Paoli, and Kremen}{Carlson et~al.}{2018}]{CarlsonEtAl17}
Carlson, K.~M., R.~Heilmayr, H.~K. Gibbs, P.~Noojipady, D.~N. Burns, D.~C.
  Morton, N.~F. Walker, G.~D. Paoli, and C.~Kremen (2018).
\newblock Effect of oil palm sustainability certification on deforestation and
  fire in indonesia.
\newblock {\em Proceedings of the National Academy of Sciences\/}~{\em
  115\/}(1), 121--126.

\bibitem[\protect\citeauthoryear{Chen and Mueller}{Chen and
  Mueller}{2018}]{ChenMueller18}
Chen, J.~J. and V.~Mueller (2018).
\newblock Coastal climate change, soil salinity and human migration in
  bangladesh.
\newblock {\em Nature Climate Change\/}~{\em 8}, 981–985.

\bibitem[\protect\citeauthoryear{Chen, Mueller, Jia, and Tseng}{Chen
  et~al.}{2017}]{ChenEtAl17}
Chen, J.~J., V.~Mueller, Y.~Jia, and S.~K.-H. Tseng (2017).
\newblock Validating migration responses to flooding using satellite and vital
  registration data.
\newblock {\em American Economic Review: Papers \& Proceedings\/}~{\em
  107\/}(5), 441--445.

\bibitem[\protect\citeauthoryear{Dar, de~Janvry, Emerick, Raitzer, and
  Sadoulet}{Dar et~al.}{2013}]{dar_flood-tolerant_2013}
Dar, M.~H., A.~de~Janvry, K.~Emerick, D.~Raitzer, and E.~Sadoulet (2013).
\newblock Flood-tolerant rice reduces yield variability and raises expected
  yield, differentially benefitting socially disadvantaged groups.
\newblock {\em Scientific Reports\/}~{\em 3}, 3315.

\bibitem[\protect\citeauthoryear{Dell}{Dell}{2010}]{Dell10}
Dell, M. (2010).
\newblock The persisten effects of {Peru's} mining \emph{mita}.
\newblock {\em Econometrica\/}~{\em 78\/}(6), 1863--1903.

\bibitem[\protect\citeauthoryear{Dell, Jones, and Olken}{Dell
  et~al.}{2014}]{DellEtAl14}
Dell, M., B.~F. Jones, and B.~A. Olken (2014).
\newblock What do we learn from the weather? the new climate–economy
  literature.
\newblock {\em Journal of Economic Literature\/}~{\em 52\/}(3), 740--798.

\bibitem[\protect\citeauthoryear{Desch{\^e}ne and Greenstone}{Desch{\^e}ne and
  Greenstone}{2007}]{DescheneGreenstone07}
Desch{\^e}ne, O. and M.~Greenstone (2007).
\newblock The economic impacts of climate change: Evidence from agricultural
  output and random fluctuations in weather.
\newblock {\em American Economic Review\/}~{\em 97\/}(1), 354--85.

\bibitem[\protect\citeauthoryear{Dolan, BenYishay, Gr\'{e}pin, Tanner, Kimmel,
  Wheeler, and McCord}{Dolan et~al.}{2019}]{DolanEtAl19}
Dolan, C.~B., A.~BenYishay, K.~A. Gr\'{e}pin, J.~C. Tanner, A.~D. Kimmel, D.~C.
  Wheeler, and G.~C. McCord (2019).
\newblock The impact of an insecticide treated bednet campaign on all-cause
  child mortality: A geospatial impact evaluation from the {Democratic}
  {Republic} of {Congo}.
\newblock {\em PLOS ONE\/}~{\em 14\/}(2), e0212890.

\bibitem[\protect\citeauthoryear{Donaldson and Storeygard}{Donaldson and
  Storeygard}{2016}]{DonaldsonStoreygard16}
Donaldson, D. and A.~Storeygard (2016).
\newblock The view from above: Applications of satellite data in economics.
\newblock {\em Journal of Economic Perspectives\/}~{\em 30\/}(4), 171--198.

\bibitem[\protect\citeauthoryear{Duflo, Banerjee, Finkelstein, Katz, Olken, and
  Sautman}{Duflo et~al.}{2020}]{DufloEtAl20}
Duflo, E., A.~Banerjee, A.~Finkelstein, L.~F. Katz, B.~A. Olken, and A.~Sautman
  (2020).
\newblock In praise of moderation: Suggestions for the scope and use of
  pre-analysis plans for {RCTs} in economics.
\newblock {NBER} Working Paper 26993.

\bibitem[\protect\citeauthoryear{Ehrlich}{Ehrlich}{1975}]{Ehrlich75}
Ehrlich, I. (1975).
\newblock The deterrent effect of capital punishment: A question of life and
  death.
\newblock {\em American Economic Review\/}~{\em 65\/}(3), 397--417.

\bibitem[\protect\citeauthoryear{Emerick, de~Janvry, Sadoulet, and Dar}{Emerick
  et~al.}{2016}]{emerick_technological_2016}
Emerick, K., A.~de~Janvry, E.~Sadoulet, and M.~H. Dar (2016).
\newblock Technological innovations, downside risk, and the modernization of
  agriculture.
\newblock {\em American Economic Review\/}~{\em 106\/}(6), 1537--1561.

\bibitem[\protect\citeauthoryear{Emerick and Ronald}{Emerick and
  Ronald}{2019}]{EmerickRonald19}
Emerick, K. and P.~C. Ronald (2019).
\newblock \emph{Sub1} rice: Engineering rice for climate change.
\newblock {\em Cold Spring Harbor Perspectives in Biology\/}~{\em 11\/}(12),
  a034637.

\bibitem[\protect\citeauthoryear{Evenson and Gollin}{Evenson and
  Gollin}{2003}]{EvensonGollin03}
Evenson, R. and D.~Gollin (2003).
\newblock {\em Crop Variety Improvement and Its Effect on Productivity: The
  Impact of International Agricultural Research}.
\newblock Wallingford, U.K.: CAB International.

\bibitem[\protect\citeauthoryear{Ferraro and Simorangkir}{Ferraro and
  Simorangkir}{2020}]{FerraroEtAl20}
Ferraro, P.~J. and R.~Simorangkir (2020).
\newblock Conditional cash transfers to alleviate poverty also reduced
  deforestation in indonesia.
\newblock {\em Science Advances\/}~{\em 6\/}(24), eaaz1298.

\bibitem[\protect\citeauthoryear{Giezendanner, Mukherjee, Purri, Thomas,
  Mauerman, Islam, and Tellman}{Giezendanner et~al.}{2023}]{GiezendannerEtAl23}
Giezendanner, J., R.~Mukherjee, M.~Purri, M.~Thomas, M.~Mauerman, A.~S. Islam,
  and B.~Tellman (2023).
\newblock Inferring the past: a combined {CNN–LSTM} deep learning framework
  to fuse satellites for historical inundation mapping.
\newblock In {\em 2023 IEEE/CVF Conference on Computer Vision and Pattern
  Recognition Workshops (CVPRW)}, pp.\  2155--2165.

\bibitem[\protect\citeauthoryear{Gollin, Hansen, and Wingender}{Gollin
  et~al.}{2021}]{GollinEtAl21}
Gollin, D., C.~W. Hansen, and A.~M. Wingender (2021).
\newblock Two blades of grass: The impact of the {Green Revolution}.
\newblock {\em Journal of Political Economy\/}~{\em 129\/}(8), 2344--2384.

\bibitem[\protect\citeauthoryear{Guiteras, Jina, and Mobarak}{Guiteras
  et~al.}{2015}]{GuiterasEtAl15}
Guiteras, R., A.~Jina, and A.~M. Mobarak (2015).
\newblock Satellites, self-reports, and submersion: Exposure to floods in
  bangladesh.
\newblock {\em American Economic Review\/}~{\em 105\/}(5), 232--236.

\bibitem[\protect\citeauthoryear{Gumma, Thenkabail, Maunahan, Islam, and
  Nelson}{Gumma et~al.}{2014}]{gumma2014}
Gumma, M.~K., P.~S. Thenkabail, A.~Maunahan, S.~Islam, and A.~Nelson (2014).
\newblock Mapping seasonal rice cropland extent and area in the high cropping
  intensity environment of {Bangladesh} using {MODIS} 500m data for the year
  2010.
\newblock {\em ISPRS Journal of Photogrammetry and Remote Sensing\/}~{\em 91},
  98--113.

\bibitem[\protect\citeauthoryear{Hawker, Uhe, Paulo, Sosa, Savage, Sampson, and
  Neal}{Hawker et~al.}{2022}]{hawker2022}
Hawker, L., P.~Uhe, L.~Paulo, J.~Sosa, J.~Savage, C.~Sampson, and J.~Neal
  (2022).
\newblock A 30 m global map of elevation with forests and buildings removed.
\newblock {\em Environmental Research Letters\/}~{\em 17\/}(2), 024016.

\bibitem[\protect\citeauthoryear{Hazell, {n}o, Nakamura, and Steinsson}{Hazell
  et~al.}{2022}]{HazellEtAl22}
Hazell, J., J.~H. {n}o, E.~Nakamura, and J.~Steinsson (2022).
\newblock The slope of the {Phillips} curve: Evidence from {U.S. States}.
\newblock {\em Quarterly Journal of Economics\/}~{\em 137\/}(3), 1299--1344.

\bibitem[\protect\citeauthoryear{Hossain, Bose, and Mustafi}{Hossain
  et~al.}{2006}]{hossain_adoption_2006}
Hossain, M., M.~L. Bose, and B.~A.~A. Mustafi (2006).
\newblock Adoption and productivity impact of modern rice varieties in
  {Bangladesh}.
\newblock {\em The Developing Economies\/}~{\em 44\/}(2), 149--166.

\bibitem[\protect\citeauthoryear{Huang, Hsiang, and Gonzalez-Navarro}{Huang
  et~al.}{2021}]{HuangEtAl21}
Huang, L.~Y., S.~M. Hsiang, and M.~Gonzalez-Navarro (2021).
\newblock Using satellite imagery and deep learning to evaluate the impact of
  anti-poverty programs.
\newblock {NBER} Working Paper 29105.

\bibitem[\protect\citeauthoryear{IRRI}{IRRI}{2018}]{irri_climate2018}
IRRI (2018, August).
\newblock Climate change - ready rice.
\newblock https://www.irri.org/climate-change-ready-rice.

\bibitem[\protect\citeauthoryear{Islam, Matsushita, Noguchi, and Ahamed}{Islam
  et~al.}{2021}]{IslamEtAl21}
Islam, M.~M., S.~Matsushita, R.~Noguchi, and T.~Ahamed (2021).
\newblock Development of remote sensing-based yield prediction models at the
  maturity stage of boro rice using parametric and nonparametric approaches.
\newblock {\em Remote Sensing Applications: Society and Environment\/}~{\em
  22}, 100494.

\bibitem[\protect\citeauthoryear{Ismail, Singh, Singh, Dar, and Mackill}{Ismail
  et~al.}{2013}]{ismail_2013}
Ismail, A.~M., U.~S. Singh, S.~Singh, M.~H. Dar, and D.~J. Mackill (2013).
\newblock The contribution of submergence-tolerant ({Sub1}) rice varieties to
  food security in flood-prone rainfed lowland areas in {Asia}.
\newblock {\em Field Crops Research\/}~{\em 152}, 83--93.

\bibitem[\protect\citeauthoryear{Jain, Singh, Rao, Srivastava, Poonia, Blesh,
  Azzari, McDonald, and Lobell}{Jain et~al.}{2019}]{JainEtAl19}
Jain, M., B.~Singh, P.~Rao, A.~K. Srivastava, S.~Poonia, J.~Blesh, G.~Azzari,
  A.~J. McDonald, and D.~B. Lobell (2019).
\newblock The impact of agricultural interventions can be doubled by using
  satellite data.
\newblock {\em Nature Sustainability\/}~{\em 2}, 931–934.

\bibitem[\protect\citeauthoryear{Janzen and Michler}{Janzen and
  Michler}{2021}]{JanzenMichler20}
Janzen, S. and J.~D. Michler (2021).
\newblock Ulysses' pact or {U}lysses' raft: Using pre-analysis plans in
  experimental and non-experimental research.
\newblock {\em Applied Economic Perspectives and Policy\/}~{\em 43\/}(4),
  1286--1304.

\bibitem[\protect\citeauthoryear{Jayachandran, de~Laat, Lambin, Stanton, Audy,
  and Thomas}{Jayachandran et~al.}{2017}]{JayachandranEtAl17}
Jayachandran, S., J.~de~Laat, E.~F. Lambin, C.~Y. Stanton, R.~Audy, and N.~E.
  Thomas (2017).
\newblock Cash for carbon: A randomized trial of payments for ecosystem
  services to reduce deforestation.
\newblock {\em Science\/}~{\em 357}, 267--273.

\bibitem[\protect\citeauthoryear{Johnson}{Johnson}{2016}]{Johnson16}
Johnson, D.~M. (2016).
\newblock A comprehensive assessment of the correlations between field crop
  yields and commonly used {MODIS} products.
\newblock {\em International Journal of Applied Earth Observation and
  Geoinformation\/}~{\em 52}, 65--81.

\bibitem[\protect\citeauthoryear{Josephson and Michler}{Josephson and
  Michler}{2023}]{JosephsonMichler23}
Josephson, A. and J.~D. Michler (2023).
\newblock {\em Research Ethics in Applied Economics: A Practical Guide}.
\newblock New York: Routledge.

\bibitem[\protect\citeauthoryear{Josephson, Michler, Kilic, and
  Murray}{Josephson et~al.}{2025}]{JosephsonEtAl25}
Josephson, A., J.~D. Michler, T.~Kilic, and S.~Murray (2025).
\newblock The mismeasure of weather: Using remotely sensed earth observation
  data in economic contexts.
\newblock World Bank Policy Research Working Paper, No. 11015.

\bibitem[\protect\citeauthoryear{Keynes}{Keynes}{1936}]{Keynes16}
Keynes, J.~M. (1936).
\newblock {\em The General Theory of Employment, Interest, and Money}.
\newblock New York: Houghton Mifflin Harcourt.
\newblock Reprint (2016).

\bibitem[\protect\citeauthoryear{Khan and Roy}{Khan and Roy}{2020}]{khan_2020}
Khan, M. and P.~Roy (2020, September).
\newblock {Aman Farming: Recurring flood ruins a season}.
\newblock
  https://www.thedailystar.net/frontpage/news/aman-farming-recurring-flood-ruins-season-1954913.

\bibitem[\protect\citeauthoryear{Khush and Virk}{Khush and
  Virk}{2000}]{KhushVirk00}
Khush, G. and P.~Virk (2000).
\newblock Rice breeding: Achievements and future strategies.
\newblock {\em Crop Improvement\/}~{\em 27\/}(2), 115--144.

\bibitem[\protect\citeauthoryear{Kilic, Moylan, Ilukor, Mtengula, and
  Pangapanga-Phiri}{Kilic et~al.}{2021}]{KilicEtAl21}
Kilic, T., H.~Moylan, J.~Ilukor, C.~Mtengula, and I.~Pangapanga-Phiri (2021).
\newblock Root for the tubers: Extended-harvest crop production and
  productivity measurement in surveys.
\newblock {\em Food Policy\/}~{\em 102}, 102033.

\bibitem[\protect\citeauthoryear{Kretzschmar, Mbanjo, Magalit, Dwiyanti, Habib,
  Diaz, Hernandez, Huelgas, Malabayabas, Das, and Yamano}{Kretzschmar
  et~al.}{2018}]{KretzschmarEtAl18}
Kretzschmar, T., E.~G.~N. Mbanjo, G.~A. Magalit, M.~S. Dwiyanti, M.~A. Habib,
  M.~G. Diaz, J.~Hernandez, Z.~Huelgas, M.~L. Malabayabas, S.~K. Das, and
  T.~Yamano (2018).
\newblock {DNA} fingerprinting at farm level maps rice biodiversity across
  {Bangladesh} and reveals regional varietal preferences.
\newblock {\em Scientific Reports\/}~{\em 8}, 14920.

\bibitem[\protect\citeauthoryear{Lichtenthaler}{Lichtenthaler}{1998}]{lichtenthaler-1998}
Lichtenthaler, H.~K. (1998, 6).
\newblock {The stress concept in plants: an introduction.}
\newblock {\em Annals of the New York Academy of Sciences\/}~{\em 851},
  187--98.

\bibitem[\protect\citeauthoryear{Liu, Huffman, Qian, Shang, Li, Dong, Davidson,
  and Jing}{Liu et~al.}{2020}]{LiuEtAl20}
Liu, J., T.~Huffman, B.~Qian, J.~Shang, Q.~Li, T.~Dong, A.~Davidson, and
  Q.~Jing (2020).
\newblock Crop yield estimation in the {Canadian Prairies} using
  {Terra/MODIS}-derived crop metrics.
\newblock {\em IEEE Journal of Selected Topics in Applied Earth Observations
  and Remote Sensing\/}~{\em 13}, 2685--2697.

\bibitem[\protect\citeauthoryear{Lobell, Azzari, Burke, Gourlay, Jin, Kilic,
  and Murray}{Lobell et~al.}{2020}]{lobell_eyes_2020}
Lobell, D.~B., G.~Azzari, M.~Burke, S.~Gourlay, Z.~Jin, T.~Kilic, and S.~Murray
  (2020).
\newblock Eyes in the sky, boots on the ground: Assessing satellite- and
  ground-based approaches to crop yield measurement and analysis.
\newblock {\em American Journal of Agricultural Economics\/}~{\em 102\/}(1),
  202--219.

\bibitem[\protect\citeauthoryear{Mackill, Ismail, Singh, Labios, and
  Paris}{Mackill et~al.}{2012}]{MackillEtAl12}
Mackill, D., A.~Ismail, U.~Singh, R.~Labios, and T.~Paris (2012).
\newblock Development and rapid adoption of submergence-tolerant (sub1) rice
  varieties.
\newblock In D.~L. Sparks (Ed.), {\em Advances in Agronomy}, Volume 115, pp.\
  299--352. Academic Press.

\bibitem[\protect\citeauthoryear{Marx}{Marx}{1867}]{Marx67}
Marx, K. (1867).
\newblock {\em Capital: A Critique of Political Economy}, Volume~1.
\newblock New York: International Publishers.
\newblock Reprint (1967).

\bibitem[\protect\citeauthoryear{Michler, Rafi, Giezendanner, Josephson, Pede,
  and Tellman}{Michler et~al.}{2024}]{popPAP}
Michler, J.~D., D.~A.~A. Rafi, J.~Giezendanner, A.~Josephson, V.~O. Pede, and
  E.~Tellman (2024).
\newblock Impact evaluations in data poor settings: The case of stress-tolerant
  rice varieties in bangladesh.
\newblock https://arxiv.org/abs/2409.02201.

\bibitem[\protect\citeauthoryear{Mishra, Mottaleb, Khanal, and Mohanty}{Mishra
  et~al.}{2015}]{mishra_abiotic_2015}
Mishra, A.~K., K.~A. Mottaleb, A.~R. Khanal, and S.~Mohanty (2015).
\newblock Abiotic stress and its impact on production efficiency: {The} case of
  rice farming in {Bangladesh}.
\newblock {\em Agriculture, Ecosystems \& Environment\/}~{\em 199}, 146--153.

\bibitem[\protect\citeauthoryear{Mkhabela, Bullock, Raj, Wang, and
  Yang}{Mkhabela et~al.}{2011}]{MkhabelaEtAl11}
Mkhabela, M., P.~Bullock, S.~Raj, S.~Wang, and Y.~Yang (2011).
\newblock Crop yield forecasting on the {Canadian Prairies} using {MODIS NDVI}
  data.
\newblock {\em Agricultural and Forest Meteorology\/}~{\em 151\/}(3), 385--393.

\bibitem[\protect\citeauthoryear{Mottaleb, Gumma, Mishra, and Mohanty}{Mottaleb
  et~al.}{2015}]{mottaleb_quantifying_2015}
Mottaleb, K.~A., M.~K. Gumma, A.~K. Mishra, and S.~Mohanty (2015).
\newblock Quantifying production losses due to drought and submergence of
  rainfed rice at the household level using remotely sensed {MODIS} data.
\newblock {\em Agricultural Systems\/}~{\em 137}, 227--235.

\bibitem[\protect\citeauthoryear{Nagin}{Nagin}{2013}]{Nagin13}
Nagin, D.~S. (2013).
\newblock Deterrence: A review of the evidence by a criminologist for
  economists.
\newblock {\em Annual Review of Economics\/}~{\em 5}, 83--105.

\bibitem[\protect\citeauthoryear{Nature-Education}{Nature-Education}{2014}]{nature_ed2014}
Nature-Education (2014).
\newblock {General Transcription factor / Transcription factor | Learn Science
  at Scitable}.
\newblock https://www.nature.com/scitable/definition/transcription-factor-167/.

\bibitem[\protect\citeauthoryear{Neumark}{Neumark}{2001}]{Neumark01}
Neumark, D. (2001).
\newblock The employment effects of minimum wages: Evidence from a prespecified
  research design.
\newblock {\em Industrial Relations\/}~{\em 40\/}(1), 121--44.

\bibitem[\protect\citeauthoryear{Pandey, Gauchan, Malabayabas, Bool-Emerick,
  and Hardy}{Pandey et~al.}{2012}]{gauchan_patterns_2012}
Pandey, S., D.~Gauchan, M.~Malabayabas, M.~Bool-Emerick, and B.~Hardy (2012).
\newblock {\em Patterns of Adoption of Improved Rice Varieties and Farm-Level
  Impacts in Stress-Prone Rainfed Areas in South Asia}.
\newblock Los Banos: International Rice Research Institute.

\bibitem[\protect\citeauthoryear{Park, Lee, Seto, Hochberg, Wong, Miller,
  Takasaki, Kubota, Oozeki, Doshi, Midzik, Hanich, Sullivan, Woods, and
  Kroodsma}{Park et~al.}{2020}]{ParkEtAl20}
Park, J., J.~Lee, K.~Seto, T.~Hochberg, B.~A. Wong, N.~A. Miller, K.~Takasaki,
  H.~Kubota, Y.~Oozeki, S.~Doshi, M.~Midzik, Q.~Hanich, B.~Sullivan, P.~Woods,
  and D.~A. Kroodsma (2020).
\newblock Illuminating dark fishing fleets in {North Korea}.
\newblock {\em Science Advances\/}~{\em 6\/}(30), eabb1197.

\bibitem[\protect\citeauthoryear{Phillips}{Phillips}{1958}]{Phillips58}
Phillips, A. (1958).
\newblock The relation between unemployment and the rate of change of money
  wage rates in the {United Kingdom}, 1861–1957.
\newblock {\em Economica\/}~{\em 25\/}(100), 283--299.

\bibitem[\protect\citeauthoryear{Ray, Panda, and Sarkar}{Ray
  et~al.}{2017}]{ray_can17}
Ray, A., D.~Panda, and R.~K. Sarkar (2017, June).
\newblock Can rice cultivar with submergence tolerant quantitative trait locus
  ( \textit{{SUB1}} ) manage submergence stress better during reproductive
  stage?
\newblock {\em Archives of Agronomy and Soil Science\/}~{\em 63\/}(7),
  998--1008.

\bibitem[\protect\citeauthoryear{Rockoff}{Rockoff}{2020}]{Rockoff20}
Rockoff, H. (2020).
\newblock Off to a good start: The {NBER} and the measurement of national
  income.
\newblock {NBER} Working Paper 26895.

\bibitem[\protect\citeauthoryear{Roth, Sant’Anna, Bilinski, and Poe}{Roth
  et~al.}{2023}]{RothEtAl23}
Roth, J., P.~H. Sant’Anna, A.~Bilinski, and J.~Poe (2023).
\newblock What’s trending in difference-in-differences? a synthesis of the
  recent econometrics literature.
\newblock {\em Journal of Econometrics\/}~{\em 235\/}(2), 2218--2244.

\bibitem[\protect\citeauthoryear{Sanglestsawai, Rejesus, and
  Yorobe}{Sanglestsawai et~al.}{2014}]{sanglestsawai_lower_2014}
Sanglestsawai, S., R.~M. Rejesus, and J.~M. Yorobe (2014).
\newblock Do lower yielding farmers benefit from {Bt} corn? {Evidence} from
  instrumental variable quantile regressions.
\newblock {\em Food Policy\/}~{\em 44}, 285--296.

\bibitem[\protect\citeauthoryear{Sarkar, Panda, Reddy, Patnaik, Mackill, and
  Ismail}{Sarkar et~al.}{2009}]{SarkarEtAl09}
Sarkar, R.~K., D.~Panda, J.~N. Reddy, S.~S.~C. Patnaik, D.~J. Mackill, and
  A.~M. Ismail (2009).
\newblock Performance of submergence tolerant rice (\emph{{O}ryza sativa})
  genotypes carrying the \emph{{S}ub1} quantitative trait locus under stressed
  and non-stressed natural field conditions.
\newblock {\em Indian Journal of Agricultural Sciences\/}~{\em 79\/}(11),
  876--883.

\bibitem[\protect\citeauthoryear{Saunders, Giezendanner, Tellman, Islam,
  Bhuyan, and Islam}{Saunders et~al.}{2023}]{SaunderEtAl23}
Saunders, A., J.~Giezendanner, B.~Tellman, A.~Islam, A.~Bhuyan, and A.~Islam
  (2023).
\newblock A comparison of remote sensing approaches to assess the devastating
  {M}ay-{J}une 2022 flooding in {S}ylhet, {B}angladesh.
\newblock In {\em IGARSS 2023 - 2023 IEEE International Geoscience and Remote
  Sensing Symposium}, pp.\  452--455.

\bibitem[\protect\citeauthoryear{Sevanthi, Prakash1, and
  Shanmugavadivel}{Sevanthi et~al.}{2019}]{sevanthi_2019}
Sevanthi, A.~M., C.~Prakash1, and P.~Shanmugavadivel (2019).
\newblock {Recent progress in rice varietal development for abiotic stress
  tolerance}.
\newblock In M.~Hasanuzzaman, M.~Fujita, K.~Nahar, and J.~K. Biswas (Eds.),
  {\em Advances in Rice Research for Abiotic Stress Tolerance}, pp.\  47--68.
  Woodhead Publishing.

\bibitem[\protect\citeauthoryear{Singh, Mackill, and Ismail}{Singh
  et~al.}{2009}]{SinghEtAl09}
Singh, S., D.~J. Mackill, and A.~M. Ismail (2009).
\newblock Responses of \emph{{SUB1}} rice introgression lines to submergence in
  the field: Yield and grain quality.
\newblock {\em Field Crops Research\/}~{\em 113\/}(1), 12--23.

\bibitem[\protect\citeauthoryear{Singh, Mackill, and Ismail}{Singh
  et~al.}{2011}]{SinghEtAl11}
Singh, S., D.~J. Mackill, and A.~M. Ismail (2011).
\newblock Tolerance of longer-term partial stagnant flooding is independent of
  the sub1 locus in rice.
\newblock {\em Field Crops Research\/}~{\em 121\/}(3), 311--323.

\bibitem[\protect\citeauthoryear{Singh, Dar, Singh, Zaidi, Bari, Mackill,
  Collard, Singh, Singh, Reddy, and Ismail}{Singh
  et~al.}{2013}]{singh_field_2013}
Singh, U., M.~Dar, S.~Singh, N.~Zaidi, M.~Bari, D.~Mackill, B.~Collard,
  V.~Singh, J.~Singh, J.~Reddy, and R.~Ismail (2013).
\newblock Field {Performance}, {Dissemination}, {Impact} and {Tracking} {Of}
  {Submergence} {Tolerant} ({Sub1}) {Rice} {Varieties} in {South} {Asia}.
\newblock {\em SABRAO journal of breeding and genetics\/}~{\em 45}, 112--131.

\bibitem[\protect\citeauthoryear{Singha, Dong, Zhang, and Xiao}{Singha
  et~al.}{2019}]{singha2019a}
Singha, M., J.~Dong, G.~Zhang, and X.~Xiao (2019).
\newblock High resolution paddy rice maps in cloud-prone {Bangladesh} and
  {Northeast} {India} using {Sentinel}-1 data.
\newblock {\em Scientific Data\/}~{\em 6\/}(1), 26.

\bibitem[\protect\citeauthoryear{Stanton}{Stanton}{2007}]{Stanton07}
Stanton, B.~F. (2007).
\newblock {\em George F. Warren: Farm Economist}.
\newblock Ithaca: Cornell University.

\bibitem[\protect\citeauthoryear{Tellman, Sullivan, Kuhn, Kettner, Doyle,
  Brakenridge, Erickson, and Slayback}{Tellman et~al.}{2021}]{tellman-2021}
Tellman, B., J.~P. Sullivan, C.~C. Kuhn, A.~J. Kettner, C.~S. Doyle, G.~R.
  Brakenridge, T.~Erickson, and D.~Slayback (2021, 8).
\newblock {Satellite imaging reveals increased proportion of population exposed
  to floods}.
\newblock {\em Nature\/}~{\em 596\/}(7870), 80--86.

\bibitem[\protect\citeauthoryear{Thomas, Tellman, Osgood, DeVries, Islam,
  Steckler, Goodman, and Billah}{Thomas et~al.}{2023}]{ThomasEtAl23}
Thomas, M., E.~Tellman, D.~Osgood, B.~DeVries, A.~S. Islam, M.~S. Steckler,
  M.~Goodman, and M.~Billah (2023).
\newblock A framework to assess remote sensing algorithms for satellite-based
  flood index insurance.
\newblock {\em IEEE Journal of Selected Topics in Applied Earth Observations
  and Remote Sensing\/}~{\em 16}, 2589--2604.

\bibitem[\protect\citeauthoryear{Weiss, Jacob, and Duveiller}{Weiss
  et~al.}{2020}]{WeissEtAl20}
Weiss, M., F.~Jacob, and G.~Duveiller (2020).
\newblock Remote sensing for agricultural applications: A meta-review.
\newblock {\em Remote Sensing of Environment\/}~{\em 236}, 111402.

\bibitem[\protect\citeauthoryear{Wollburg, Tiberti, and Zezza}{Wollburg
  et~al.}{2021}]{Wollburg21}
Wollburg, P., M.~Tiberti, and A.~Zezza (2021).
\newblock Recall length and measurement error in agricultural surveys.
\newblock {\em Food Policy\/}~{\em 100}, 102003.

\bibitem[\protect\citeauthoryear{Wooldridge}{Wooldridge}{2021}]{Wooldridge21}
Wooldridge, J.~M. (2021).
\newblock Two-way fixed effects, the two-way mundlak regression, and
  difference-in-differences estimators.
\newblock {SSRN}
  \href{http://dx.doi.org/10.2139/ssrn.3906345}{http://dx.doi.org/10.2139/ssrn.3906345}.

\bibitem[\protect\citeauthoryear{Xiao, Boles, Frolking, Li, Babu, Salas, and
  Moore~III}{Xiao et~al.}{2006}]{XiaoEtAl06}
Xiao, X., S.~Boles, S.~Frolking, C.~Li, J.~Y. Babu, W.~Salas, and B.~Moore~III
  (2006).
\newblock Mapping paddy rice agriculture in {S}outh and {S}outheast {A}sia
  using multi-temporal {MODIS} images.
\newblock {\em Remote Sensing of Environment\/}~{\em 100\/}(1), 95--113.

\bibitem[\protect\citeauthoryear{Xu}{Xu}{2006}]{Xu06}
Xu, H. (2006).
\newblock Modification of normalised difference water index ({NDWI}) to enhance
  open water features in remotely sensed imagery.
\newblock {\em International Journal of Remote Sensing\/}~{\em 27\/}(14),
  3025--3033.

\bibitem[\protect\citeauthoryear{Xu and Mackill}{Xu and
  Mackill}{1996}]{XuMackill96}
Xu, K. and D.~J. Mackill (1996).
\newblock A major locus for submergence tolerance mapped on rice chromosome 9.
\newblock {\em Molecular Breeding\/}~{\em 2}, 219--224.

\bibitem[\protect\citeauthoryear{Yamano}{Yamano}{2017}]{rms-2017}
Yamano, T. (2017).
\newblock {Rice Monitoring Survey: South Asia}.
\newblock \url{https://doi.org/10.7910/DVN/0VPRGD}.

\bibitem[\protect\citeauthoryear{Yamano, Malabayabas, Habib, and Das}{Yamano
  et~al.}{2018}]{yamano_neighbors_2018}
Yamano, T., M.~L. Malabayabas, M.~A. Habib, and S.~K. Das (2018).
\newblock Neighbors follow early adopters under stress: panel data analysis of
  submergence-tolerant rice in northern {Bangladesh}.
\newblock {\em Agricultural Economics\/}~{\em 49\/}(3), 313--323.

\bibitem[\protect\citeauthoryear{Yamauchi, Shimamura, Nakazono, and
  Mochizuki}{Yamauchi et~al.}{2013}]{yamauchi_2013}
Yamauchi, T., S.~Shimamura, M.~Nakazono, and T.~Mochizuki (2013).
\newblock Aerenchyma formation in crop species: A review.
\newblock {\em Field Crops Research\/}~{\em 152}, 8--16.

\end{thebibliography}


\newpage 
\FloatBarrier

\begin{landscape}
\begin{figure}[!htbp]
	\begin{minipage}{\linewidth}		
		\caption{Construction of Rice Area Maps and Workflow for Flood Maps}
		\label{fig:rice_flood}
		\begin{center}
			\includegraphics[width=.59\linewidth,keepaspectratio]{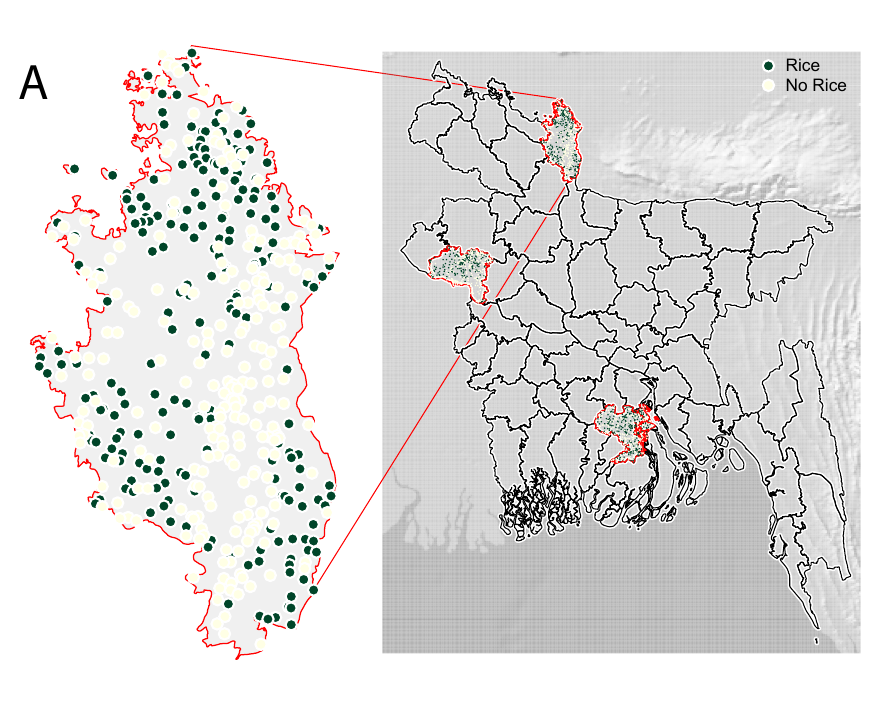} \includegraphics[width=.39\linewidth,keepaspectratio]{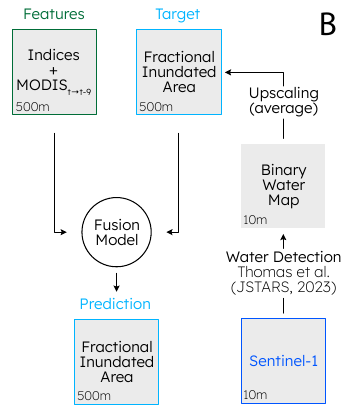}
		\end{center}
		\footnotesize  \textit{Note}: Panel A shows an example of the ``ground truth'' data from a single district in a single year. For the three districts (Barisal, Kurigam, and Rajshahi) we selected a stratified random sample of 150 points (75 points each for rice and non-rice areas) in each district for a total of 450 points. The map shows the example of rice (green) and not rice (white) for a single year in the district of Kurigram. Panel B summarizes the fusion model used to create historic flood maps for Bangladesh as described in \cite{GiezendannerEtAl23}. The target flood maps are generated from Sentinel-1 data. Water is detected with a dynamic thresholding algorithm tailored to Bangladesh and then upscaled (averaged) to MODIS resolution. The fusion model uses hydrologically relevant indices (elevation, slope, and height above nearest drainage) and a time series of 10 MODIS Terra 8-day composite images to regress the target fractional inundated area.
	\end{minipage}	
\end{figure}
\end{landscape}

\clearpage
\newpage
\appendix
\onehalfspacing

\begin{center}
	\section*{Online-Only Appendix to ``Impact Evaluations in Data Poor Settings: The Case of Stress Tolerant Rice Varieties in Bangladesh''} \label{sec:app}
\end{center}


\FloatBarrier
\setcounter{table}{0}
\renewcommand{\thetable}{A\arabic{table}}
\setcounter{figure}{0}
\renewcommand{\thefigure}{A\arabic{figure}}


\section{\emph{Sub1} Varietal Development and Dissemination} \label{sec:Sub1_agron_app}

In the paper, we briefly summarize the development, agronomic characteristics, and dissemination of STRVs. In this Appendix we expand on those details and provide additional evidence regarding the when and how STRVs will generate non-zero treatment effects relative to non-STRVs.

The process of developing high yielding STRVs began when researchers identified a rice variety in India named \textit{Dhalputtia}, which, despite its poor grain quality and yield, possesses an unusual ability to survive complete submergence for over 14 days. In 1980, scientists identified the \textit{Sub1} locus and its associated gene, which subsequently led to the isolation of the \textit{Sub1} gene \citep{XuMackill96}. This process of genetic mapping began with two parent varieties: a japonica rice from California and a submergence-tolerant line derived from the donor variety \textit{Dhalputtia}. 

All STRVs exhibit a similar cellular and molecular mechanism. The \textit{Sub1} sequence closely resembles a protein that functions as a transcription factor.\footnote{Transcription factors are proteins that regulate genetic expression by binding to specific DNA sequences \citep{nature_ed2014}} This transcription factor facilitates the accumulation of the \textit{ethylene} hormone in response to submergence stress. The accumulation of \textit{ethylene} also triggers the production of \textit{gibberellic acid}. The \textit{ethylene} hormone is critical for the plant's vegetative growth, while \textit{gibberellic acid} promotes the elongation of plant shoots. During complete submergence, however, shoot elongation is reduced. Additionally, other plant metabolic processes, such as carbohydrate consumption and chlorophyll breakdown, are slowed down, activating alternative energy pathways. As a result, the plant enters a ``hold your breath'' state, conserving energy until the flooding subsides \citep{EmerickRonald19}. In contrast, normal rice varieties attempt to elongate their stems and leaves to escape deep submergence. This strategy often results in excessive energy expenditure, leaving the plant unable to recover once the submergence is over. So, the flood-escaping strategy for conventional high-yielding rice varieties differs from that of STRVs, making the latter more adaptable and economically viable \citep{irri_climate2018}.

Having isolated \emph{Sub1} and understood how it conveys submergence tolerance, researchers at IRRI were then able to use a combination of traditional cross-breeding methods and marker assisted backcross breeding to create Swarna-Sub1, a hybrid with the popular high yielding Swarna rice variety \citep{yamauchi_2013}. \cite{ismail_2013}, \cite{irri_climate2018}, and \cite{EmerickRonald19} have estimated that there is no yield penalty for STRVs under normal condition but a yield advantage of around 1 to 3 t/ha during flooding situation. Additionally, \cite{ray_can17} highlight that the \textit{Sub1} cultivar outperforms after complete submergence at reproductive stage compared to traditional varieties in terms of carbohydrate quantities and total dehydrogenase.

Numerous studies have been conducted to assess the yield of STRVs at various stages of their growth cycle. Researchers have experimented with different levels of submergence, including variations in duration and water depth, yielding distinctive results. For instance, \cite{singh_field_2013} report that after five days of complete submergence, Swarna-Sub1 rice achieved a yield of 3.98 t/ha, compared to 2.68 t/ha for the traditional Swarna. When submergence extended to 7 to 9 days with water depths of 6 to 9 feet, traditional rice varieties failed to produce any yield. Conversely, the \textit{Sub1} variety experienced only a 10-30\% reduction in yield. 

In a separate study, \cite{ismail_2013} found that brief, late flooding during the panicle initiation stage could diminish the yield of all traditional rice varieties to 1.7 t/ha, whereas the \textit{Sub1} variety yielded 4.75 t/ha. However, if flooding persisted for more than 10 days at this stage, the average yield of the \textit{Sub1} variety dropped to 3 t/ha. This observation suggests that while mortality during the reproductive stage is not a significant concern, maintaining yield remains a critical issue \citep{ray_can17}.

Given the success of the \emph{Sub1} varieties, the Bill and Melinda Gates Foundation (BMGF) supported production and dissemination of the seeds under the IRRI-Africa Rice collaboration Stress-Tolerant Rice for Africa and South Asia (STRASA) program \citep{EmerickRonald19}. As part of the dissemination effort, BMGF funded, through a variety of initiatives, both an RCT on STRV impact in Odisha (2011-2012) \citep{dar_flood-tolerant_2013} and the Rice Monitoring Survey (RMS) in Bangladesh, India and Nepal (2014-2017) \citep{KretzschmarEtAl18}. The RCT produced evidence of the impact of STRVs in the 64 treated villages for farmers who were given seed. STRVs reduced downside risk and resulted in farmers investing more in production through increased cultivation area, fertilizer use, and credit utilization \citep{emerick_technological_2016}. The RMS produced evidence of the extent of adoption in South Asia. Farmers who experienced flooding in the previous year were more likely to adopt and the adoption impact was larger for neighbors of early adopters \citep{yamano_neighbors_2018}. There are also numerous studies coming out of the STRASA program in Asia and Africa that examine the correlation between adoption of STRVs and various outcomes, such as yield, profit, and rice consumption. A non-exhaustive list includes \cite{Asfawa2016, BairagiEtAl21, gauchan_patterns_2012, hossain_adoption_2006, mottaleb_quantifying_2015, sanglestsawai_lower_2014}.


\newpage
\FloatBarrier
\setcounter{table}{0}
\renewcommand{\thetable}{B\arabic{table}}
\setcounter{figure}{0}
\renewcommand{\thefigure}{B\arabic{figure}}


\section{Pre-Analysis Plan} \label{sec:framework}

Using EO data as proxies for on-the-ground truth is common in geography and becoming common in economics. However, for many variables, there is disagreement on the best way to construct EO proxy metrics. This includes the best way to quantify seasonal EVI to proxy for yields and the best way to quantify flooding that is agronomically relevant. This issue is heightened for flooding in our case because of the Goldilocks Problem. In particular, the plausible variation in the flood window creates an opportunity for an unscrupulous researcher to $p$-hack or data mine their way to significant results, testing a large number of flood metrics but only reporting ones that produce the desired outcomes. Our pre-analysis plan OSF was designed to combat the temptation to find and report results from only a subset of proxies that we might try \citep{PAP}.\footnote{Note that the PAP was registered on 9 February 2023. While we refer to the third wave of panel data as 2022, this is because we refer to all cropping season data by the year of planting not the year of harvest. Thus data from 2014, 2017, and 2022 all come from the \emph{Aman} season in which planting occurred in 2014, 2017, or 2022 and harvest at the end of that calendar year, extending into the subsequent calendar year. Data collection for 2022 was not completed until 12 March 2023. Timestamps on the data are available for verification.} However, in pre-specifying the analysis, we under-estimated the sensitivity of our results to precisely defined flood measures. All of our pre-specified flood measures produced null results. It was only after the fact that we began to explore this sensitivity and test new flood metrics that were not pre-specified. Results from the original, pre-specified approach are publicly available in our populated pre-analysis plan \citep{popPAP}.

While pre-analysis plans (PAPs) are common in lab and field experiments, they remain relatively uncommon for observational studies. However, \cite{JosephsonMichler23} argue that any studying collecting new data, regardless of the type of data, or trying to establish causal effects should develop and register a PAP. They point out that the very first PAP in economics was observational in nature and was created by \cite{Neumark01}, who was studying the impacts of changes to minimum wage on employment - a particularly contentious issue - and one in which a PAP could be developed ahead of an announced wage increase.

Our study, though observational, is well suited to a PAP given that the research relies on a newly collected round of data and that it uses DID and TWFE methods to establish causal effects of STRV adoption. In addition to preventing the $p$-hacking that \cite{BrodeurEtAl18} shows can be common to DID methods, the plan also helps limit research degrees of freedom - the ability of researchers to make many reasonable or justifiable decision during the data cleaning and analysis process. In our context, it is unclear what the best EO proxy is for agronomically relevant flooding or for yield. There are also a large number of reasonable criteria to use for deciding how to deal with outliers, how to convert, scale, and aggregate data, and how to deal with missing values. In the absence of a PAP, a researcher might be tempted to try many alternative options in these domains and then only present the choices that result it strongly significant results. Our PAP ties our hands so that we can avoid the temptation to selective report results favorable to our hypothesis.

The initial draft of this paper, which is 
\href{https://doi.org/10.48550/arXiv.2409.02201}{available on arXiv}, followed the PAP with two substantial deviations. First, we initially planned to measure flooding as the total number of distinct flood events. Since we filed our PAP, the research team developed the fusion model described in \cite{GiezendannerEtAl23} that provides a fractional index of inundation in a pixel. This new approach provides much greater opportunity to develop more precise measures of flooding than the original count of flood events. Second, the event study regression was not pre-specified but was added at the suggestion of Kyle Emerick during a seminar. We view the version of the manuscript on arXiv as our populated pre-analysis plan that reports on all pre-specified results

During the review process, we received recommendations from two anonymous reviewers and the handling editor, Ariel BenYishay, regarding substantial changes to both data and method. These suggestions included dropping the pre-specified IV analysis and focusing results on the large set of potential Goldilocks floods. This involved a major revision and refocus of the paper, with a number of pre-specified results being dropped or relegated to this online appendix. These changes were made in the spirit of \cite{DufloEtAl20} and \cite{JanzenMichler20}, who advocate authors to not stick to an inferior pre-specified method when a new, superior method becomes available. In moving away from our pre-specified approach, we have endeavored to be as transparent as possible, which is why we report all regressions that we run in the specification charts in the paper and why we have included an explicit discussion of our dropping our pre-specified analysis to focus on what many would consider exploratory analysis.


\newpage
\FloatBarrier
\setcounter{table}{0}
\renewcommand{\thetable}{C\arabic{table}}
\setcounter{figure}{0}
\renewcommand{\thefigure}{C\arabic{figure}}


\section{Why No Other Data} \label{sec:data_app}

In the paper we make the claim that no socioeconomic data exists to allow for a traditional impact evaluation of the long-term, large scale effects of STRV adoption in Bangladesh. In this Appendix, we provide evidence to support that claim by discussing the existing nationally or sub-nationally representative surveys, along with their strengths and weakness for answering our research question.

We searched through government, non-government, and university databases to collect a set of possible socioeconomic survey data that would allow us to conduct a strongly identified impact evaluation of STRVs. Recall, our criteria encompass the following prerequisites:

    \begin{enumerate}
        \item must be panel data,
        \item must have observations before and after 2010m
        \item must be representative at the national level or at least of rice cultivation, and
        \item must have data on rice varieties grown and yields. 
    \end{enumerate}

\noindent Many candidate datasets satisfied none or only one of the criteria. Frequently a dataset would have multiple observations over time but not contain information on rice variety, the critical data need. Only a few satisfied three of the criteria and none satisfied all four. The best three data sources, other than the RMS data we used in this study, are:

\begin{itemize}
    \item \href{https://vdsa.icrisat.org/vdsa-database.aspx}{Village Dynamics in South Asia (VDSA)}: This dataset is an outcome of a research initiative by the International Crop Research Institute for the Semi-Arid Tropics (ICRISAT).. Spanning 42 villages in Bangladesh, the dataset comprises survey data from 1,831 households. The survey is designed to be representative of agricultural households in Bangladesh, is a panel, and has data on varieties and yields. However, there is one critical limitation:
        \begin{itemize}
            \item The temporal scope of the data panel is limited, spanning from 2009 to 2014, in contrast to the RMS dataset, which extends from 2014 to 2022. Given that seed availability remained very low as late as 2015, the VDSA does not extend far enough past 2010 to capture adoption. In fact, adoption of STRVs in the 2014 round of the VDSA is $0\%$, making the VDSA unusable for studying STRV adoption.
        \end{itemize}

    \item \href{https://dataverse.harvard.edu/dataverse/IFPRI/?q=BIHS}{Bangladesh Integrated Household Survey (BIHS)}: This survey was implemented by the International Food Policy Research Institute (IFPRI). It is nationally representative, a panel, and contains a host of information on plot-level agricultural production and practices, dietary intake of individual household members, anthropometric measurements (height and weight) of all household members, and data to measure women’s empowerment in agriculture index (WEAI). However, it faces the same limitation as the RMS dataset:
        \begin{itemize}
            \item The panel is constrained to three rounds of data (2011/12, 2015, and 2018/19), all post-STRV introduction. The absence of pre-introduction data limits our capacity to evaluate the impact of STRV adoption in the same ways as the RMS.
        \end{itemize}

    \item The \href{https://bit.ly/47McOjI}{Bangladesh Rice Research Institute-Rice Database (BRRI-RD)}: As a government-owned rice database it provides comprehensive data across all years and districts regarding total production. It is nationally representative and contains production information. Nevertheless, this dataset is encumbered by certain limitations:
        \begin{itemize}
            \item While a panel, it is not a micro-level panel (plot, household, farm) but constructed as an aggregate district or division-level panel. Thus, it only has data on total production and acreage data by divisions and districts of Bangladesh. Thus, there is no variety-specific production, which restricts the use of this dataset.
        \end{itemize}
\end{itemize}


\subsection{The Rice Monitoring Survey (RMS)} \label{subsec:data_rms}

The dataset that closest matches what we require is the Rice Monitoring Survey (RMS), which we use as a robustness check. The RMS is a Gates Foundation-funded project designed to capture varietal turnovers over time in Bangladesh, India, and Nepal. The data was originally collected as two waves of a panel in 2014 and 2017 \citep{rms-2017}. Households were selected following a clustered random sampling procedure to ensure the overall survey was representative of rice growing regions of each country. The plan was to collect a third wave in 2020, but data collection efforts were delayed due to COVID until 2022 and only Bangladesh was revisited. In total, 1,500 households were part of the initial sampling frame in Bangladesh. The RMS was able to follow-up with 1,490 households in 2017 and 1,484 households in 2022. This gives us an attrition rate of $1.7\%$ over a span of eight years. 

As the RMS was designed to monitor varietal turnover, the survey instrument is focused on collecting information on rice production at the crop and plot level. While the data is a panel of households, limited information was collected about household well-being. These include demographics, asset ownership, membership with different farming organizations, farmer opinion on the prevalence of different stresses, such as flood, drought, and salinity, and the availability of different rice varieties. No information was collected on consumption, income, or other measures of welfare, such as food security or women's empowerment.
                
In terms of production data, the survey collected information by both plot and crop so as to be able to distinguish between different varieties of rice grown on the same plot. Plot data includes information on plot size, ownership, and land type. Crop data, within a plot, includes which rice variety was planted in which season, methods of planting, damage of crops by different abiotic shocks, input use, total production, and disposition of the harvest. 
            
We construct an unbalanced panel of households that cultivated rice in \emph{Aman} season. This gives us a total of 3,865 observations from 1,488 households. Our primary outcome of interest is rice yield at the household level, which we construct by summing up rice harvest (kg) on all plots and dividing by the sum of area (ha) for all rice plot. We consider a household as having adopted STRVs if they report planting any flood tolerant rice variety on at least one plot. To measure the incidence of flooding faced by the household, we combine household GPS location with the EO flood maps we generate.\footnote{While we would ideally use plot locations to match with flood data, only household GPS locations were collected in 2014 and 2017. Thus, we only know the precise location of plots in 2022.} We then calculate five measures of flooding for each survey year: cumulative, maximum, mean, median, and a measure of the flood experienced by the village. Our village measure includes cumulative flooding for all households in the village, minus the household in question. We also construct flood time series for each household going back to 2002. In some specifications we use the household's historic experience with flooding to calculate the probability of a household experiencing a flood in a given year as an instrument for STRV adoption.


\newpage
\FloatBarrier
\setcounter{table}{0}
\renewcommand{\thetable}{D\arabic{table}}
\setcounter{figure}{0}
\renewcommand{\thefigure}{D\arabic{figure}}


\section{Remote Sensing Details} \label{sec:rs_app}

In the paper we provide a non-technical, intuitive summary of the methods used to generate the EO data. In this appendix we provide the technical details sufficient to implement our procedures in alternative contexts or point to the existing publications, data, and code to replicate the deep learning and fusion model work.


\subsection{Rice Area Map}\label{sec:rs_app:rice_mapping}

To train the rice area mapping algorithm, ground truth data first needed to be generated. Three districts were selected to be the base for the study to extrapolate to the country: Barisal, Kurigram, and Rajshahi). The three districts come from the north, central, and south of the country and experience substantial flooding.

Google Earth imagery were inspected as the basis in determining the land cover of each cell. Pixels were categorized as rice if more than 70\% of the cell was rice and categorized as non-rice otherwise. In total 150 points were in each district (75 rice, 75 non-rice). This generated 450 rice-no-rice (RNR) points for each year (2002, 2004, 2006, 2009, 2015, 2016, 2018 -  2020) or 4,050 RNR points as a training dataset (see Figure \ref{fig:rice_flood}).

We developed a RF model using the training data generated above and used a leave-one-out cross-validation scheme. The input data of our model is: a PCA is applied on the median band values, calculated from the time series covering the growth period of \emph{Aman} rice. The two principal components are used as input to the model. Additionally, we use the median, $5^{th}$ and $95^{th}$ quantile EVI values over the growth period for each pixel, as well as elevation and slope as static features.

We apply the following process in Google Earth Engine to prepare the data before applying the RF algorithm:

\begin{enumerate}
    \item Extract the MOD09Q1.061 Terra Surface Reflectance 8-Day Global 250m data (band 1 and 2) images over the \emph{Aman} rice growth period.
    \item Calculate the median value for each pixel.
    \item Compute the PCA over the bands.
    \item Extract the time series of EVI from MOD13Q1.061 Terra Vegetation Indices 16-Day Global 250m, and compute the median, $5^{th}$ and $95^{th}$ quantiles over the period for each pixel.
    \item Extract the elevation from FABDEM \citep{hawker2022}, and compute the slope for each pixel.
    \item Stack all layers together to create the input to the RF.
\end{enumerate}

The RF model consists of a 1,000 trees, with each tree having a minimum of 5 leaves. We used the Statistical Machine Intelligence and Learning Engine (SMILE) RF implementation of Google Earth Engine. The random forest is trained with a leave-one-out cross validation scheme, with an additional separated test set. The leave-one-out scheme consists of rotating the districts used for training and leaving one of the districts out each time. The data used here consists of all years except 2020, which is removed and kept as test set. The leave-one-out validation scheme ensures the model is generalizable in space, meaning that it performs similarly outside of the area used for training. The hold out test set of 2020 allows us to validate the model in time, meaning that it will perform similarly outside of the observed years. The accuracies are reported in Table~\ref{tab:SI:EO:Accuracy}.

This way of validating and training the model naturally creates as many models as leave-one-out sets. To run the inference on the whole country, we implemented a majority voting scheme, i.e. each model infers the presence/absence of rice for the whole country, and the category with the most votes receives the final classification. This means that here, with three regions, we have three models, and each pixels needs at least two of the models to classify the pixel as rice, for the pixel to be set as rice in the final map.

After reviewing the result, and comparing them with a study performed by IRRI in 2012 \citep{gumma2014} on \emph{Aman} rice presence for 2010, we realized that submergence-rice, i.e. rice irrigated by submergence, was not being detected by our RF. According to \cite{gumma2014}, submergence-rice is only present in the north east of Bangladesh, and our ground sampling did not covering any of the north east, the algorithm could not pick up on it. We thus decided to add the submergence rice area from \cite{gumma2014} as additional training data and extended the time frame used in the training data at point 1 and 4 to capture post-flooding periods. This allowed us to successfully capture the submergence rice in the northeast.
After generating the rice maps for all of Bangladesh with MODIS, we had to investigate the necessity of generating them with Landsat. The same process as described at the district level cannot be reproduced: the satellite images were hand-picked for each of the three districts, as cloud free as possible, and representative of the growth period. This process cannot be automated at the country scale. Additionally, because of the technical challenges of working with Landsat 7 (gap filling due to SLC failure), it made sense to investigate the necessity to generate such data before producing it.
In this study, the rice field data is primarily used to mask the inundation and EVI data, but is then aggregated at the district level. Additionally, the data being masked is given at 500 meters (inundation), and 250 meters (EVI). In order to understand the necessity of the higher resolution map, we downloaded a high resolution map based on Sentinel-1 (10 meters) generated by \cite{singha2019a}, of all of Bangladesh for 2017. We then masked the inundation data with this high resolution mask, and with the 250 meters mask generated through the RF algorithm, and compared the two at the district level.

The figure shows that the two datasets do not significantly differ from each other. Given the complexity of generating high resolution masks with Landsat, the source of errors would be relatively high, and would not significantly improve the accuracy of the data at the district level.


\subsection{Flood Map}\label{sec:rs_app:flood}

As with the rice maps, building reliable flood maps that look so far back into the past raises data availability challenges. While EO data allows for analysis geographical areas, particularly in remote or inaccessible regions it suffers from a recency bias, at least in terms of high spatial and temporal resolution products that can see beyond the visible spectrum. For flood mapping, MODIS is often used because of its long history, its availability on a daily basis, and its relatively cloud free image derived from an 8-days composite image. Products like Landsat, which come at a higher spatial resolution and have a longer time series, do not provide such a consistent image series. In particular, Landsat has a low revisit frequency, harmonization issues between Landsat versions, and instrument failure in Landsat 7.

Further, MODIS is not without its own limitations. It is an optical sensor that acquires two images daily. But, being an optical sensor, the content of these images is often obscured by clouds. This means that during the tiime of the year when floods are most frequent (the rainy season), MODIS is at its least useful because cloud cover often makes it difficult to see flooding. Furthermore, MODIS's coarse spatial resolution (250m, 500m, and 1km) compared to LANDSAT, makes it unsuitable for accurately extracting flooded information in urban regions or other complex and heterogeneous landscapes. For us, studying the flat rice growing plains in rural Bangladesh, this last issue is not a major concern.

In recent years, radar sensors have overcome the limitations of optical sensors by penetrating cloud cover to consistently detect water signals. Sentinel-1 is an ideal satellite for flood detection as it captures imagines at a high spatial resolution (10m) and is equipped with a radar sensor so that even in cloudy conditions it can capture accurate and consistent flood events. However, Sentinel-1's weakness, relative to MODIS and Landsat, is that it only became available since 2017 and is thus unsuitable for historical flood mapping.

Given the strengths and weaknesses of each product, we first run comparisons of food detection between MODIS and Landsat. While Landsat has greater spatial resolution, which is extremely beneficial in pinpointing floods on small rice plots, its coarse temporal resolution was a critical detriment. Landsat often completely missed flood events detected by MODIS because of Landsat's infrequency of images and the extent of cloud over. Because of this, we decided to use MODIS as are base product for both flood detection and rice mapping.

Having settled on MODIS, we then developed a method to fuse together the shorter time series of Sentinel-1 with the longer MODIS time series as a potential solution to the disadvantages of either sensor. This fusion model allows us to leverage the advantages of high resolution modern flood detection and historical data availability. The data used for building our flood maps are the product of a deep learning model trained on Sentinel-1 derived fractional inundated area to predict historical MODIS satellite data. We utilize a Convolutional Neural - Long Short-Term Memory Network (CNN-LSTM) to take advantage of both the spatial and temporal feature representation of inundation and compare it to a traditional Convolutional Neural Network (CNN). We apply the CNN-LSTM model from 2002 to 2020 over all of Bangladesh at 500m resolution. From this inundation dataset, we extract one map per year with various statistics over the monsoon season.

\begin{figure}[htbp]
	\begin{minipage}{\linewidth}		
		\caption{Correlation of Flooding with High- and Low-Resolution Masking}
    \label{fig:SI:EO:resolution-comparison}
		\begin{center}
    \includegraphics[width=1\columnwidth]{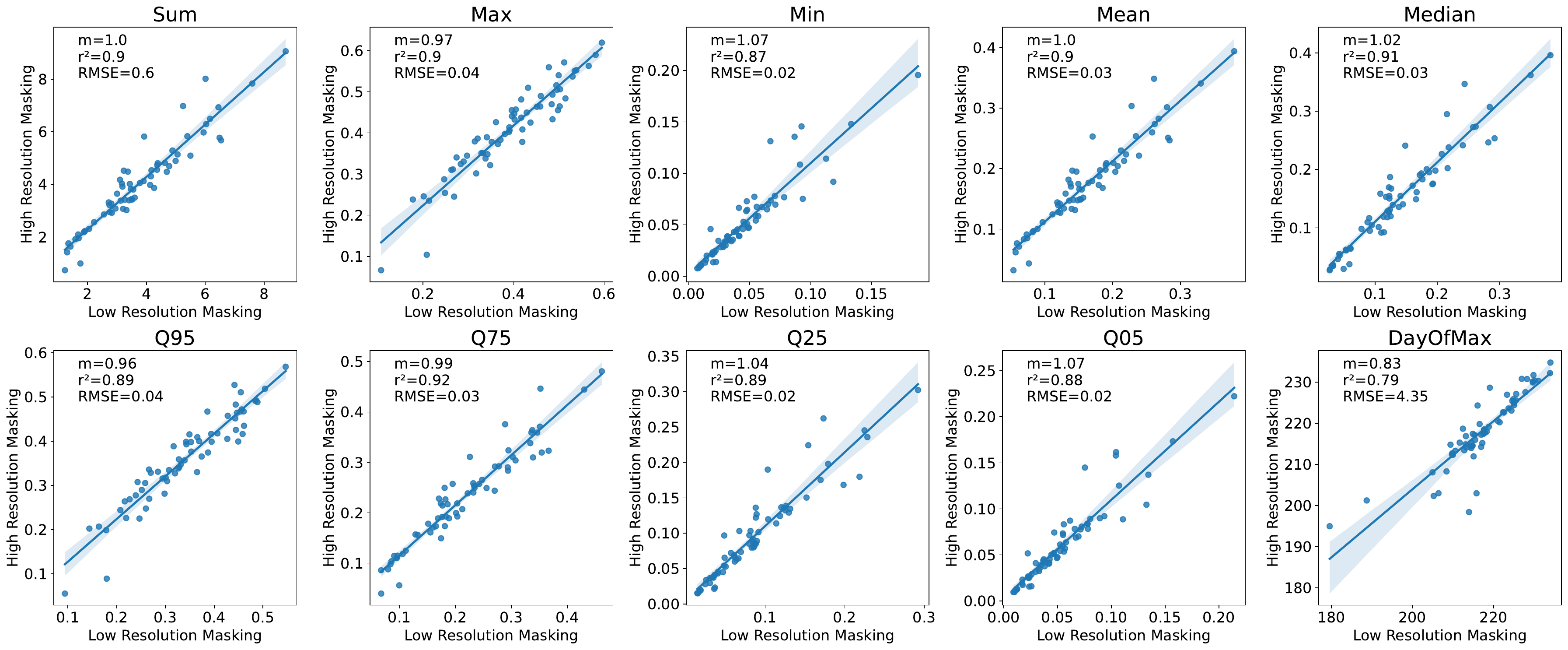}
		\end{center}
		\footnotesize  \textit{Note}: Comparison between inundation data aggregated at the district level with high-resolution masking, and low resolution masking
	\end{minipage}	
\end{figure}

Figure~\ref{fig:SI:EO:resolution-comparison} provides tests for accuracy of the fusion model by comparing results with the high resolution Sentinel-1 data. Our model produces results that match the high resolution data with a high degree of accuracy, ranging from .79 to .92. Additional details and validation tests, along with reproducible data and code, has been published by the research team in \cite{GiezendannerEtAl23}.


\newpage

\begin{figure}[htbp]
	\begin{minipage}{\linewidth}		
		\caption{2010 Rice Area in Bangladesh}
    \label{fig:SI:EO:output}
		\begin{center}
    \includegraphics[width=.6\columnwidth]{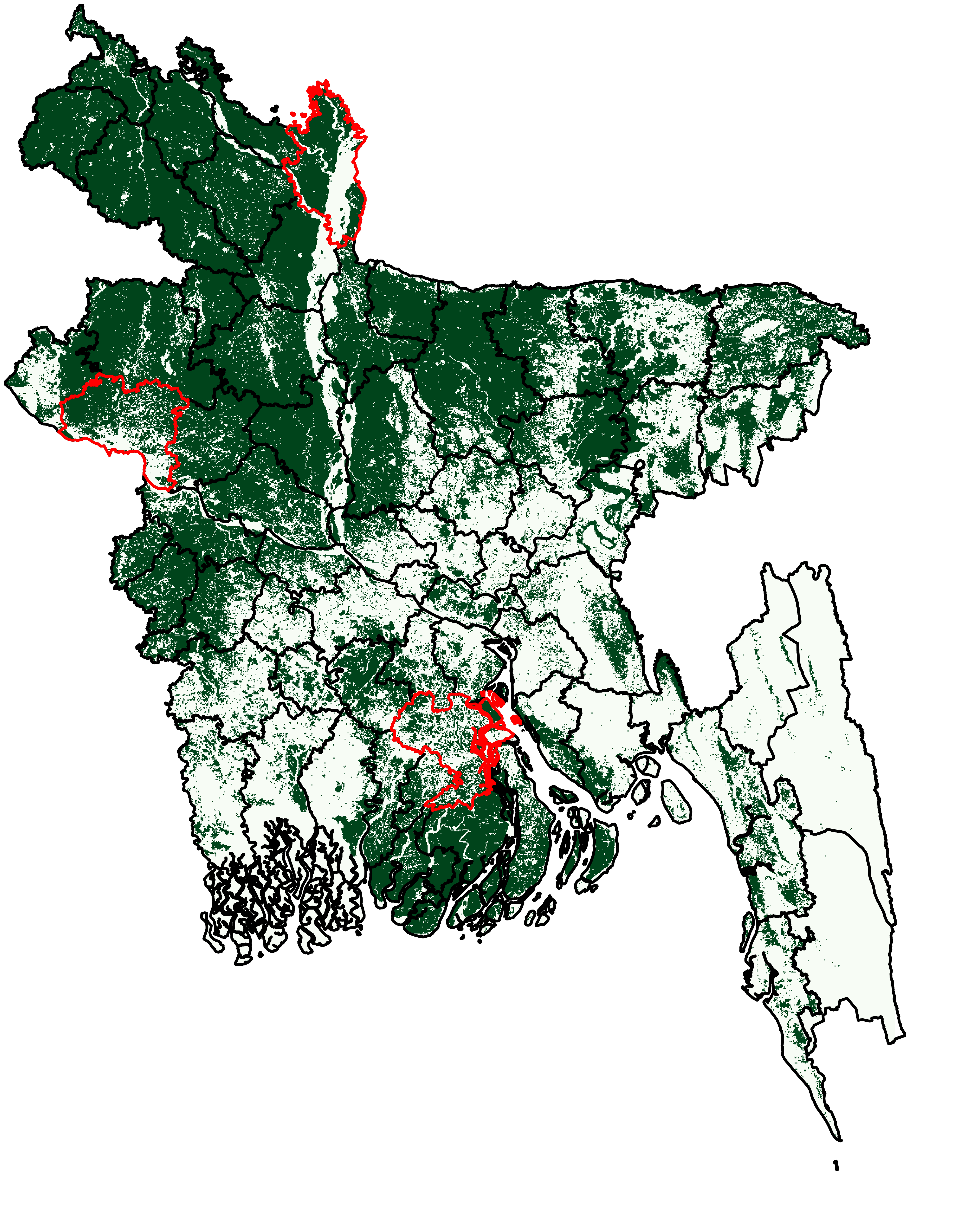}
		\end{center}
		\footnotesize  \textit{Note}: Output of RF algorithm for Bangladesh for 2010: presence (dark green)/ absence (light green) of rice fields. Overlayed are the districts, with the three training districts in red.
	\end{minipage}	
\end{figure}

\begin{table}[htbp]
    \centering
    \caption{Accuracies reported from the leave-one-district-out validation scheme, as well as the test set (2020, all districts)}
    \begin{tabular}{lc}
    \hline \hline \\[-1.8ex]
         Leave-one-out cross validation &  Accuracy\\\hline
         Barisal & 0.65\\
        Kurigram & 0.72\\
        Rajshahi & 0.61\\\hline
        Test Set (2020) & 0.82\\\hline \hline
    \end{tabular}
    \label{tab:SI:EO:Accuracy}
\end{table}


\newpage
\FloatBarrier
\setcounter{table}{0}
\renewcommand{\thetable}{E\arabic{table}}
\setcounter{figure}{0}
\renewcommand{\thefigure}{E\arabic{figure}}


\section{Remote Sensing Robustness Checks} \label{sec:gis_result_app}


\subsection{District-Level Results Without Coastal Districts} \label{sec:nocoast_app}

One concern regarding the main results in our analysis is that we might be picking up the impact of other abiotic stressors, and the adoption of seed designed to mitigate those stresses. Key among these is salinity tolerant rice designed to protect against crop loss due to salt intrusion into the water, either through the water table or through coastal flooding. Work by \cite{ChenEtAl17} and \cite{ChenMueller18} measure the impact of saline tolerant rice in Bangladesh. In their work they find very little impact of flooding on rice yields because farmers adjust to floods by delaying planting date. The concern regarding our results is that in coastal regions the effect that we are capturing is the impact of salinity tolerant rice, biasing our estimates of the impact of flood tolerant rice. Absent coastal districts our results might be more muted or their significance may go away completely.

To evaluate the severity of this concern, we conduct our event study, DID, and TWFE analysis on our district-level data but exclude 19 coastal districts where salinity may be a problem. These 19 include all districts in the division of Barisal plus Chandpur, Chittagoing, Cox's Bazar, Feni, Lakshmipur, and Noakhali (Chittagong division); Gopalganj and Shariatpur (Dhaka division); and Bagerhat, Jessore, Khulna, Narail, and Sakhira (Khulna division). 

\begin{figure}[!htbp]
	\begin{minipage}{\linewidth}		
		\caption{Event Study Results Without Coastal Districts}
		\label{fig:event_nc}
		\begin{center}
			\includegraphics[width=.49\linewidth,keepaspectratio]{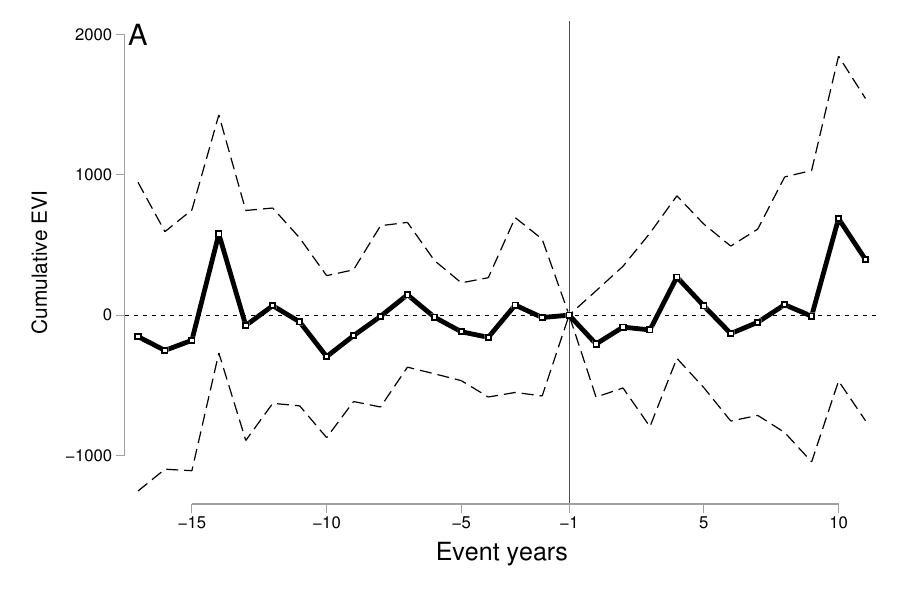}
			\includegraphics[width=.49\linewidth,keepaspectratio]{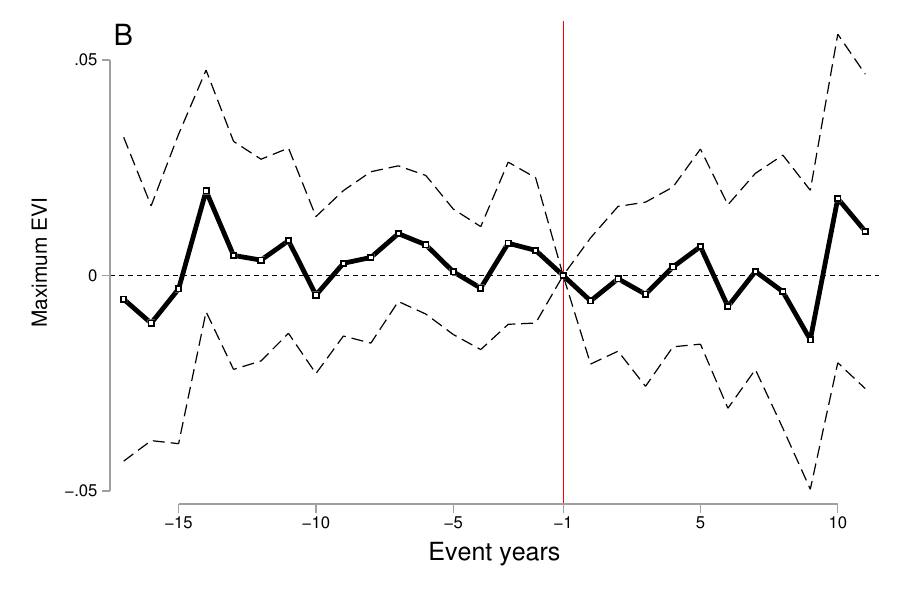}
			\includegraphics[width=.49\linewidth,keepaspectratio]{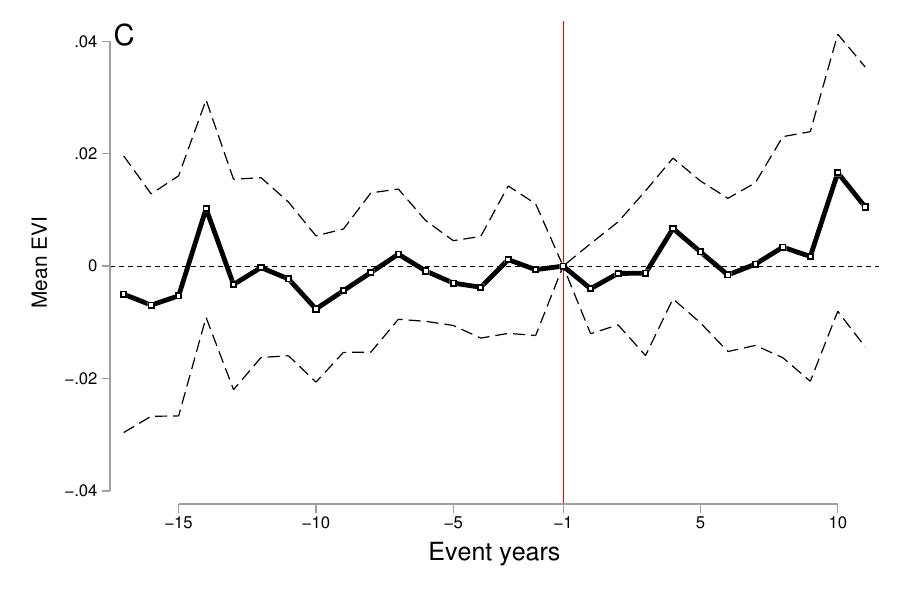}
			\includegraphics[width=.49\linewidth,keepaspectratio]{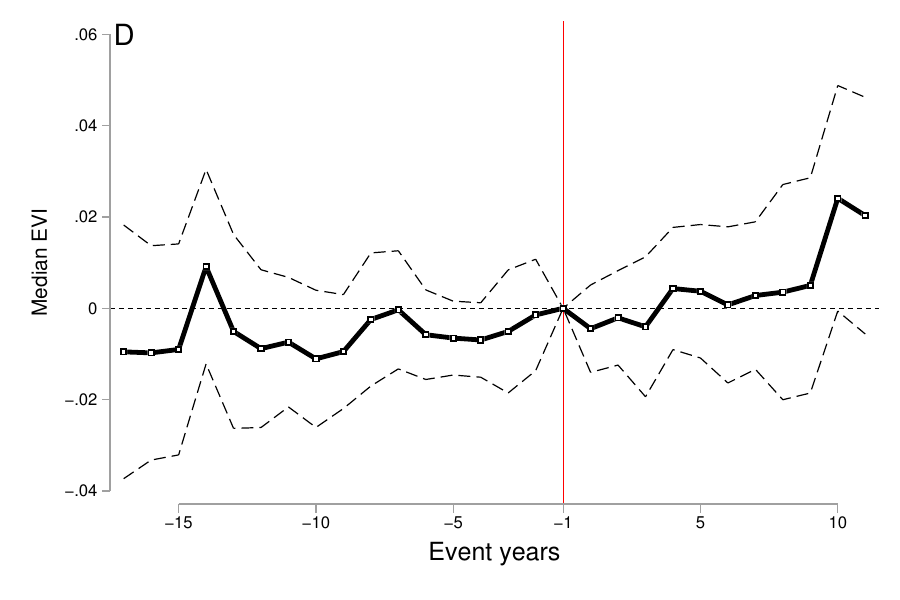}
		\end{center}
		\footnotesize  \textit{Note}: Figure displays coefficients from district-level event study regressions that excludes 19 coastal districts. The dependent variable for Panel A is cumulative EVI, for Panel B is max EVI, for Panel C is mean EVI, and Panel D reproduces the results in the paper using median EVI values. The solid line shows coefficient estimates from the model, with the event year (the year immediately prior to the upazila having access to STRVs, indicated as -1) as the excluded category. Dotted lines represent $95\%$ confidence intervals calculated using standard errors clustered at the district-level.
	\end{minipage}	
\end{figure}

Figure~\ref{fig:event_nc} presents results from the event study regressions. These are indeed more muted in their measured impact of STRVs. For cumulative, maximum, and mean EVI there appears to be no upward trend except in districts that have had STRVs for $10+$ years. Only when using median EVI do we see a trend that starts earlier but even here the impact is not significant at standard levels of significance. 

\begin{landscape}
\begin{figure}[!htbp]
	\begin{minipage}{\linewidth}		
		\caption{Specification Charts of DID Results Without Coastal Districts}
		\label{fig:flood_days_did_nc}
		\begin{center}
			\includegraphics[width=.48\linewidth,keepaspectratio]{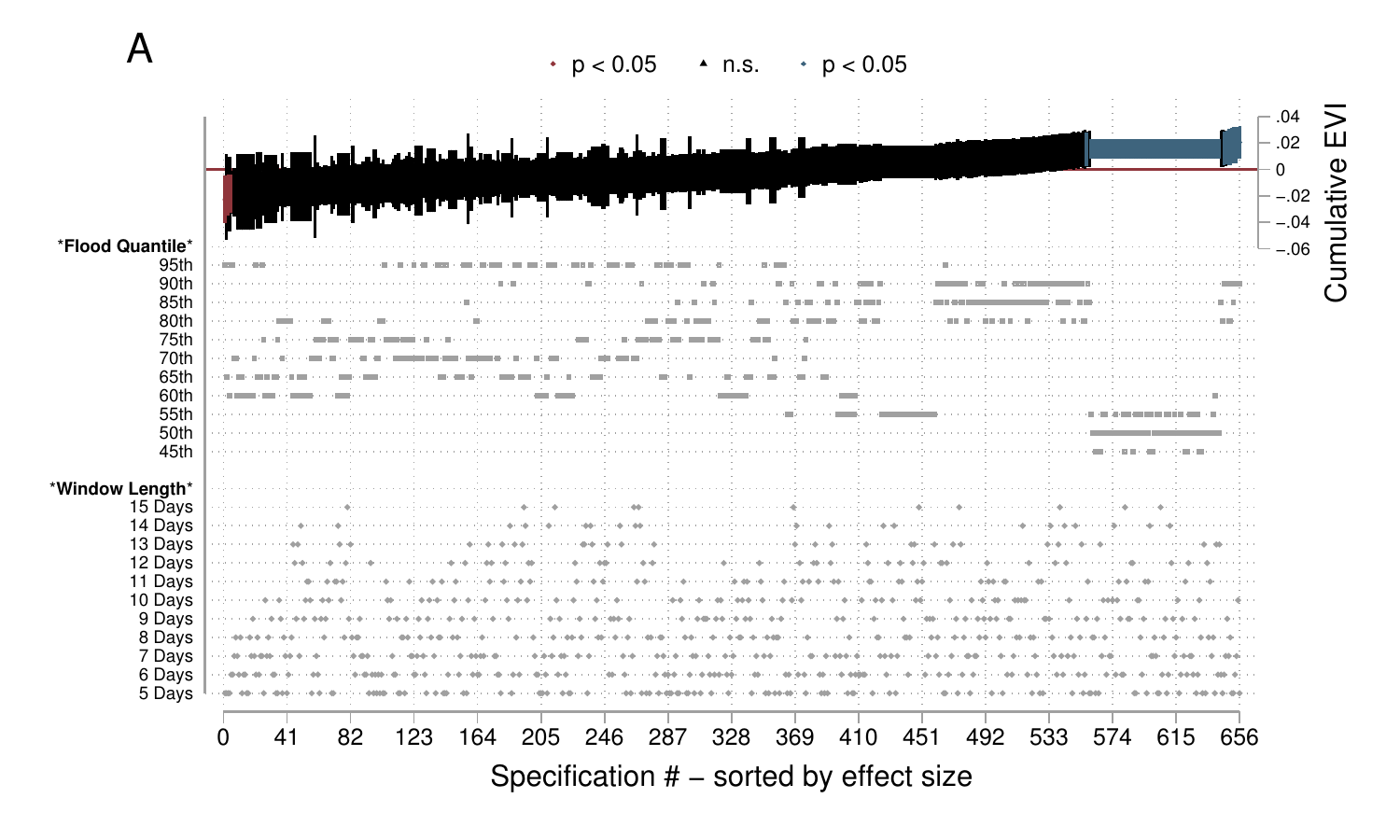}
			\includegraphics[width=.48\linewidth,keepaspectratio]{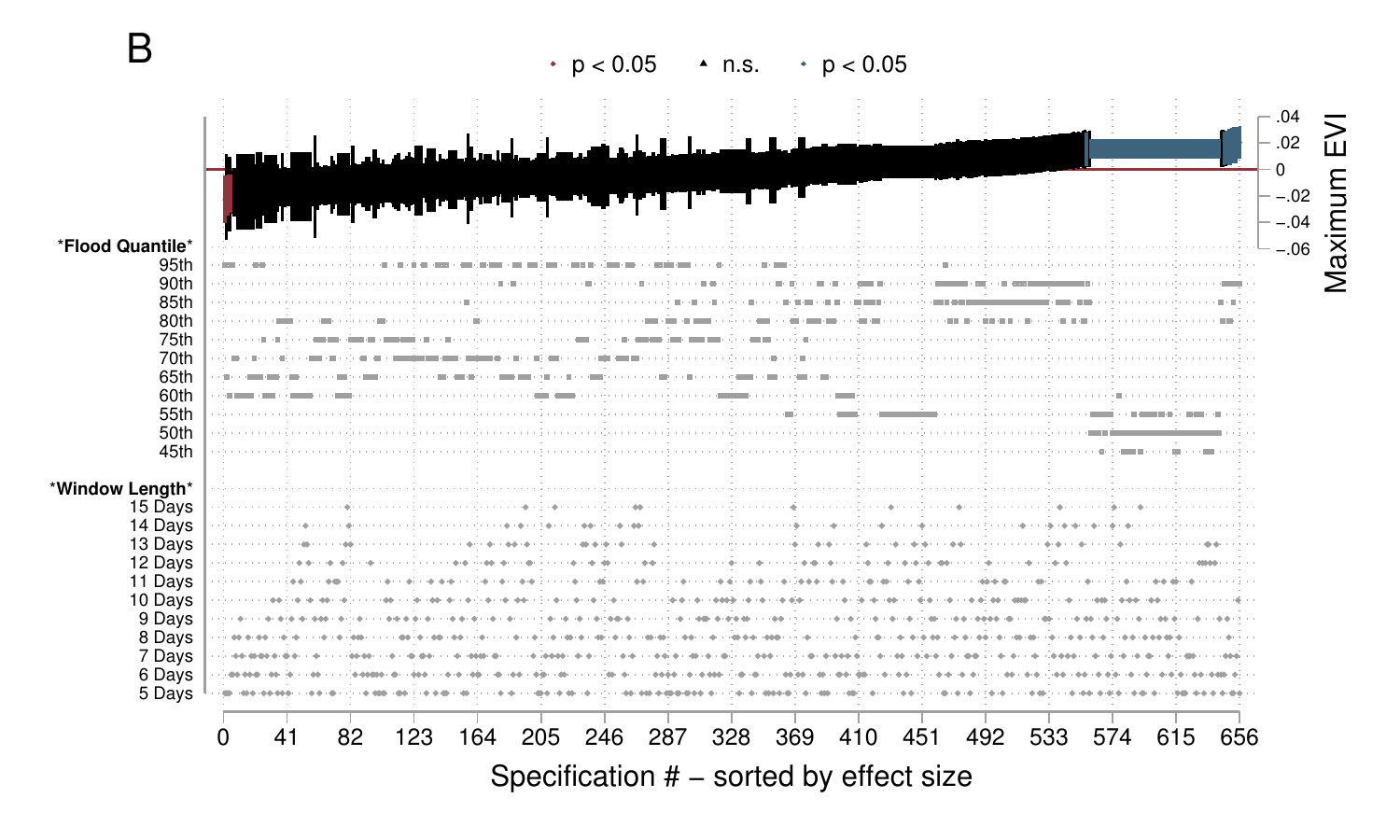}
			\includegraphics[width=.48\linewidth,keepaspectratio]{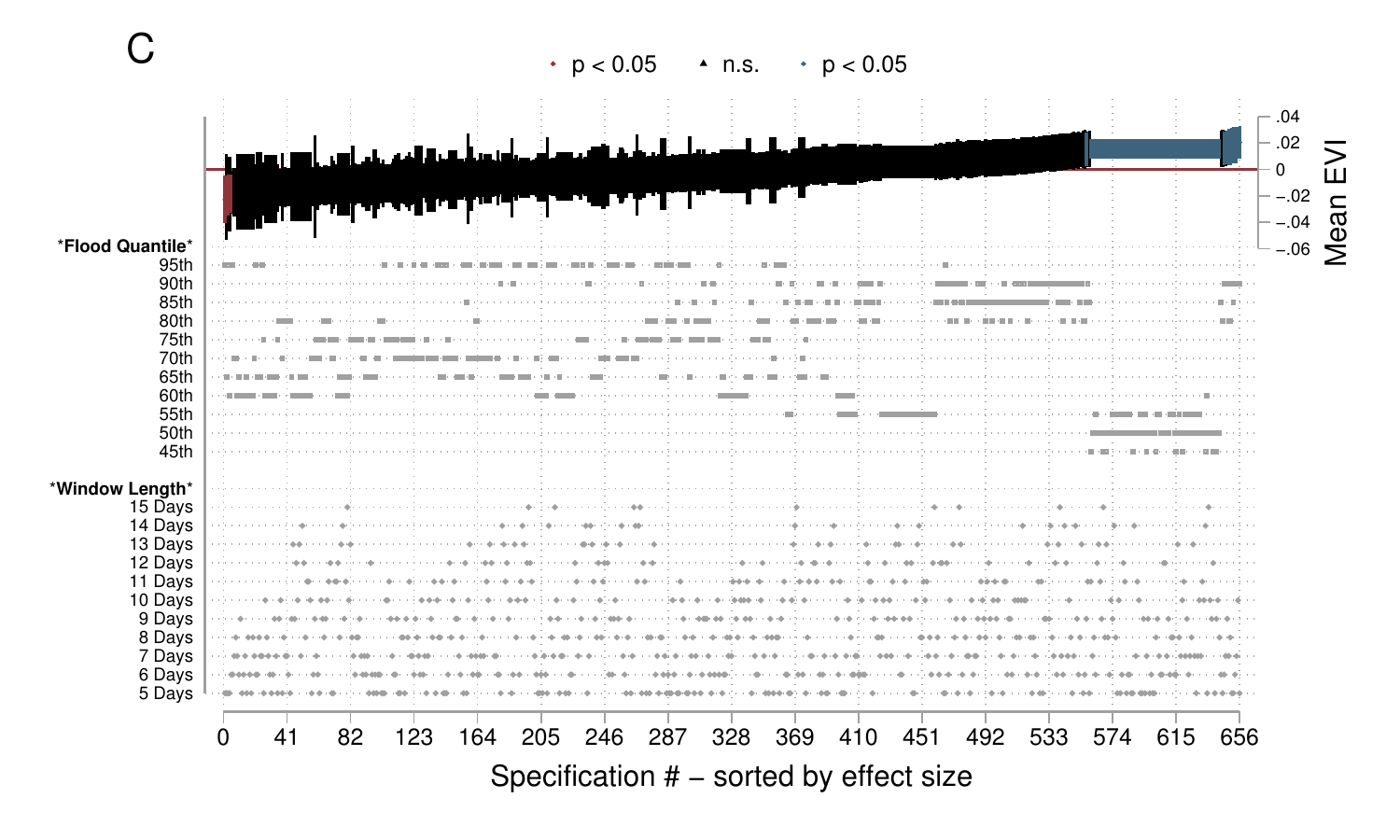}
			\includegraphics[width=.48\linewidth,keepaspectratio]{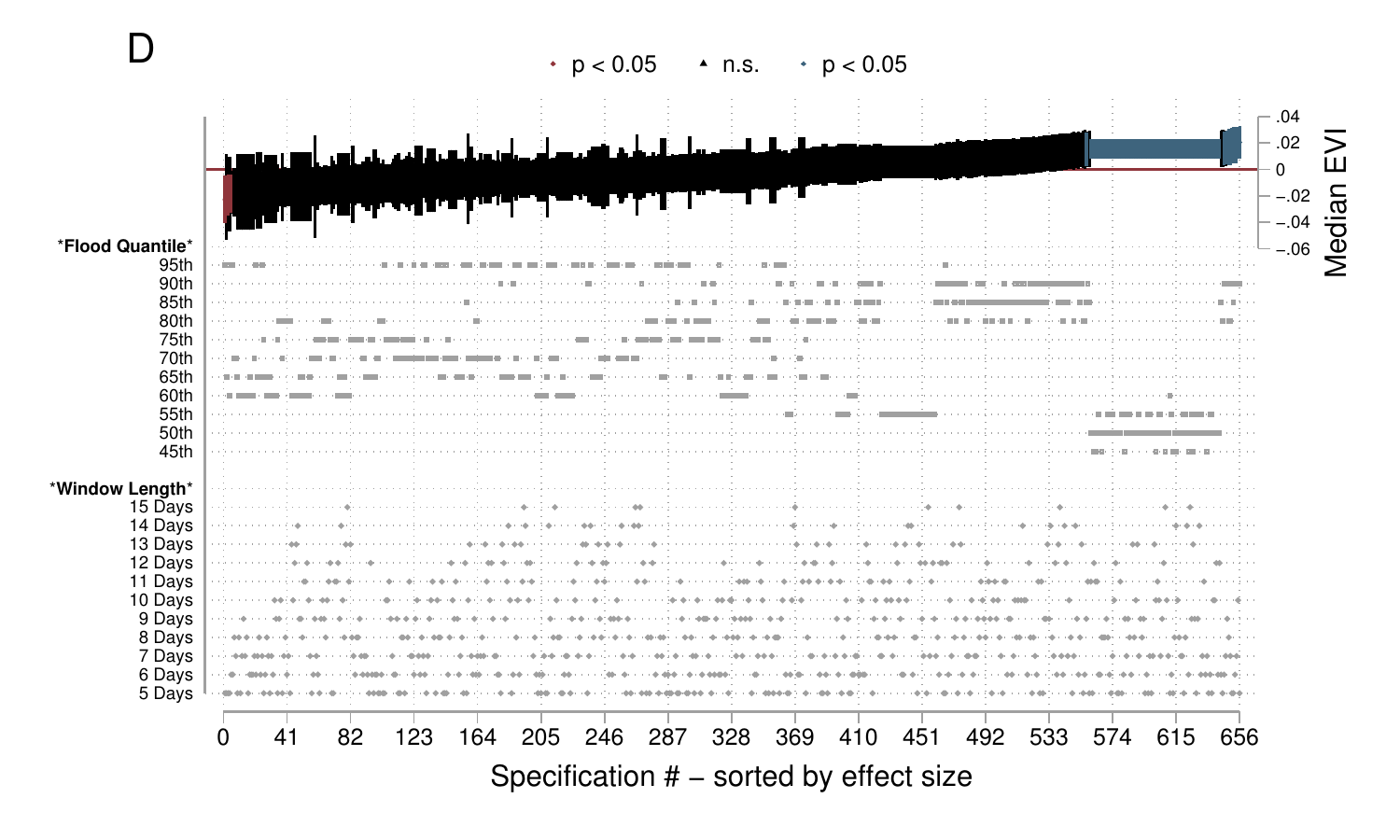}
		\end{center}
		\footnotesize  \textit{Note}: The figure presents results from district-level TWFE regressions that excludes 19 coastal districts. Each panel in the figure displays coefficient estimates and $95\%$ confidence intervals on the interaction of STRVs and a flood metric for a specific EVI measure (cumulative, max, mean, median). Specifications are sorted based on coefficient size from smallest (left) to largest (right). The gray diamonds below the coefficients indicate which combination of quantile and flood window was used in the regression.
	\end{minipage}	
\end{figure}
\end{landscape}

Figure~\ref{fig:flood_days_did_nc} presents results from the DID specification. Results using this method are actually stronger when we exclude coastal districts. While in the main results $13\%$ of potential flood metrics produce positive and significant results (using cumulative EVI), here $15\%$ of results are positive and significant. Most importantly for the robustness of our main results, Goldilocks floods come from the same quantiles as before. The majority of positive and significant results are from floods measured using the $45^{th}$ to $55^{th}$ quantile. We do however see a few Goldilocks floods at very low quantiles when floods are of very short duration, which we do not see in the main results

\begin{landscape}
\begin{figure}[!htbp]
	\begin{minipage}{\linewidth}		
		\caption{Specification Charts of TWFE Results Using Binary if Flooding Was in Window Without Coastal Districts}
		\label{fig:flood_days_bin_nc}
		\begin{center}
			\includegraphics[width=.48\linewidth,keepaspectratio]{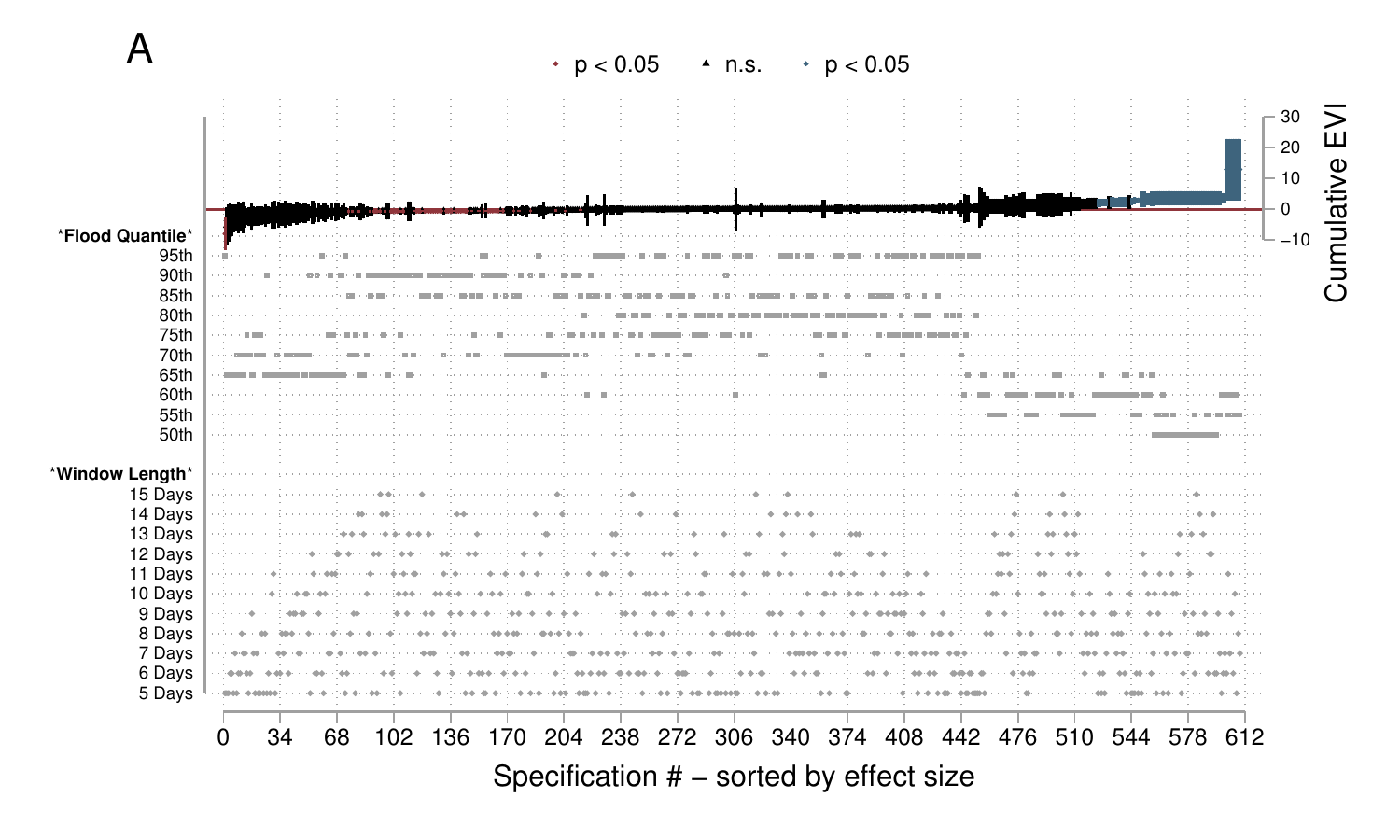}
			\includegraphics[width=.48\linewidth,keepaspectratio]{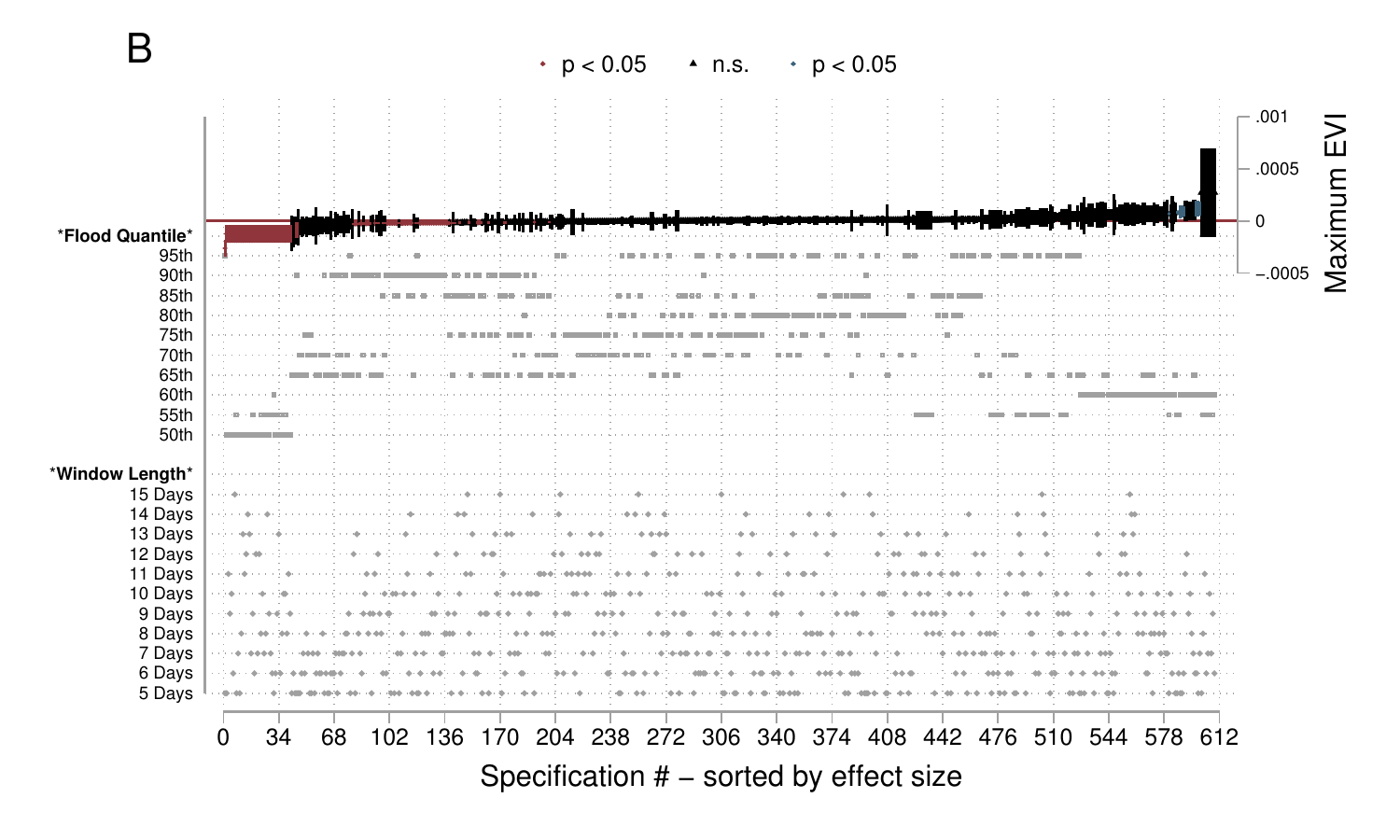}
			\includegraphics[width=.48\linewidth,keepaspectratio]{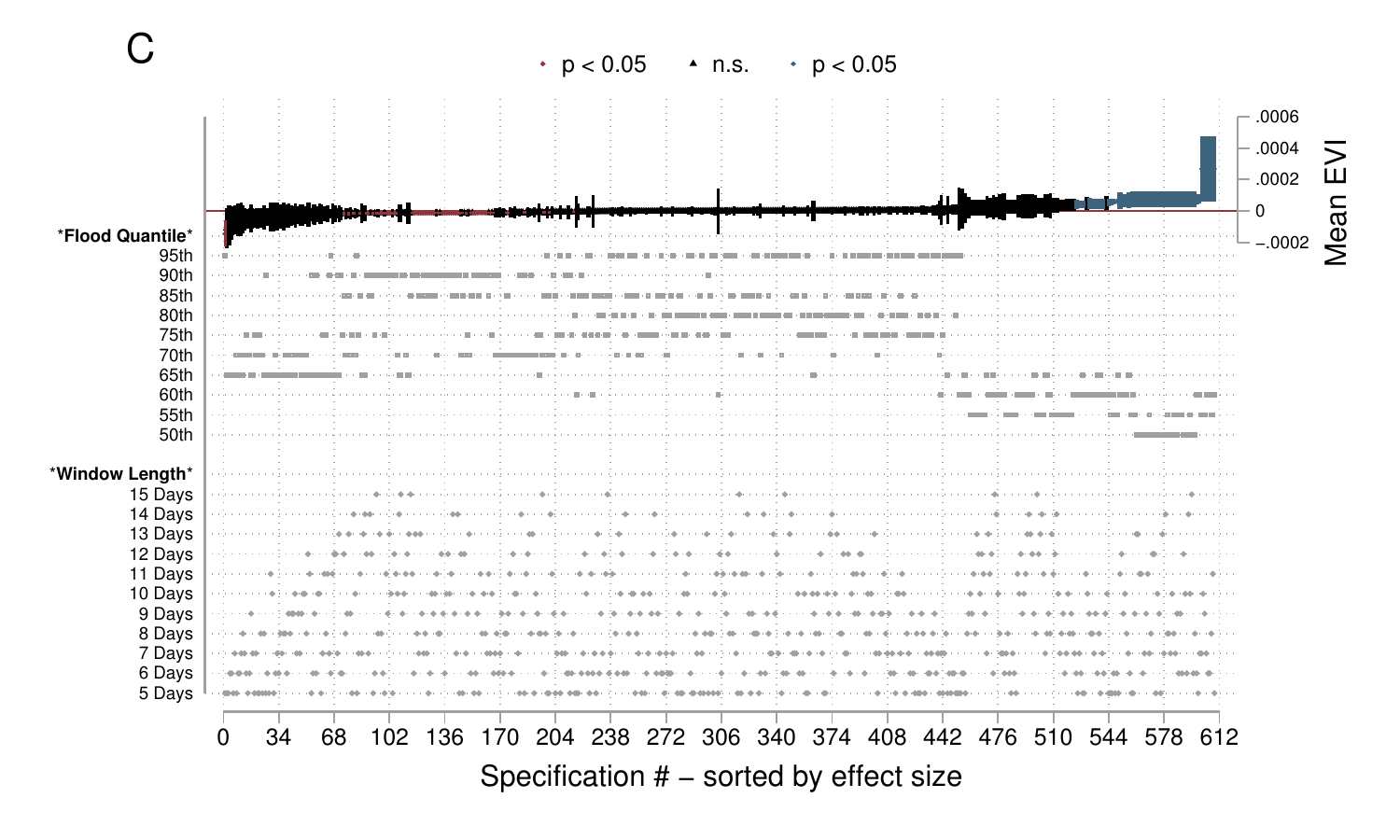}
			\includegraphics[width=.48\linewidth,keepaspectratio]{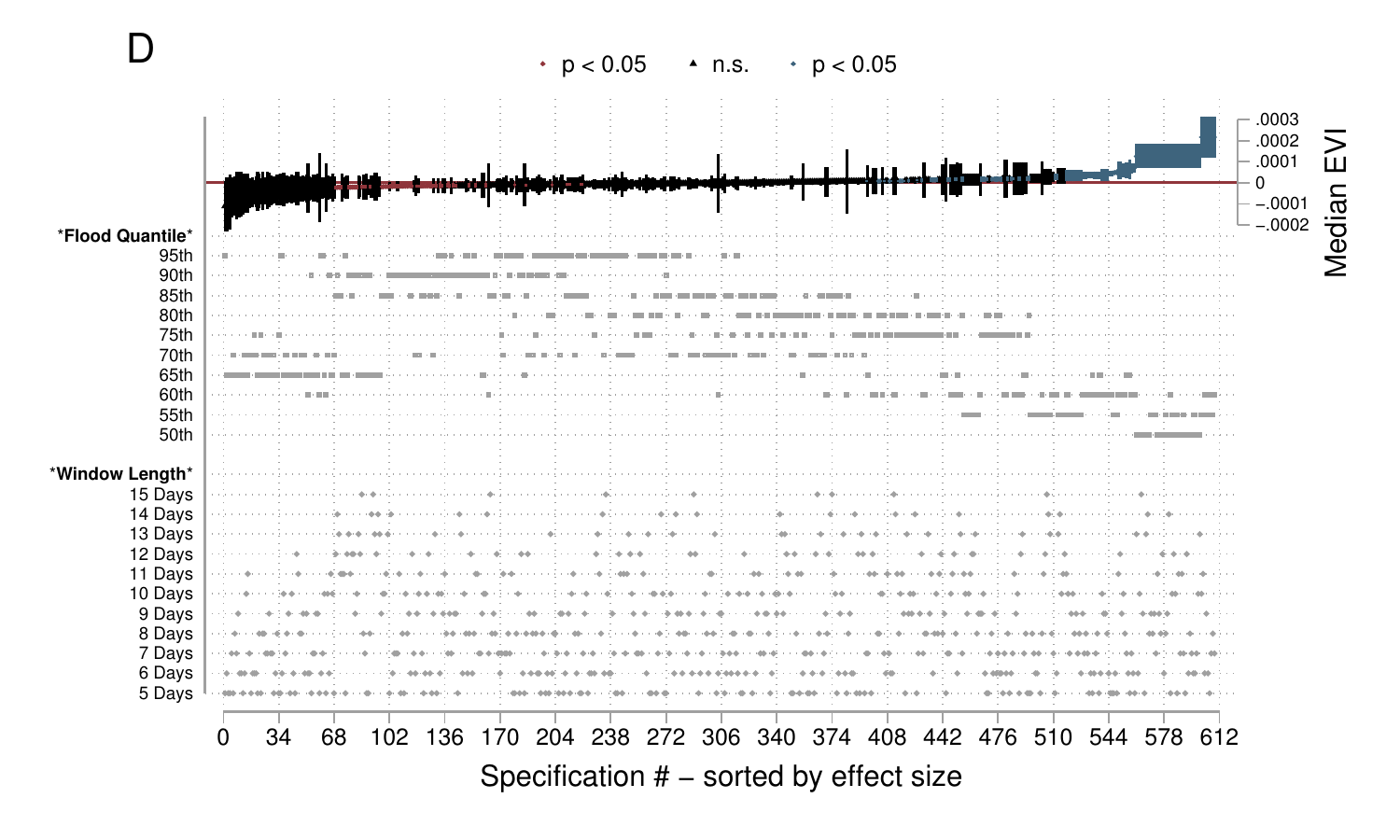}
		\end{center}
		\footnotesize  \textit{Note}: The figure presents results from district-level TWFE regressions that excludes 19 coastal districts. Each panel in the figure displays coefficient estimates and $95\%$ confidence intervals on the interaction of STRVs and a flood metric for a specific EVI measure (cumulative, max, mean, median). Specifications are sorted based on coefficient size from smallest (left) to largest (right). The gray diamonds below the coefficients indicate which combination of quantile and flood window was used in the regression.
	\end{minipage}	
\end{figure}
\end{landscape}

\begin{landscape}
\begin{figure}[!htbp]
	\begin{minipage}{\linewidth}		
		\caption{Specification Charts of TWFE Results Using Days in Flood Window Without Coastal Districts}
		\label{fig:flood_days_win_nc}
		\begin{center}
			\includegraphics[width=.48\linewidth,keepaspectratio]{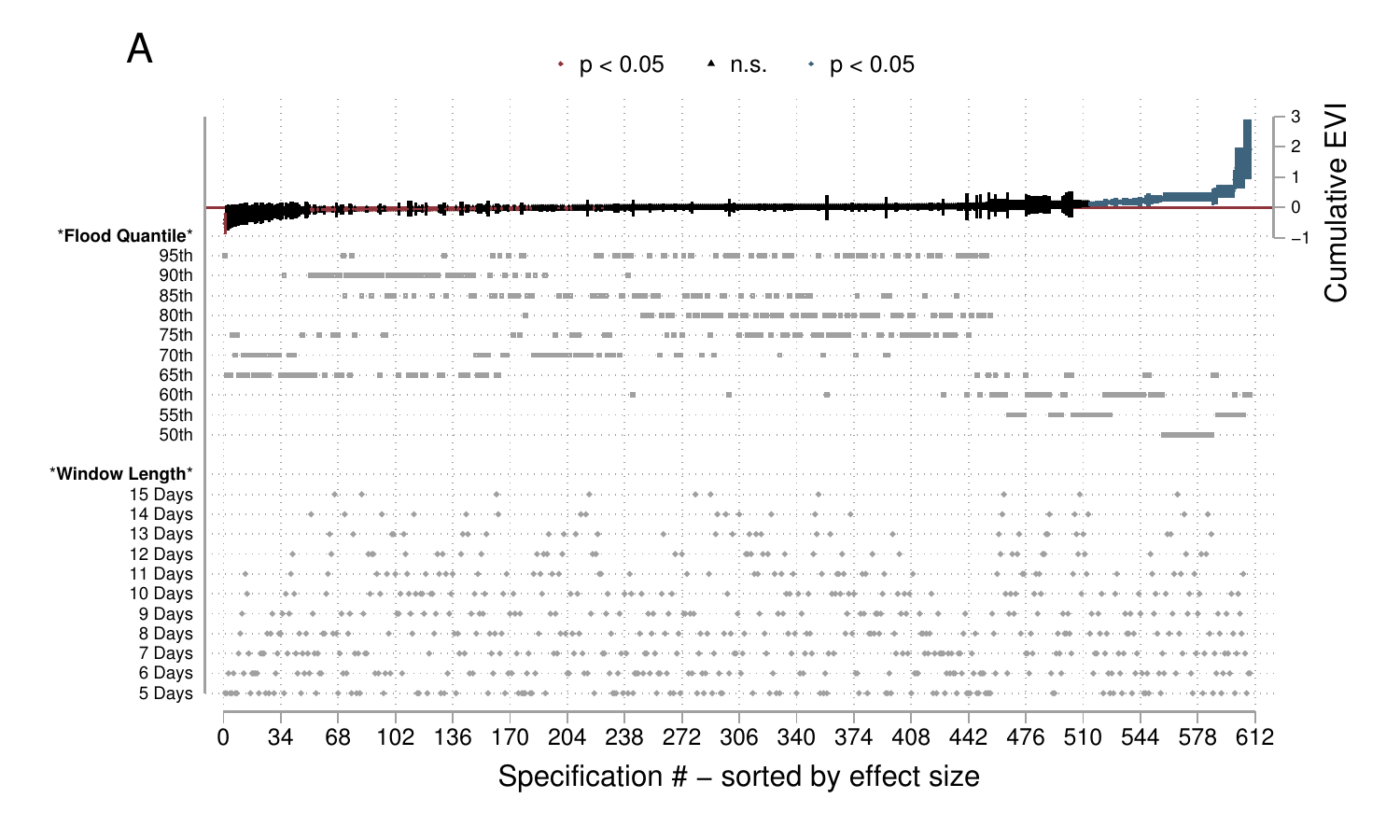}
			\includegraphics[width=.48\linewidth,keepaspectratio]{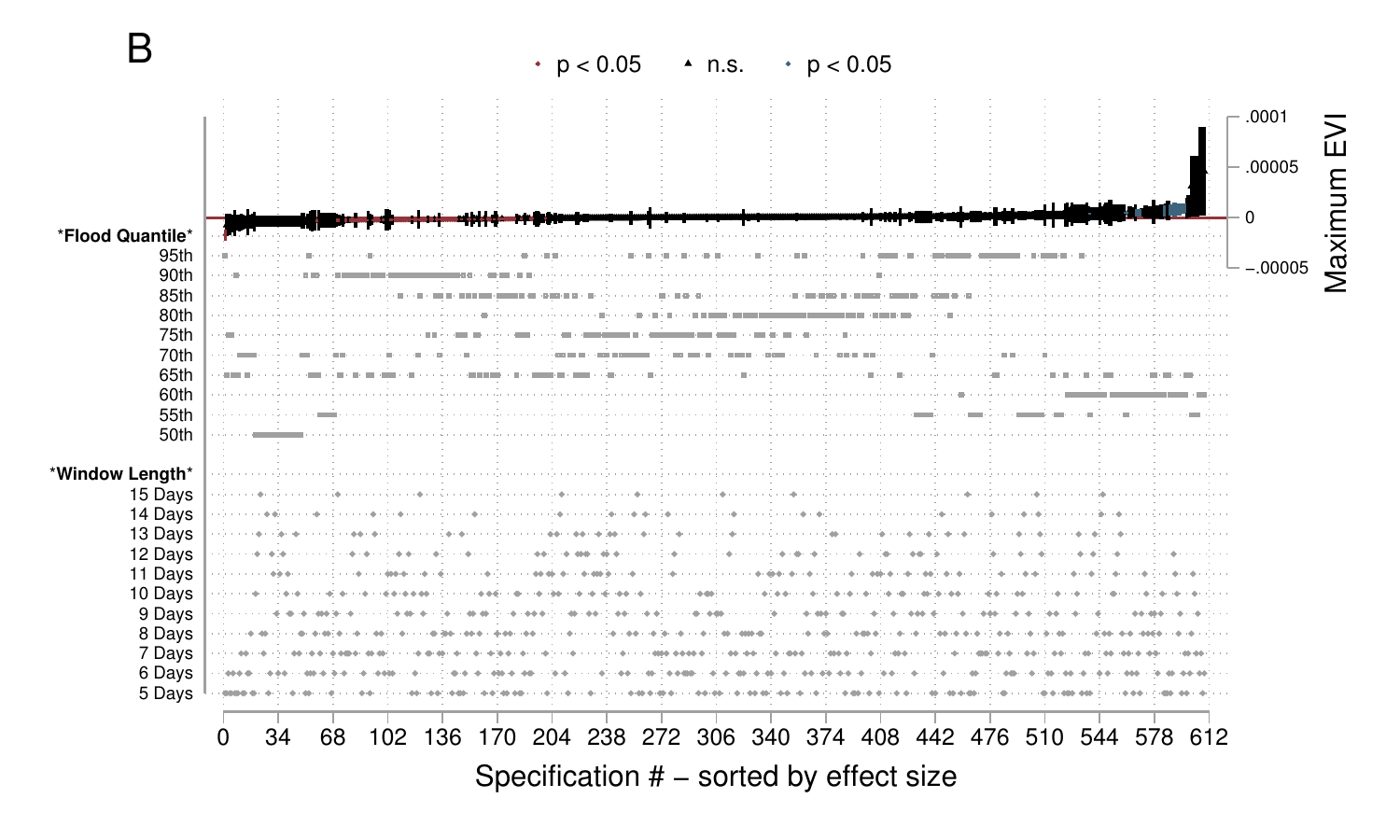}
			\includegraphics[width=.48\linewidth,keepaspectratio]{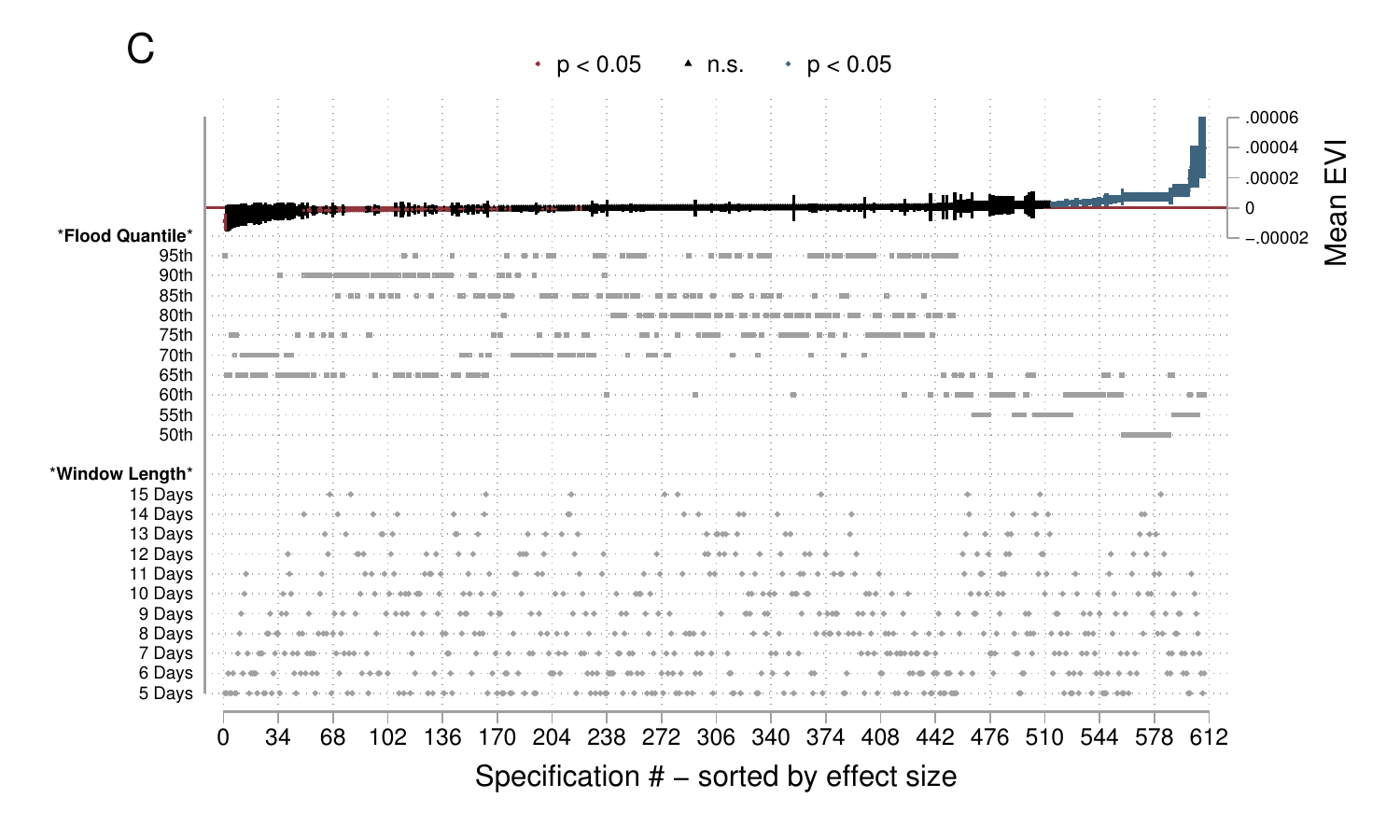}
			\includegraphics[width=.48\linewidth,keepaspectratio]{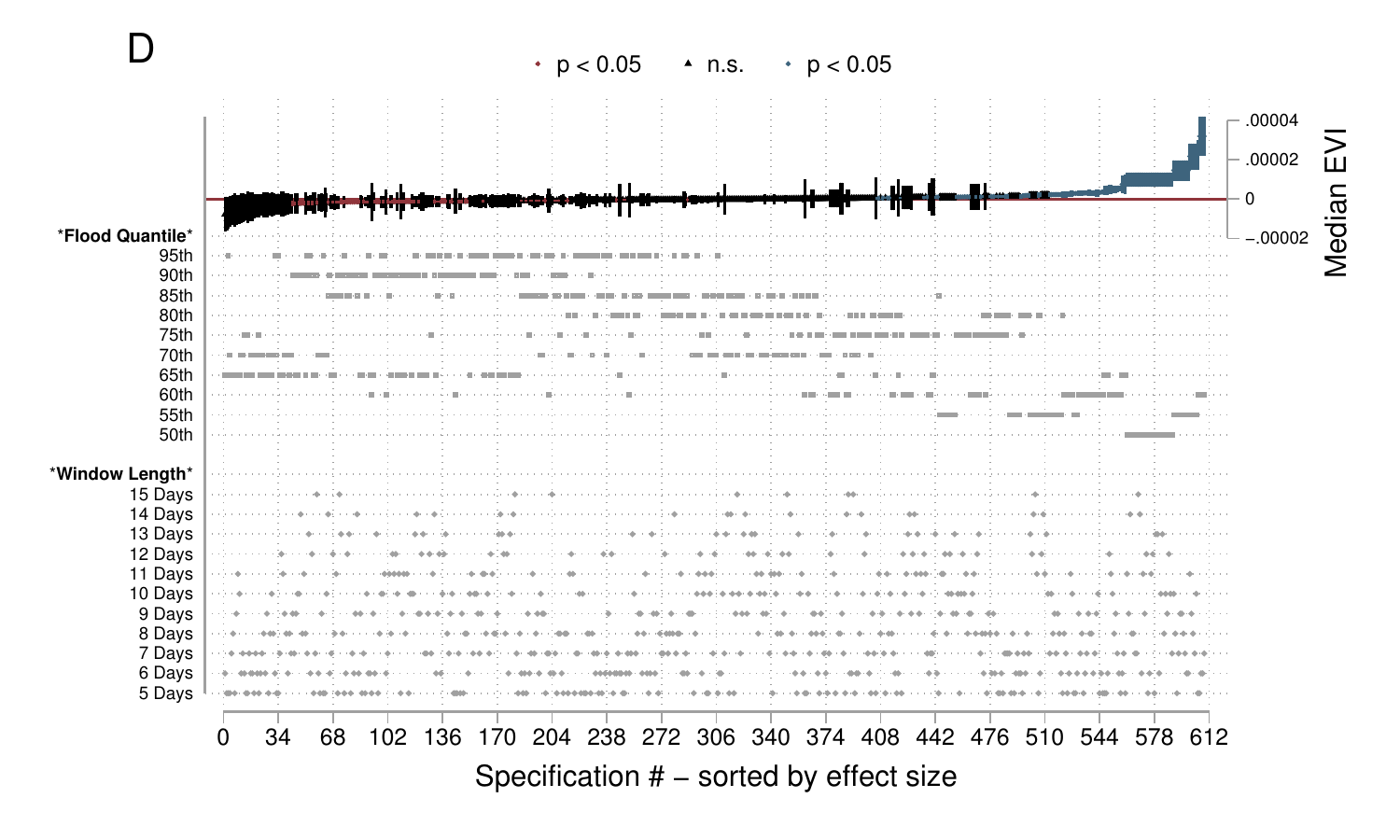}
		\end{center}
		\footnotesize  \textit{Note}: The figure presents results from district-level TWFE regressions that excludes 19 coastal districts. Each panel in the figure displays coefficient estimates and $95\%$ confidence intervals on the interaction of STRVs and a flood metric for a specific EVI measure (cumulative, max, mean, median). Specifications are sorted based on coefficient size from smallest (left) to largest (right). The gray diamonds below the coefficients indicate which combination of quantile and flood window was used in the regression.
	\end{minipage}	
\end{figure}
\end{landscape}

Turning to our TWFE regressions, Figure~\ref{fig:flood_days_bin_nc} reports results using a binary indicator for if flooding fell within the window and Figure~\ref{fig:flood_days_win_nc} reports results using the number of days in the flood window. Results look remarkably similar to our main results. We even see the same share of positive and significant results using the binary indicator and excluding coastal districts and when we include coastal districts: both $13\%$. We see fewer positive and significant results using days in the window ($15\%$) then when we include coastal districts ($18\%$).
 
Based on the preponderance of evidence, we conclude that our results are driven by the effect of submergence tolerant rice on EVI during floods and not by salinity tolerant rice.


\subsection{Upazila-Level Results} \label{sec:zil_app}

A second concern regarding the main results is that our choice regarding the unit of analysis (districts) is driving the results and that, had we conducted the analysis at a more disaggregated level, our results would not be significant. All things being equal, we would prefer to use a smaller, less aggregated unit for analysis. This is both for power reasons (there are over 500 upazilas but only 64 districts) and because averaging over the rice growing area in an entire district likely obscures important variation in EVI's response to flooding. However, the STRV seed data is only available at the district-level and this drives our choice of districts as the unit of analysis.

\begin{figure}[!htbp]
	\begin{minipage}{\linewidth}		
		\caption{Event Study Results for Upazila}
		\label{fig:event_zil}
		\begin{center}
			\includegraphics[width=.49\linewidth,keepaspectratio]{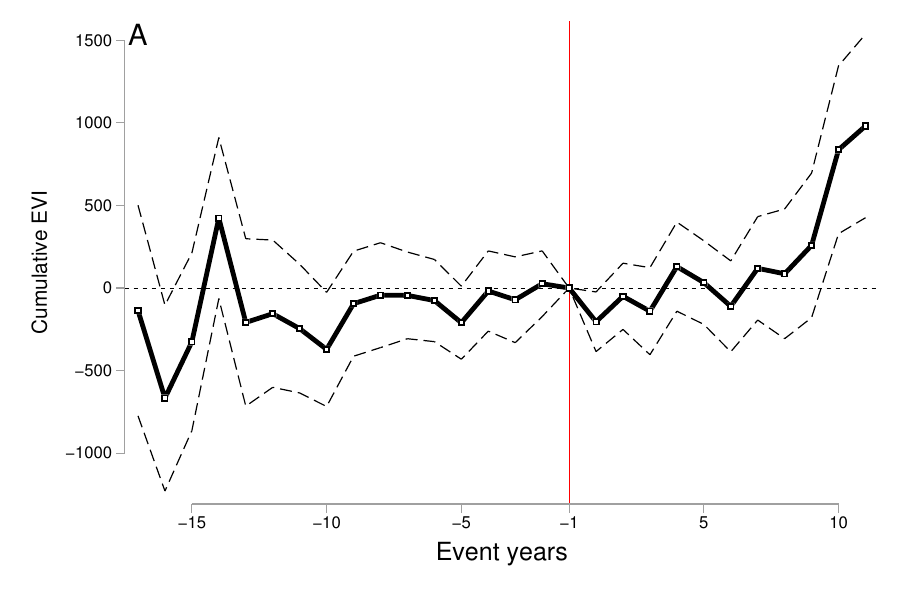}
			\includegraphics[width=.49\linewidth,keepaspectratio]{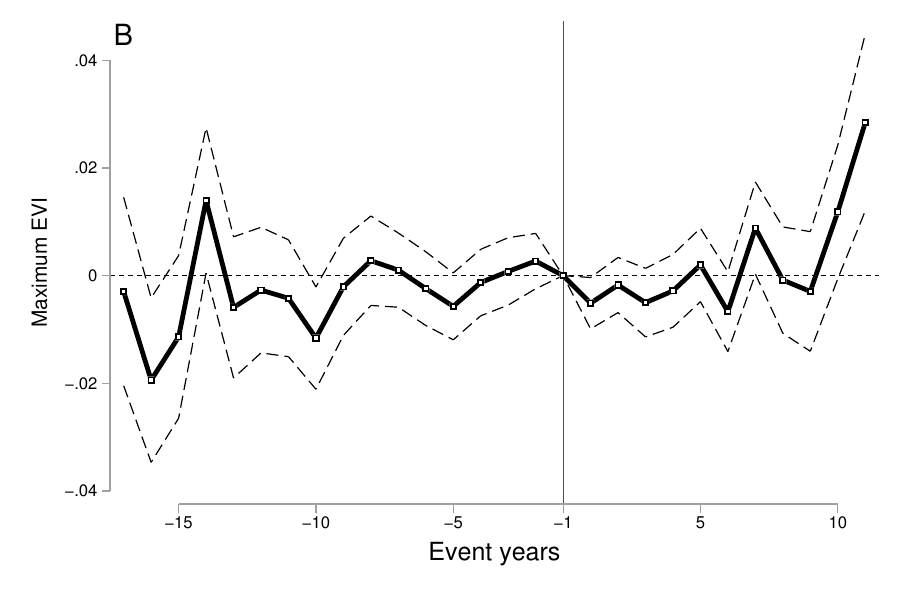}
			\includegraphics[width=.49\linewidth,keepaspectratio]{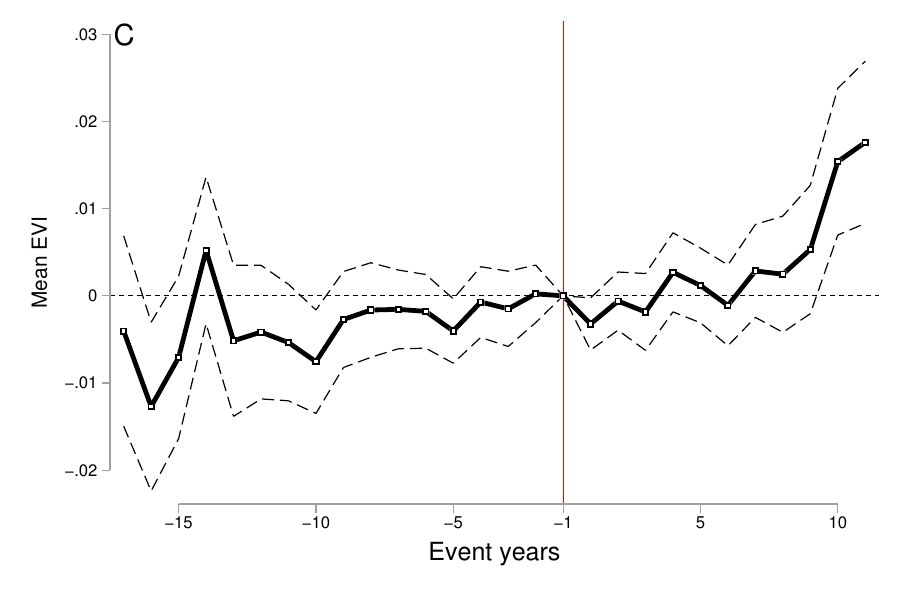}
			\includegraphics[width=.49\linewidth,keepaspectratio]{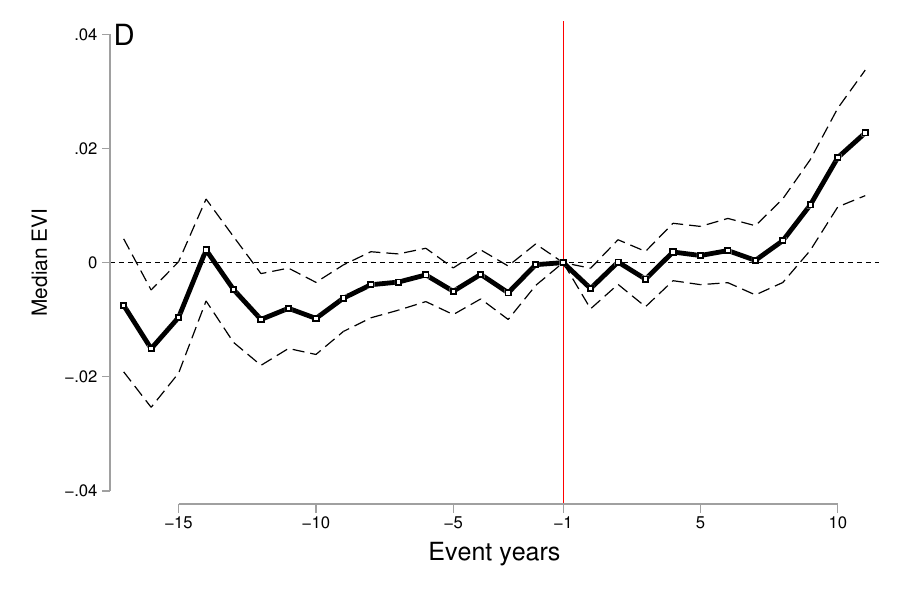}
		\end{center}
		\footnotesize  \textit{Note}: Figure displays coefficients from upazila-level event study regressions. The dependent variable for Panel A is cumulative EVI, for Panel B is max EVI, for Panel C is mean EVI, and Panel D reproduces the results in the paper using median EVI values. The solid line shows coefficient estimates from the model, with the event year (the year immediately prior to the upazila having access to STRVs, indicated as -1) as the excluded category. Dotted lines represent $95\%$ confidence intervals calculated using standard errors clustered at the district-level.
	\end{minipage}	
\end{figure}

Figure~\ref{fig:event_zil} presents results from the event study regressions. Relative to the district-level results, the upazila-level results show a stronger impact of STRVs. For all the EVI measures, a strong upward trend begins around six years after the introduction of STRVs. Results are particularly strong for cumulative, mean, and median EVI. 

\begin{landscape}
\begin{figure}[!htbp]
	\begin{minipage}{\linewidth}		
		\caption{Specification Charts of DID Results for Upazila}
		\label{fig:flood_days_did_zil}
		\begin{center}
			\includegraphics[width=.48\linewidth,keepaspectratio]{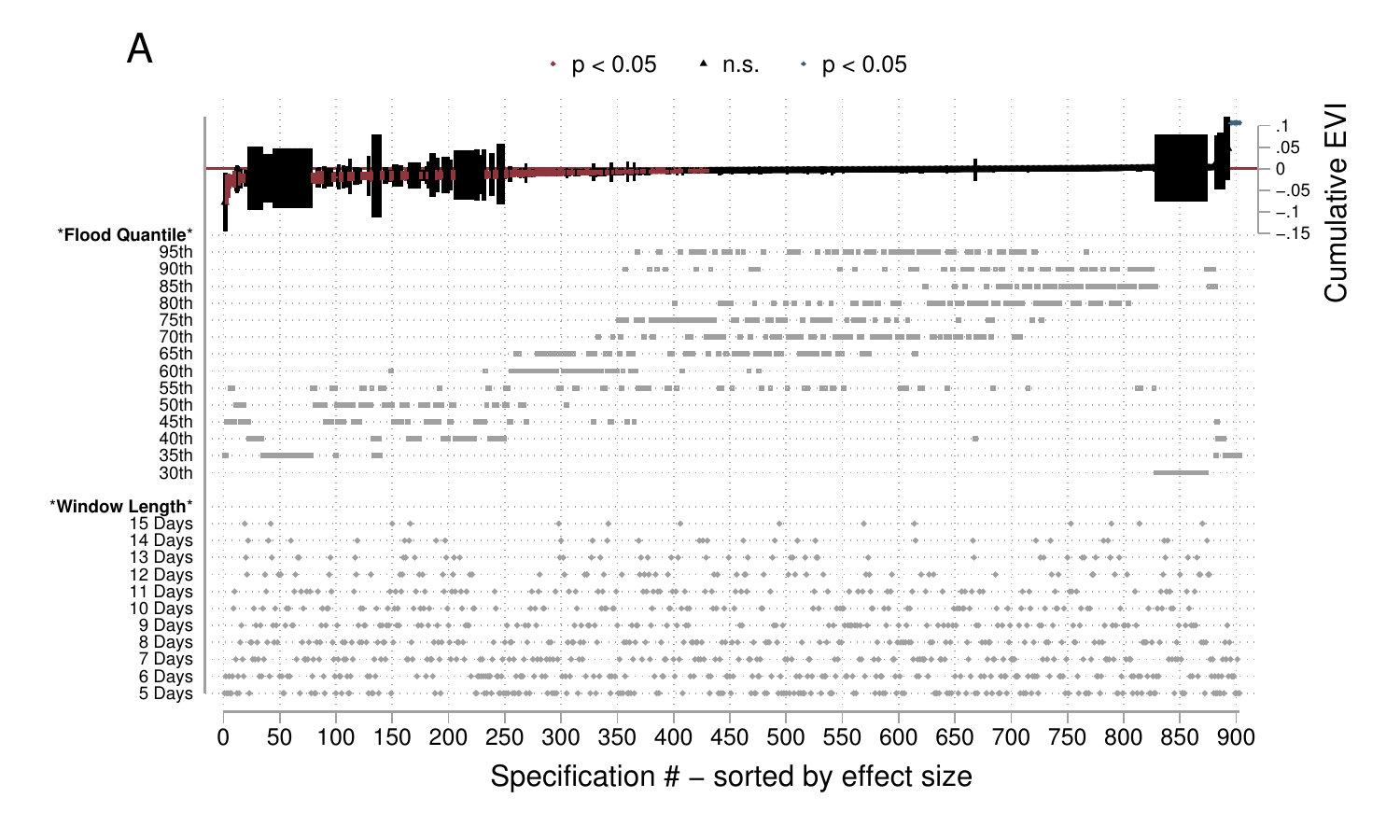}
			\includegraphics[width=.48\linewidth,keepaspectratio]{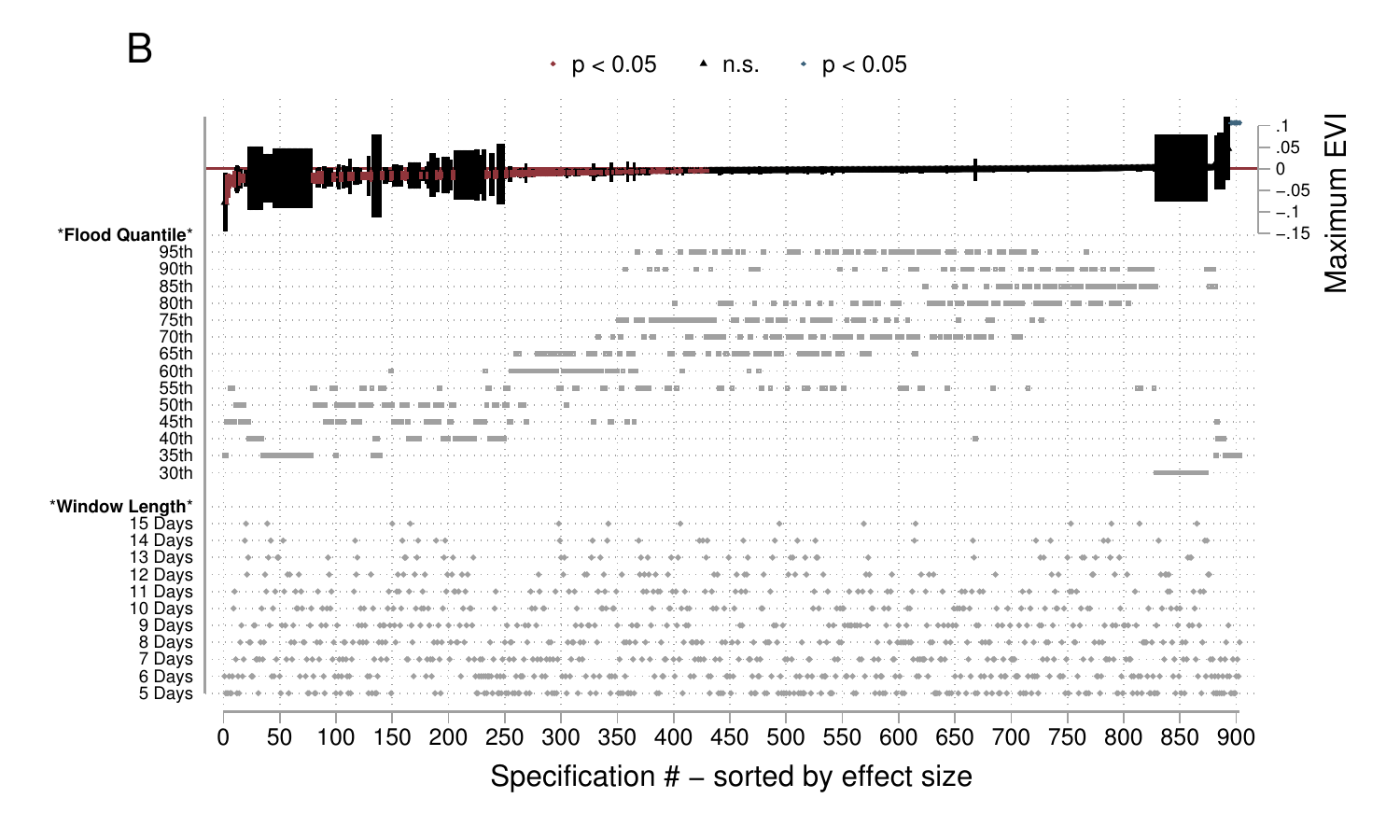}
			\includegraphics[width=.48\linewidth,keepaspectratio]{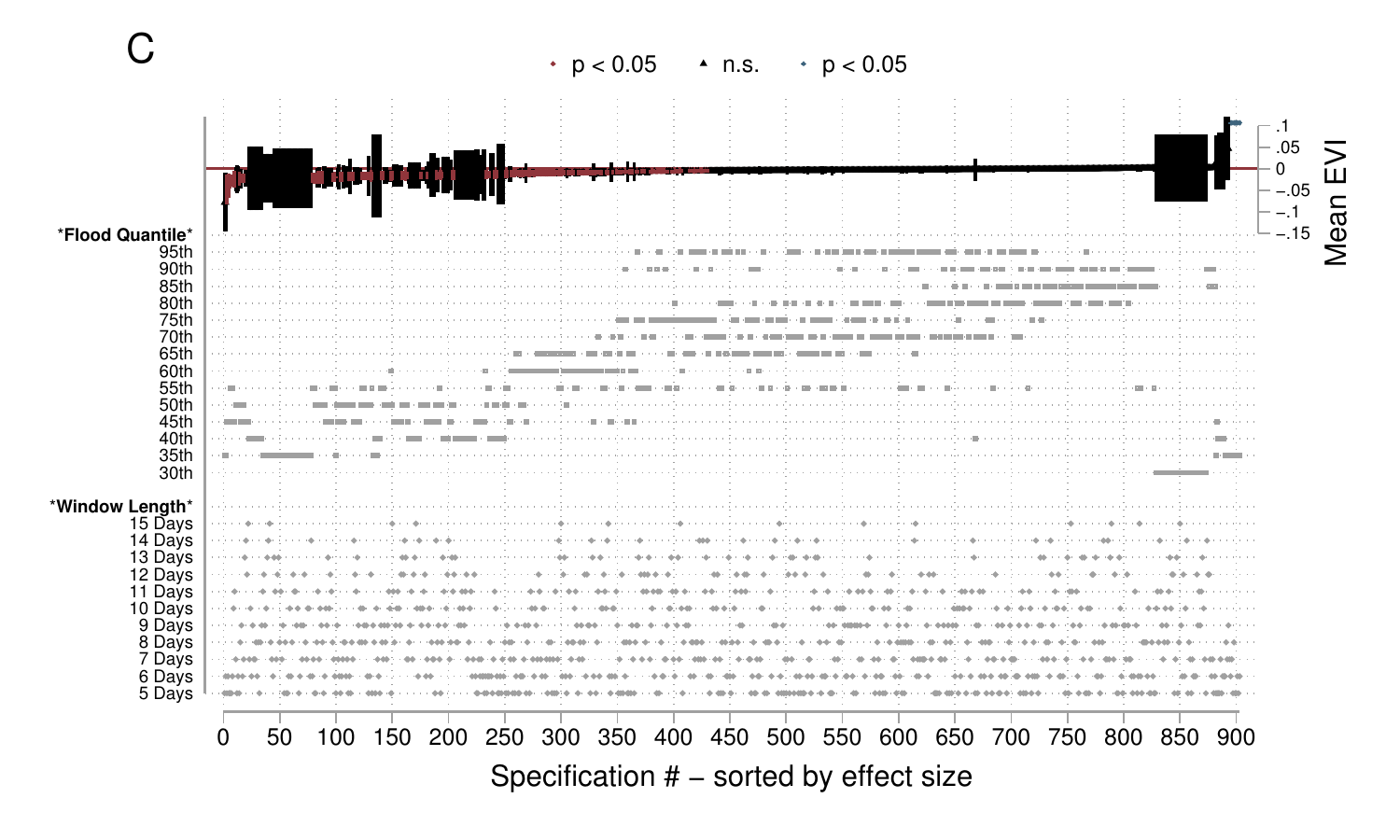}
			\includegraphics[width=.48\linewidth,keepaspectratio]{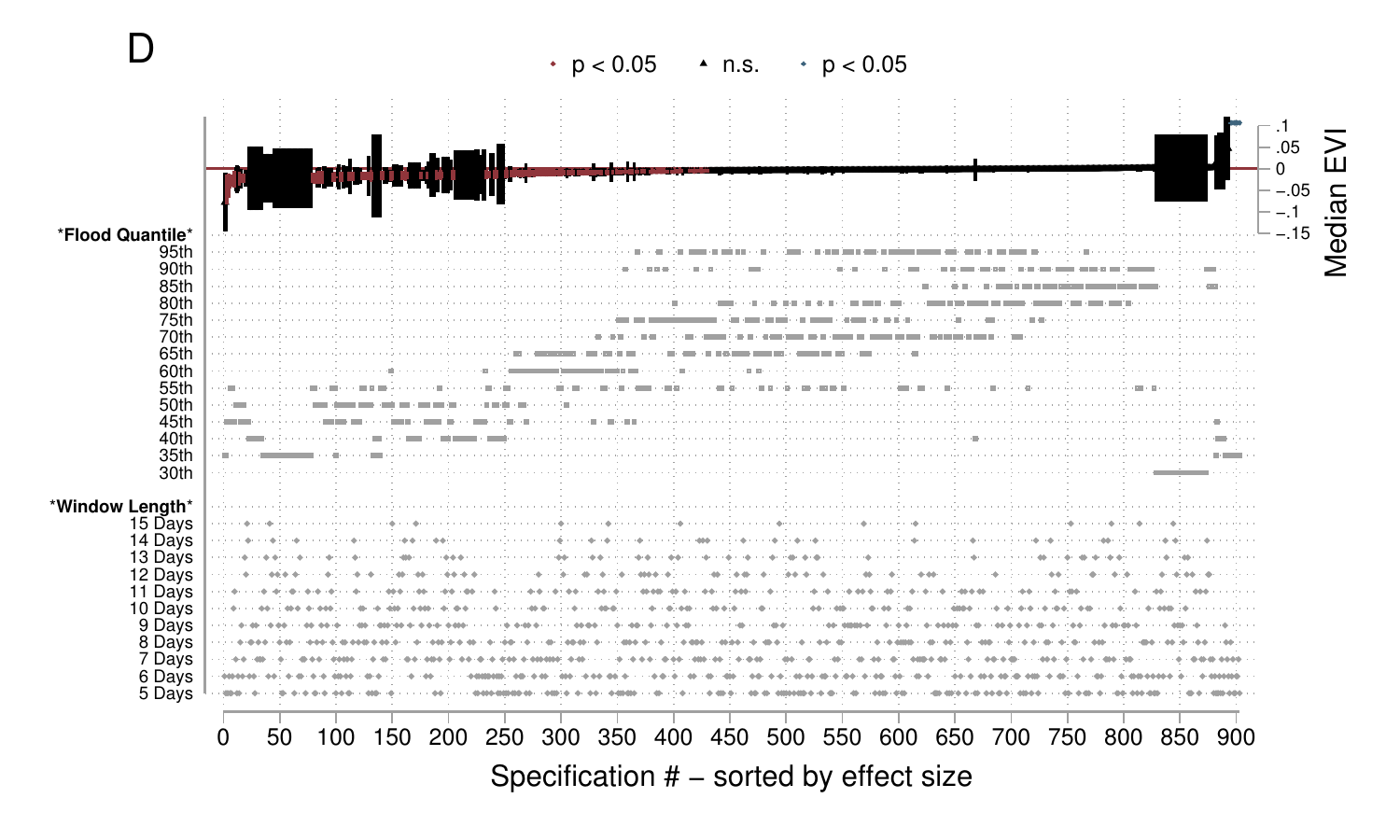}
		\end{center}
		\footnotesize  \textit{Note}: The figure presents results from upazila-level TWFE regressions. Each panel in the figure displays coefficient estimates and $95\%$ confidence intervals on the interaction of STRVs and a flood metric for a specific EVI measure (cumulative, max, mean, median). Specifications are sorted based on coefficient size from smallest (left) to largest (right). The gray diamonds below the coefficients indicate which combination of quantile and flood window was used in the regression.
	\end{minipage}	
\end{figure}
\end{landscape}

Figure~\ref{fig:flood_days_did_zil} presents results from the DID specification. Results using this method are much weaker at the upazila-level when compared to the district-level. While in the main results $13\%$ of potential flood metrics produce positive and significant results (using cumulative EVI), here less than $5\%$ of results are positive and significant, meaning we cannot discount that those significant results are purely due to chance. The few results that are positive and significant do still come from the same quantiles that produced the Goldilocks floods in our main results. But the clear take away is that our district-level DID results are not robust the disaggregation to the upazila-level. 

\begin{landscape}
\begin{figure}[!htbp]
	\begin{minipage}{\linewidth}		
		\caption{Specification Charts of TWFE Results Using Binary if Flooding Was in Window for Upazila}
		\label{fig:flood_days_bin_zil}
		\begin{center}
			\includegraphics[width=.48\linewidth,keepaspectratio]{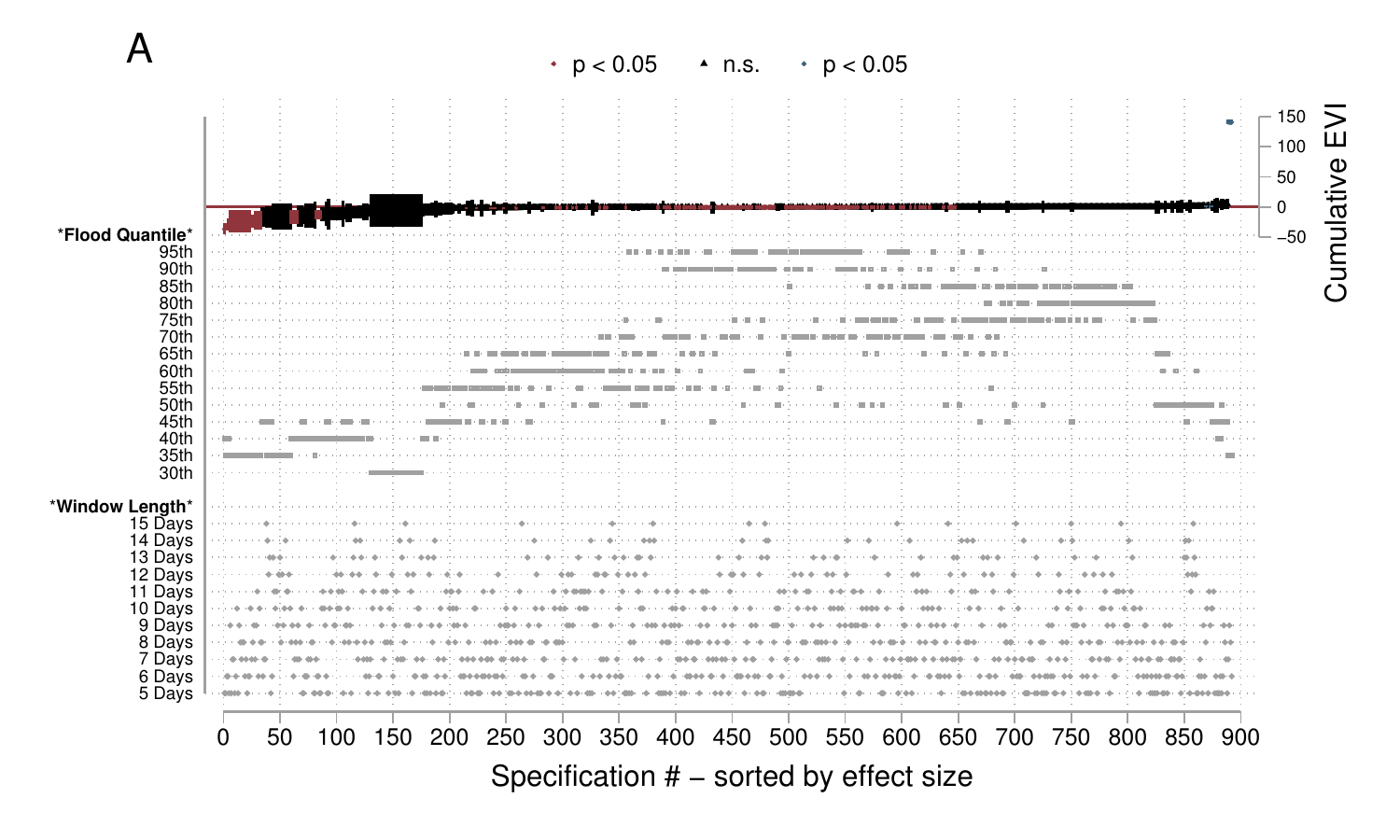}
			\includegraphics[width=.48\linewidth,keepaspectratio]{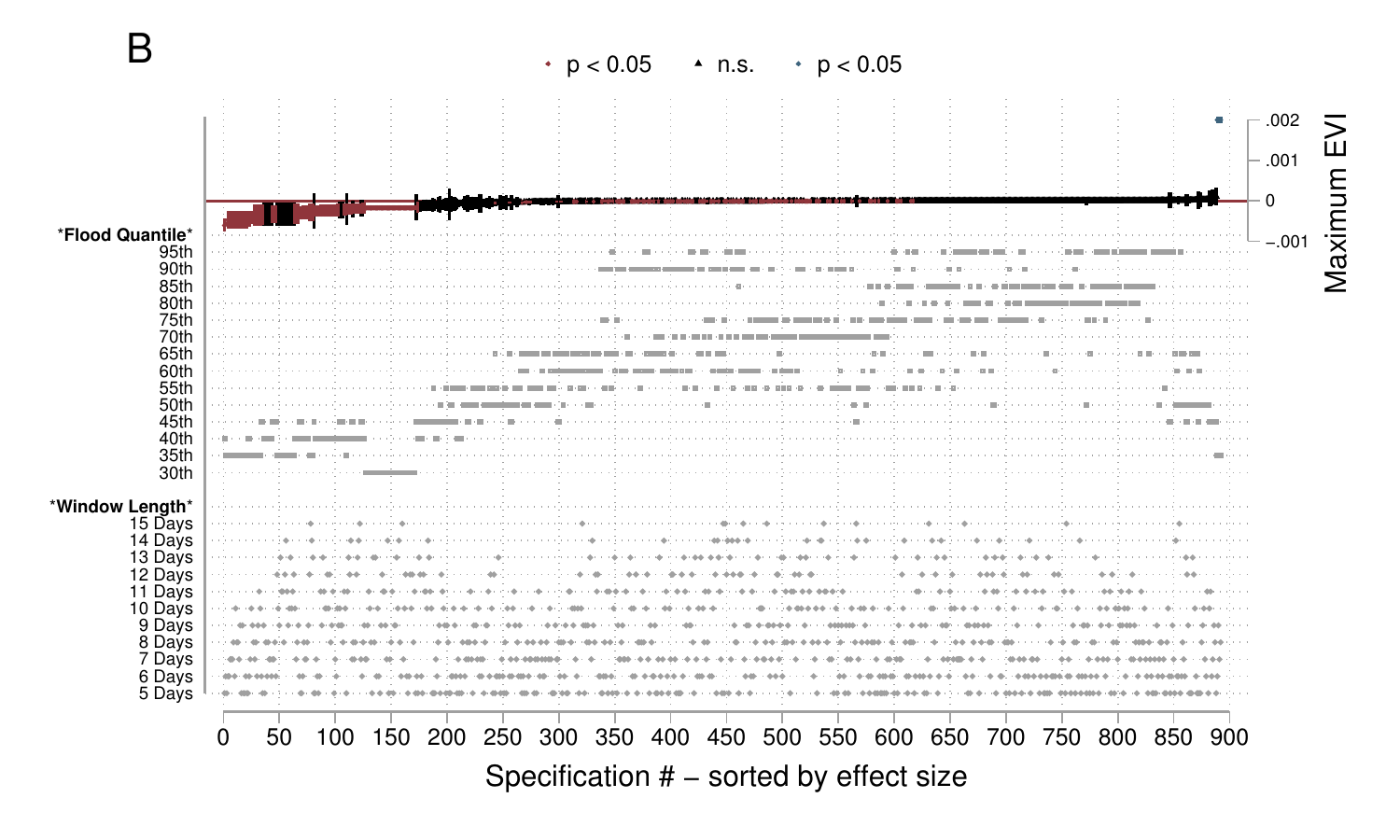}
			\includegraphics[width=.48\linewidth,keepaspectratio]{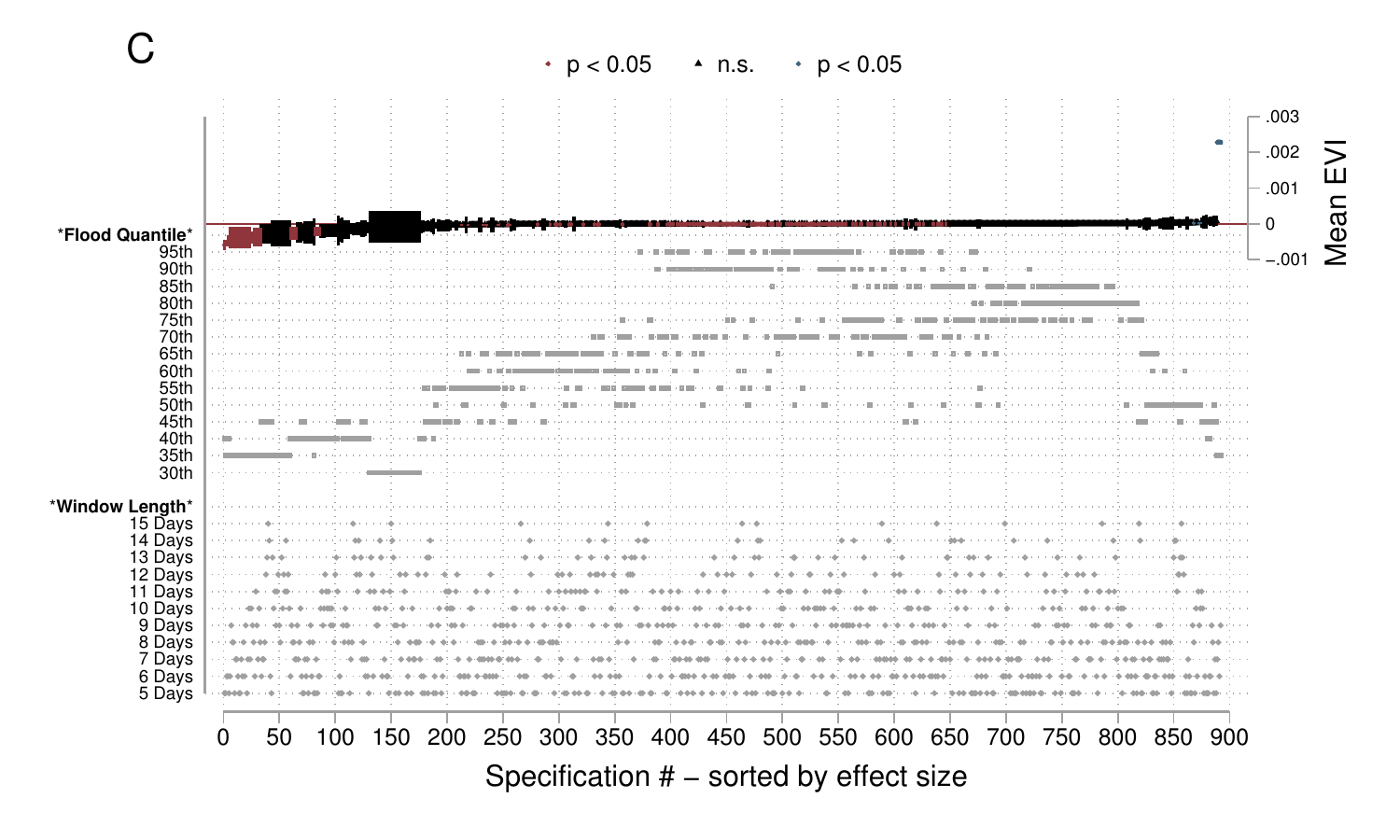}
			\includegraphics[width=.48\linewidth,keepaspectratio]{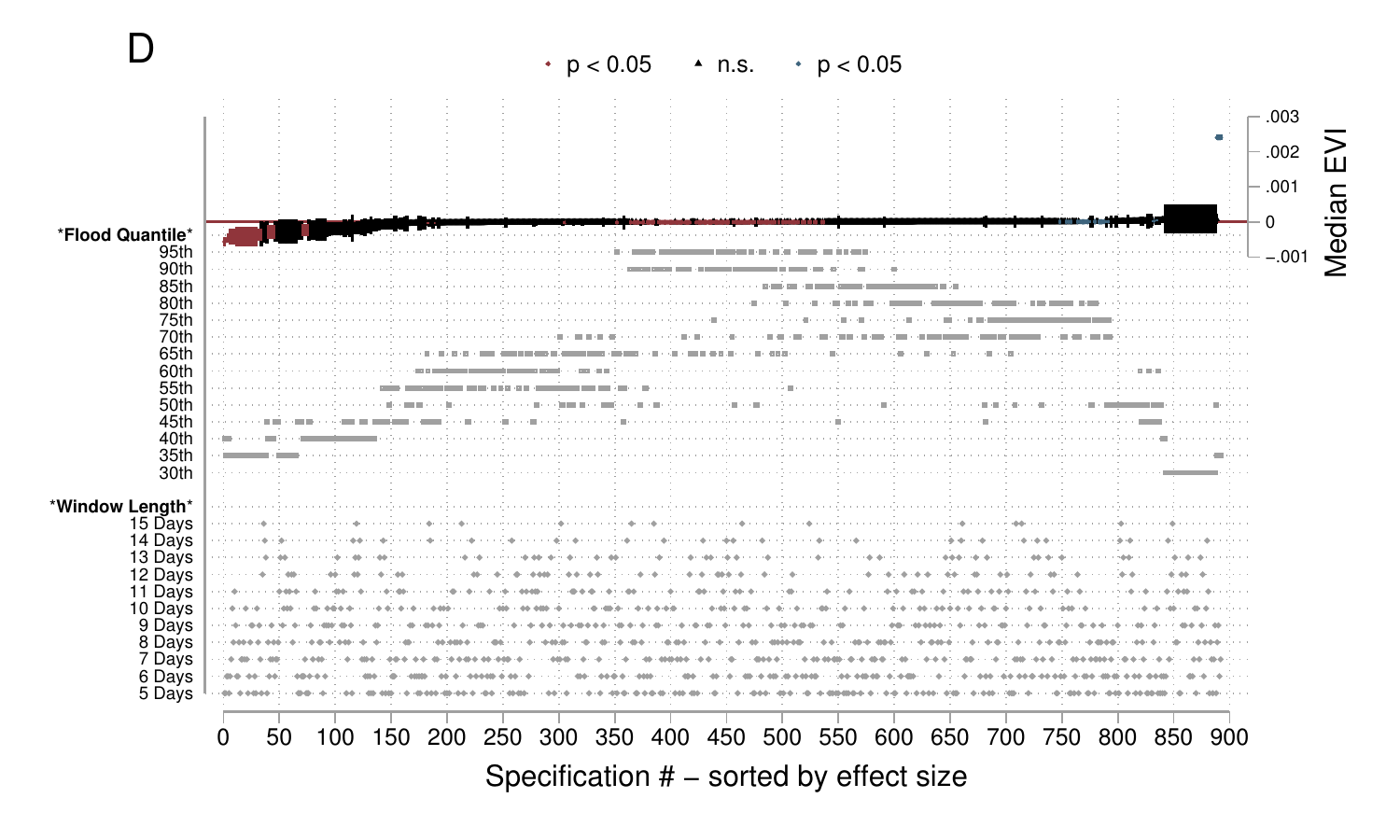}
		\end{center}
		\footnotesize  \textit{Note}: The figure presents results from upazila-level TWFE regressions. Each panel in the figure displays coefficient estimates and $95\%$ confidence intervals on the interaction of STRVs and a flood metric for a specific EVI measure (cumulative, max, mean, median). Specifications are sorted based on coefficient size from smallest (left) to largest (right). The gray diamonds below the coefficients indicate which combination of quantile and flood window was used in the regression.
	\end{minipage}	
\end{figure}
\end{landscape}

\begin{landscape}
\begin{figure}[!htbp]
	\begin{minipage}{\linewidth}		
		\caption{Specification Charts of TWFE Results Using Days in Flood Window for Upazila}
		\label{fig:flood_days_win_zil}
		\begin{center}
			\includegraphics[width=.48\linewidth,keepaspectratio]{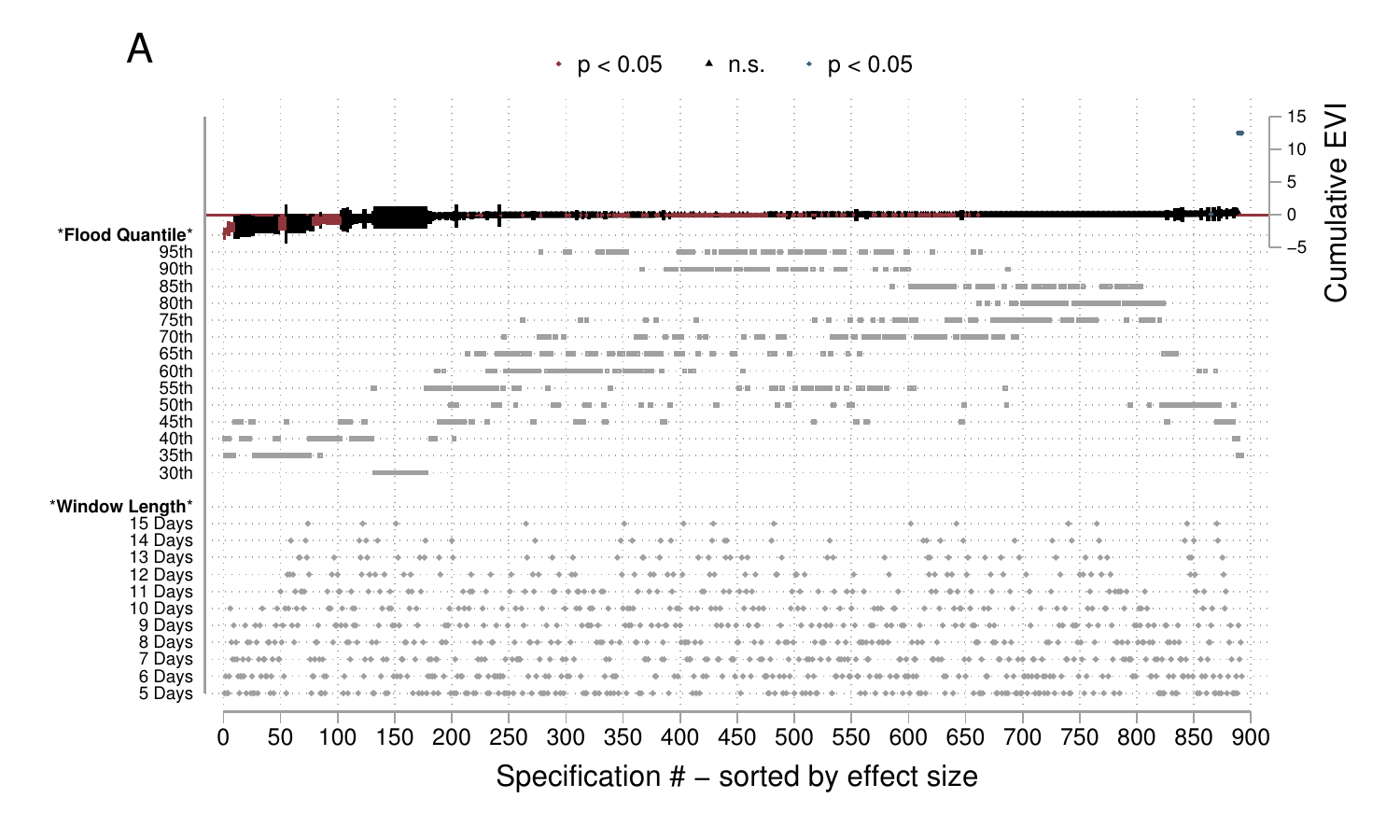}
			\includegraphics[width=.48\linewidth,keepaspectratio]{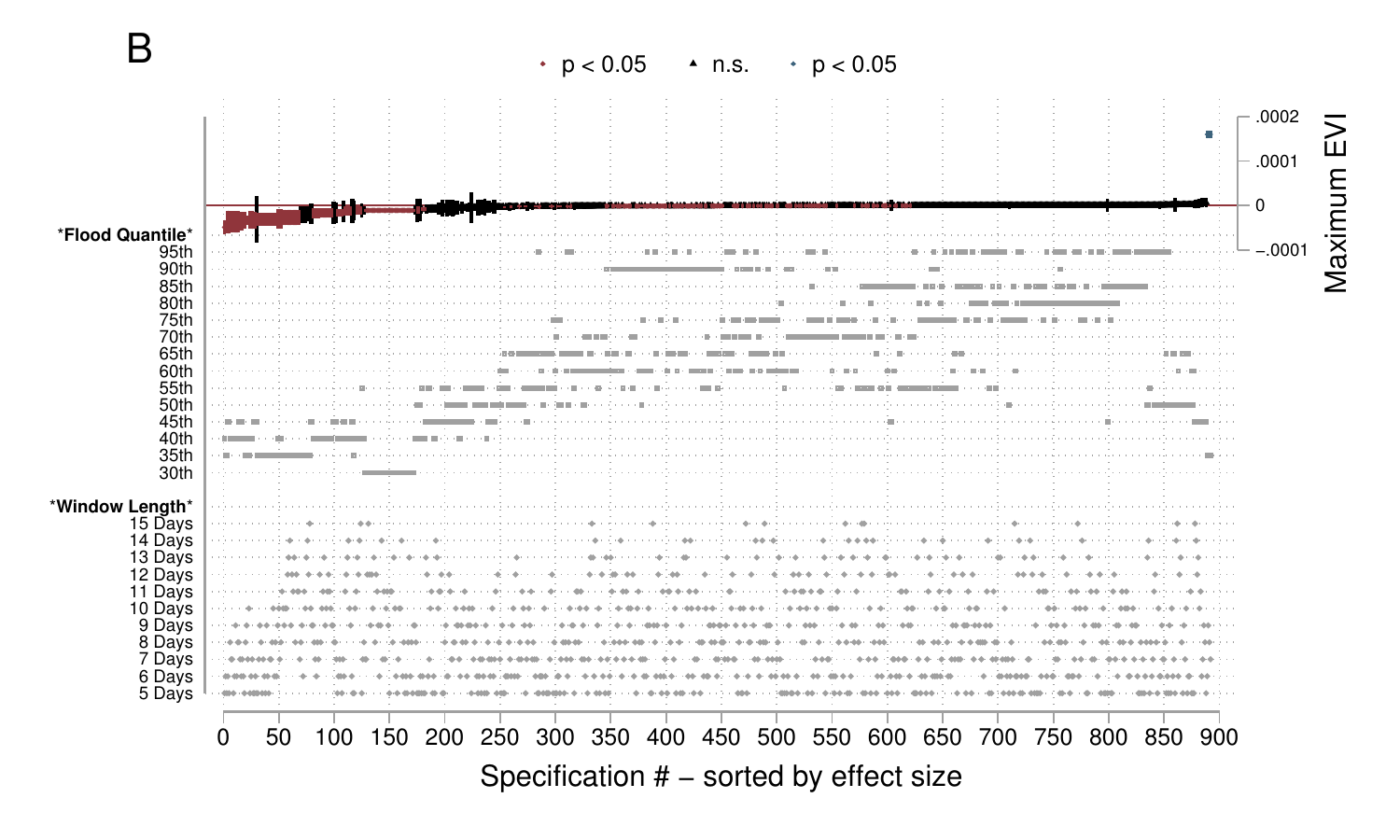}
			\includegraphics[width=.48\linewidth,keepaspectratio]{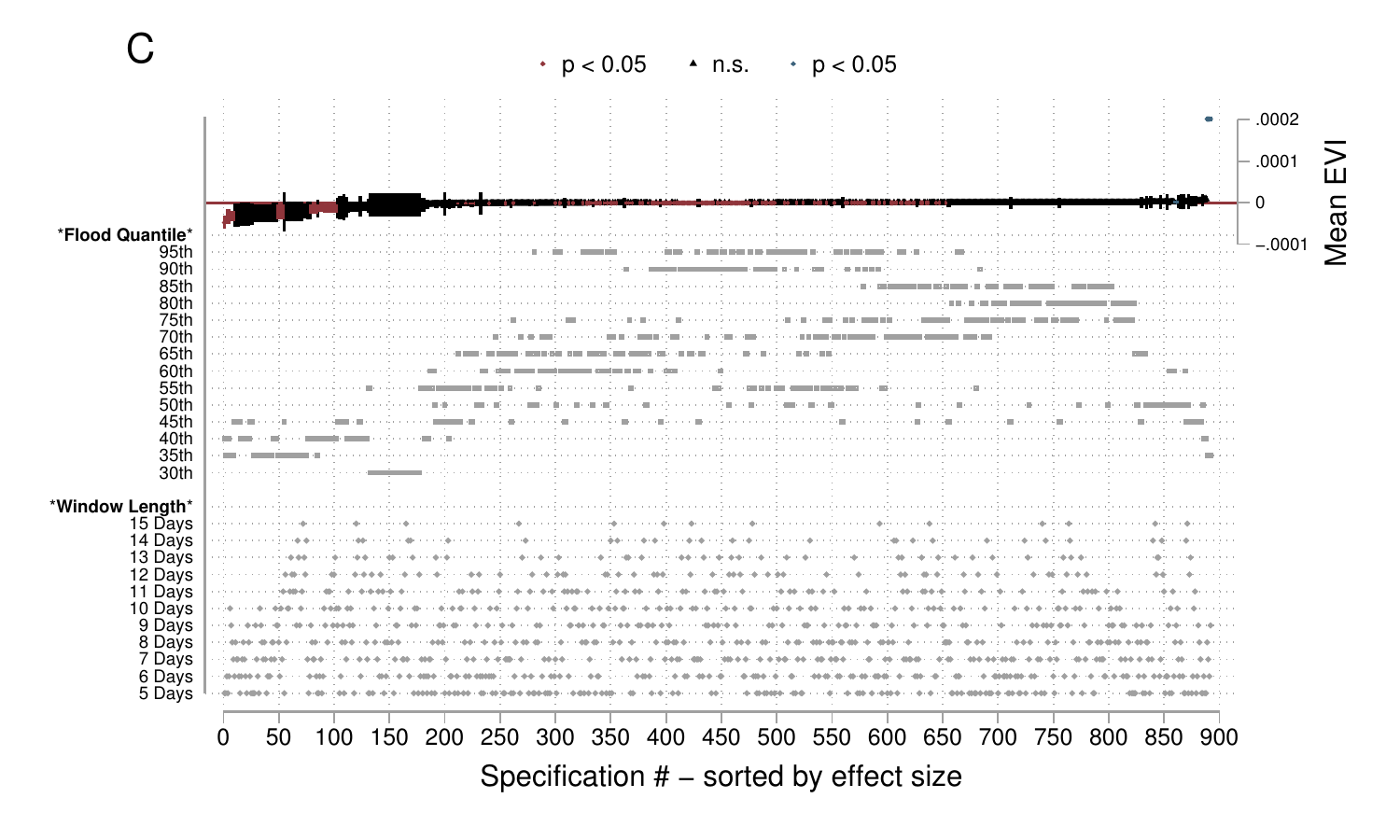}
			\includegraphics[width=.48\linewidth,keepaspectratio]{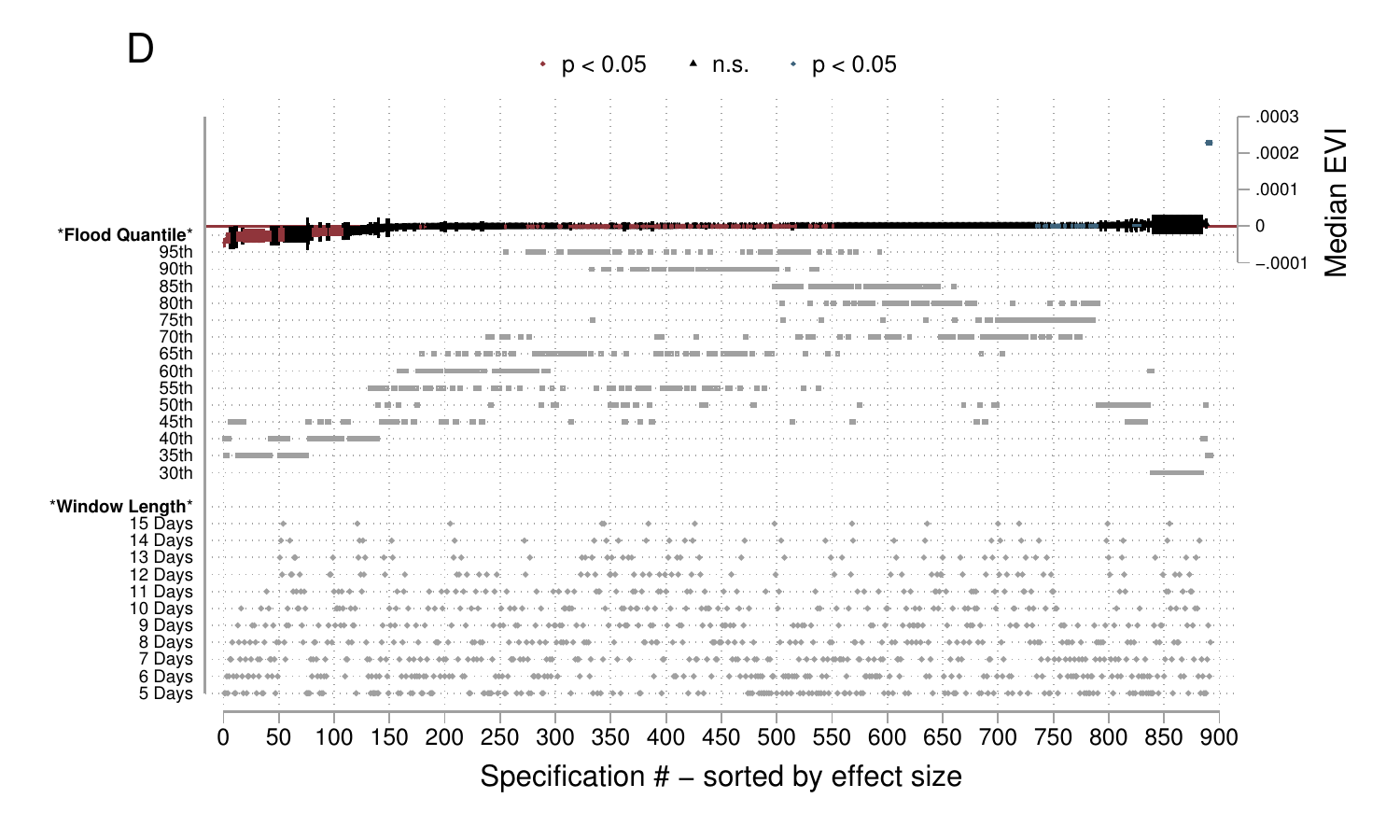}
		\end{center}
		\footnotesize  \textit{Note}: The figure presents results from upazila-level TWFE regressions. Each panel in the figure displays coefficient estimates and $95\%$ confidence intervals on the interaction of STRVs and a flood metric for a specific EVI measure (cumulative, max, mean, median). Specifications are sorted based on coefficient size from smallest (left) to largest (right). The gray diamonds below the coefficients indicate which combination of quantile and flood window was used in the regression.
	\end{minipage}	
\end{figure}
\end{landscape}

Turning to our TWFE regressions, Figure~\ref{fig:flood_days_bin_zil} reports results using a binary indicator for if flooding fell within the window and Figure~\ref{fig:flood_days_win_zil} reports results using the number of days in the flood window. Results are very similar to the DID results. The vast majority of regressions are not significant or negative and significant. As we mention in the main paper, one reason the non-event study results might not be robust to using upazila-level EO EVI and flood data is that the seed data is still at the district-level. This mismatch of units means that we are trying to explain intra-district variation in EVI using a single district-wide seed value. This type of mismatch can introduce noise into estimates and in fact we do find confidence intervals to be much larger in the upazila-level results. Coefficients of similar size in district-level and upazila-level specification charts have much larger confidence intervals in the upazila-level results than in the district-level results.

Whatever the reason, based on the preponderance of evidence, we conclude that our results are not robust to using district-level seed data but upazila-level EVI and flood data. Our main results should be interpreted with this lack of robustness in mind.


\newpage

\newpage

\newpage
\FloatBarrier
\setcounter{table}{0}
\renewcommand{\thetable}{F\arabic{table}}
\setcounter{figure}{0}
\renewcommand{\thefigure}{F\arabic{figure}}


\section{Household-Level Robustness Checks} \label{sec:plot_app}

A third potential concern regarding our results may be driven by our use of EVI as a proxy for yields and seed dissemination data as a proxy for adoption. Rice is not cultivated at the district-level but by individual households on small fields. District-level EO measures may be noisy or biased proxies of the on-the-ground household reality. To demonstrate robustness of our district-level results, we estimate TWFE regressions using household data on yields and adoption and EO data on floods experienced by that households. In this appendix we discuss the value and limitations of the three-year Rice Monitoring Survey (RMS) panel as well as summary statistics from the data and finally results of robustness checks. 


\subsection{Household Survey Data}

The RMS was a Gates Foundation-funded project designed to capture varietal turnovers over time in Bangladesh, India, and Nepal. The data was originally collected as two waves of a panel in 2014 and 2017 \citep{rms-2017}. Households were selected following a clustered random sampling procedure to ensure the overall survey was representative of rice growing regions of each country. The plan was to collect a third wave in 2020, but data collection efforts were delayed due to COVID until 2022 and only Bangladesh was revisited. In total, 1,500 households were part of the initial sampling frame in Bangladesh. The RMS was able to follow-up with 1,490 households in 2017 and 1,484 households in 2022. This gives us an attrition rate of $1.7\%$ over eight years. 

As the RMS was designed to monitor varietal turnover, the survey instrument is heavily focused on collecting information on rice production at the crop and plot level. While the data is a panel of households, limited information was collected about household well-being. These include demographics, asset ownership, membership with different farming organizations, farmer opinion on the prevalence of different stresses, such as flood, drought, and salinity, and the availability of different rice varieties. No information was collected on consumption, income, or other measures of welfare, such as food security or women's empowerment.
                
In terms of production data, the survey collected information by both plot and crop so as to be able to distinguish between different varieties of rice grown on the same plot. Plot data includes information on plot size, ownership, and land type. Crop data, within a plot, includes which rice variety was planted in which season, methods of planting, damage of crops by different abiotic shocks, input use, total production, and disposition of the harvest. 
            
We construct an unbalanced panel of households that cultivated rice in \emph{Aman} season. This gives us a total of 3,865 observations from 1,488 households. Our primary outcome of interest is rice yield at the household level, which we construct by summing up rice harvest (kg) on all plots and dividing by the sum of area (ha) for all rice plot. We consider a household as having adopted STRVs if they report planting any flood tolerant rice variety on at least one plot. To measure the incidence of flooding faced by the household, we combine household GPS location with the EO flood maps we generate. While we would ideally use plot locations to match with flood data, only household GPS locations were collected in 2014 and 2017. Thus, we only know the precise location of plots in 2022.

Table~\ref{tab:strv_sum_hh_year} provides summary statistics for key variables across the three years of the panel. The first three columns of the table report means and standard deviations for key variables in each year. The final three columns of the table report $p$-values on year-to-year comparisons of means using a $t$-test that allows for unequal variance. Statistically significant differences exist for almost every variable in almost every year-to-year comparison. Yields in 2014 and 2017 were around $3.7$ tons per hectare and increased to $4.1$ tons per hectare in 2022. Adoption was flat at eight percent of households for the first two years in the panel (in 2014, $111$ households adopted STRVs, in 2017, $112$ households adopted). Adoption grew substantially in the intervening five years, with $21\%$ ($234$ households) adopting STRVs in 2022. Rice area has steadily declined over time and while statistically significant, the magnitude of the decline is not very large. In 2014, households had $0.72$ hectares in rice, which fell by a tenth of a hectare to $0.62$ in 2022.

Table~\ref{tab:strv_sum_hh} presents summary statistics by adoption status and for the dataset overall. Households that adopt STRVs report higher yields in the years that they adopt compared to households not growing STRVs in a given year. Households can and do dis-adopt STRVs, though given how low adoption rates were in 2014 and 2017, dis-adoption is much less common that adoption. Recall also that we classify a household as adopting if they use STRVs on at least one of their rice plot. Thus, yields for adopters includes yields from both STRV and non-STRV seeds. The difference in yields between adopters and non-adopters suggests that the adoption decision is not random and may be correlated with household or farm unobservables, such as farmer ability or quality and location of plots. That said, we do not see significant differences in terms of rice area between adopters and not adopters.


\subsection{Household Empirical Method} 

In the case of the household data, a event study or DID does not make sense because treatment (adoption) can turn on and off again as households choose to adopt or dis-adopt. So we must limit our analysis to a TWFE regression similar to our district-level framework. We estimate the effects of STRV adoption by a household on rice yields as:

    \begin{equation} \label{eq:twfe_hh}
        \ln \textrm{yield}_{it} = \beta(\textrm{STRV}_{it} \cdot \textrm{flood}_{it}) + \rho \textrm{STRV}_{it} + \gamma \textrm{flood}_{it} + \mu_{i} + \mu_{t} + \epsilon_{it}
    \end{equation}

\noindent where $i$ indexes households and $t$ indexes time. The two terms $\mu_{i}$ and $\mu_{t}$ denote household and time fixed effects, meaning that only within-household time variation in relative yields remains. The household fixed effects control for all household-specific time-invariant variation and the time fixed effects control for all time-specific household-invariant variation.

In terms of identification, model~\eqref{eq:twfe_hh}, like model~\eqref{eq:twfe} relies on household and time fixed effects to identify changes in a household's yields resulting from a change in the household's adoption status. The model is identified only if the adoption decision is not drive by household-specific, time-varying events. We believe that adoption is plausibly exogenous and primarily driven either by time-invariant household characteristics like risk preferences and farming skill or by availability of seed, which changes over time but not uniquely for a household. In all models we cluster standard errors at the level of the unit of observation.


\subsection{Econometric Evidence} \label{sec:twfe}

Figure~\ref{fig:flood_days_hh} reports results using a binary indicator for if flooding fell within the window (panel A) and using the number of days in the flood window (panel B). Unlike results from our disaggregation to upazila, the household-level results are similar to the district-level TWFE results. Again, the vast majority of regressions are not significant or negative and significant. In both panels, slightly more than $7\%$ of regressions produce positive and significant results. The flood metrics that matter for the household, though, are from lower quantiles than at the district level, meaning that when we use the pixel-level fractional flood index (as opposed to district-level averages of the pixel-level data) a greater degree of inundation above baseflow is required in order to generate a Goldilocks flood.

In terms of checking the robustness of our main results, the preponderance of evidence from the household-level analysis supports our conclusions from the district-level EO analysis. STRVs have a positive and significant effect on yields when flooding is ``just right.'' At a practical level, for those looking to address the Goldilocks problem for other agricultural technologies, our household- v.s. district-level findings suggest that what qualifies as a ``just right'' stress event will vary based on the unit of analysis and the aggregation (or disaggregation) of the data.


\begin{table}[htbp]	\centering
    \caption{Summary Statistics by Different Year in Household Panel} \label{tab:strv_sum_hh_year}
	\scalebox{1}
	{ \setlength{\linewidth}{.2cm}\newcommand{\contents}
		{\begin{tabular}{l*{7}{c}} \\ [-1.8ex]\hline \hline \\[-1.8ex] 
& \multicolumn{1}{c}{2014} & \multicolumn{1}{c}{2017} &  
\multicolumn{1}{c}{2022} & \multicolumn{1}{c}{2014-2017}  &
\multicolumn{1}{c}{2017-2022} & \multicolumn{1}{c}{2014-2022}  \\ 
\midrule
Yield (kg/ha)       &        3787&        3637&        4172&       0.009&       0.000&       0.000\\
                    &      (1378)&      (1636)&      (1623)&            &            &            \\
STRV Adoption       &       0.081&       0.081&       0.213&       0.997&       0.000&       0.000\\
                    &     (0.272)&     (0.272)&     (0.409)&            &            &            \\
Rice Area (ha)      &       0.720&       0.697&       0.619&       0.468&       0.005&       0.003\\
                    &     (0.950)&     (0.619)&     (0.750)&            &            &            \\
 \midrule Observations & 1376 & 1389 & 1100 & & & &  \\ 
Households & 1376 & 1389 & 1100 & & & & \\ 
\hline \hline \\[-1.8ex] \multicolumn{8}{J{\linewidth}}{\small 
\noindent \textit{Note}: The table displays means for key variables 
by panel year. Standard deviations are in 
parentheses. We also report $p$-values 
on a $t$-test for the equality of means between adopters and non-adopters.}  \end{tabular}
}
	\setbox0=\hbox{\begin{tabular}{l*{4}{c}} \\ [-1.8ex]\hline \hline \\[-1.8ex] 
& \multicolumn{1}{c}{Non-Adoption} & \multicolumn{1}{c}{STRV Adoption} &  
\multicolumn{1}{c}{p-value} & \multicolumn{1}{c}{Total}  \\ 
\midrule
Yield (kg/ha)       &        3758&        4469&       0.000&        3842\\
                    &      (1562)&      (1395)&            &      (1560)\\
Rice Area (ha)      &       0.678&       0.719&       0.256&       0.683\\
                    &     (0.797)&     (0.708)&            &     (0.787)\\
 \midrule Observations & 3408 &  457 & & 3865 \\ 
Households & 1460 &  370 & & 1488 \\ 
\hline \hline \\[-1.8ex] \multicolumn{5}{J{\linewidth}}{\small 
\noindent \textit{Note}: The table displays means for key variables 
by adoption status and for the total sample. Standard deviations are in 
parentheses. We also report $p$-values 
on a $t$-test for the equality of means between adopters and non-adopters.}  \end{tabular}
}
    \setlength{\linewidth}{\wd0-2\tabcolsep-.25em}
    \begin{tabular}{l*{4}{c}} \\ [-1.8ex]\hline \hline \\[-1.8ex] 
& \multicolumn{1}{c}{Non-Adoption} & \multicolumn{1}{c}{STRV Adoption} &  
\multicolumn{1}{c}{p-value} & \multicolumn{1}{c}{Total}  \\ 
\midrule
Yield (kg/ha)       &        3758&        4469&       0.000&        3842\\
                    &      (1562)&      (1395)&            &      (1560)\\
Rice Area (ha)      &       0.678&       0.719&       0.256&       0.683\\
                    &     (0.797)&     (0.708)&            &     (0.787)\\
 \midrule Observations & 3408 &  457 & & 3865 \\ 
Households & 1460 &  370 & & 1488 \\ 
\hline \hline \\[-1.8ex] \multicolumn{5}{J{\linewidth}}{\small 
\noindent \textit{Note}: The table displays means for key variables 
by adoption status and for the total sample. Standard deviations are in 
parentheses. We also report $p$-values 
on a $t$-test for the equality of means between adopters and non-adopters.}  \end{tabular}
}
\end{table}  

\begin{table}[htbp]	\centering
    \caption{Summary Statistics by STRV Adoption in Household Panel} \label{tab:strv_sum_hh}
	\scalebox{1}
	{ \setlength{\linewidth}{.2cm}\newcommand{\contents}
		{\begin{tabular}{l*{4}{c}} \\ [-1.8ex]\hline \hline \\[-1.8ex] 
& \multicolumn{1}{c}{Non-Adoption} & \multicolumn{1}{c}{STRV Adoption} &  
\multicolumn{1}{c}{p-value} & \multicolumn{1}{c}{Total}  \\ 
\midrule
Yield (kg/ha)       &        3758&        4469&       0.000&        3842\\
                    &      (1562)&      (1395)&            &      (1560)\\
Rice Area (ha)      &       0.678&       0.719&       0.256&       0.683\\
                    &     (0.797)&     (0.708)&            &     (0.787)\\
 \midrule Observations & 3408 &  457 & & 3865 \\ 
Households & 1460 &  370 & & 1488 \\ 
\hline \hline \\[-1.8ex] \multicolumn{5}{J{\linewidth}}{\small 
\noindent \textit{Note}: The table displays means for key variables 
by adoption status and for the total sample. Standard deviations are in 
parentheses. We also report $p$-values 
on a $t$-test for the equality of means between adopters and non-adopters.}  \end{tabular}
}
	\setbox0=\hbox{\begin{tabular}{l*{4}{c}} \\ [-1.8ex]\hline \hline \\[-1.8ex] 
& \multicolumn{1}{c}{Non-Adoption} & \multicolumn{1}{c}{STRV Adoption} &  
\multicolumn{1}{c}{p-value} & \multicolumn{1}{c}{Total}  \\ 
\midrule
Yield (kg/ha)       &        3758&        4469&       0.000&        3842\\
                    &      (1562)&      (1395)&            &      (1560)\\
Rice Area (ha)      &       0.678&       0.719&       0.256&       0.683\\
                    &     (0.797)&     (0.708)&            &     (0.787)\\
 \midrule Observations & 3408 &  457 & & 3865 \\ 
Households & 1460 &  370 & & 1488 \\ 
\hline \hline \\[-1.8ex] \multicolumn{5}{J{\linewidth}}{\small 
\noindent \textit{Note}: The table displays means for key variables 
by adoption status and for the total sample. Standard deviations are in 
parentheses. We also report $p$-values 
on a $t$-test for the equality of means between adopters and non-adopters.}  \end{tabular}
}
    \setlength{\linewidth}{\wd0-2\tabcolsep-.25em}
    \begin{tabular}{l*{4}{c}} \\ [-1.8ex]\hline \hline \\[-1.8ex] 
& \multicolumn{1}{c}{Non-Adoption} & \multicolumn{1}{c}{STRV Adoption} &  
\multicolumn{1}{c}{p-value} & \multicolumn{1}{c}{Total}  \\ 
\midrule
Yield (kg/ha)       &        3758&        4469&       0.000&        3842\\
                    &      (1562)&      (1395)&            &      (1560)\\
Rice Area (ha)      &       0.678&       0.719&       0.256&       0.683\\
                    &     (0.797)&     (0.708)&            &     (0.787)\\
 \midrule Observations & 3408 &  457 & & 3865 \\ 
Households & 1460 &  370 & & 1488 \\ 
\hline \hline \\[-1.8ex] \multicolumn{5}{J{\linewidth}}{\small 
\noindent \textit{Note}: The table displays means for key variables 
by adoption status and for the total sample. Standard deviations are in 
parentheses. We also report $p$-values 
on a $t$-test for the equality of means between adopters and non-adopters.}  \end{tabular}
}
\end{table}

\begin{figure}[!htbp]
	\begin{minipage}{\linewidth}		
		\caption{Specification Charts of TWFE Results}
		\label{fig:flood_days_hh}
		\begin{center}
			\includegraphics[width=.85\linewidth,keepaspectratio]{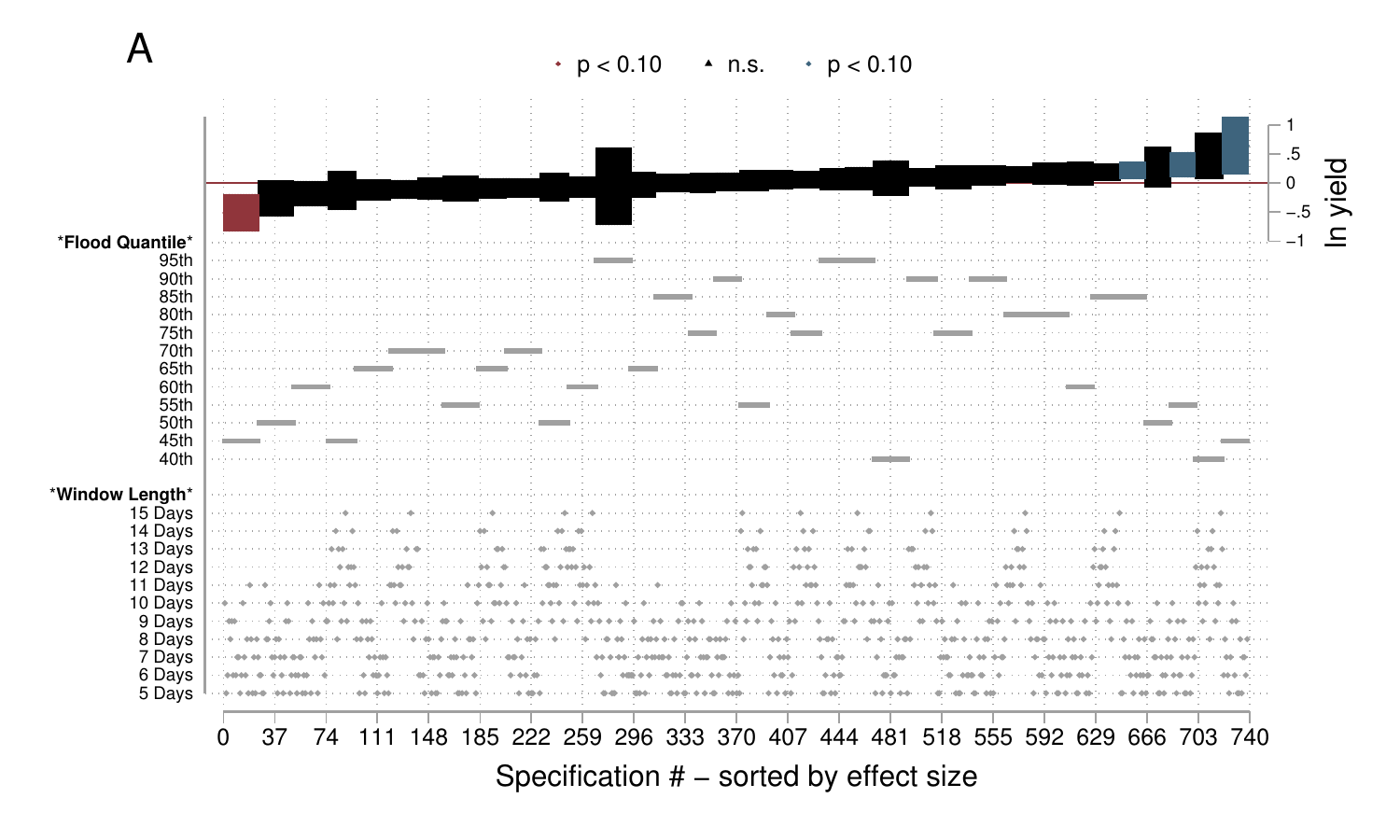}
			\includegraphics[width=.85\linewidth,keepaspectratio]{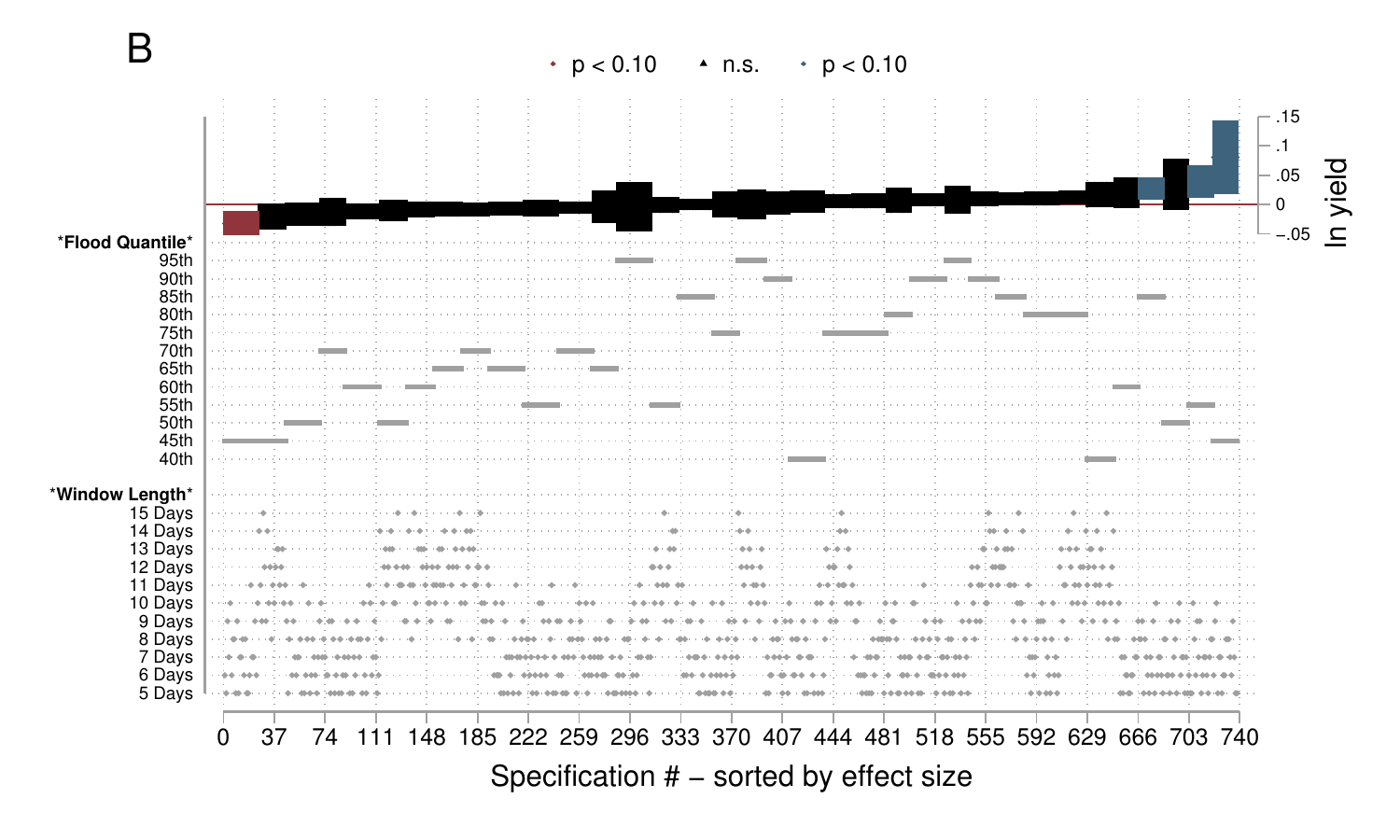}
		\end{center}
		\footnotesize  \textit{Note}: The figure presents results from household-level TWFE regressions. The figure displays coefficient estimates and $95\%$ confidence intervals on the interaction of household adoption of STRVs and a flood metric on log yield. The top panel reports on results using a binary if flooding was in the window while the bottom panel reports on results using the number of days in the window. Specifications are sorted based on coefficient size from smallest (left) to largest (right). The gray diamonds below the coefficients indicate which combination of quantile and flood window was used in the regression.
	\end{minipage}	
\end{figure}

\end{document}